\documentclass[twocolumn,preprintnumbers,superscriptaddress,amsmath,amssymb,10pt,a4paper]{revtex4-1}

\usepackage{geometry}
\usepackage{geometry}
\usepackage{url}
\usepackage{mathtools}
\usepackage{graphicx}
\usepackage{dcolumn}
\usepackage{bm}
\usepackage{makecell}
\usepackage{xcolor}
\usepackage{enumerate}
\usepackage[caption=false]{subfig}
\usepackage{float}
\usepackage{xr-hyper} 
\usepackage[unicode=true,pdfusetitle,
 bookmarks=true,bookmarksnumbered=false,bookmarksopen=false,
 breaklinks=false,pdfborder={0 0 1},backref=false,colorlinks=true]
 {hyperref}
\hypersetup{linkcolor=blue,urlcolor=blue,citecolor=blue}
\usepackage{natbib}
\usepackage{physics}
\usepackage{placeins}
\usepackage{titlesec}
\usepackage{multirow}

\usepackage{hyperref}
\usepackage{float}
\usepackage{geometry}
\usepackage{amsmath}

\usepackage{physics}

\begin{document}

\preprint{}
\title{Higher-order interactions in random Lotka--Volterra communities}

\author{Laura Sidhom}
\email{laura.sidhom@postgrad.manchester.ac.uk}
\affiliation{Theoretical Physics, Department of Physics and Astronomy, The University of Manchester, Manchester M13 9PL, United Kingdom}

\author{Tobias Galla}
\email{tobias.galla@ifisc.uib-csic.es}
\affiliation{Instituto de F\'isica Interdisciplinar y Sistemas Complejos, IFISC (CSIC-UIB), Campus Universitat Illes Balears, E-07122 Palma de Mallorca, Spain}

\date{\today}

\begin{abstract}
We use generating functionals to derive a dynamic mean-field description for generalised Lotka--Volterra systems with higher-order quenched random interactions. We use the resulting single effective species process to determine the stability diagram in the space of parameters specifying the statistics of interactions, and to calculate the properties of the surviving community in the stable phase. We find that the behaviour as a function of the model parameters is often similar to the pairwise model. For example, the presence of more exploitative interactions increases stability. However we also find differences. For instance, we confirm in more general settings an observation made previously in model with third-order interactions that more competition between species can increase linear stability, and the diversity in the community, an effect not seen in the pairwise model. The phase diagram of the model with higher-order interactions is more complex than that of the model with pairwise interactions. We identify a new mathematical condition for a sudden onset of diverging abundances.
\end{abstract}

\maketitle

\section{Introduction}
The study of so-called `complex' ecosystems started in the 1970s, sparked initially by the work of Robert May \cite{may1}. The word `complex' in this context is taken to mean ecosystems with a large number of species and governed by interaction coefficients which are drawn at random from some underlying probability distribution. May's initial approach was based on relatively simple ensembles of random community matrices, without correlation between matrix elements. He then used established results from random-matrix theory to determine the spectra of these matrices, and to decide when a complex ecosystem would be stable. Broadly speaking his stability criterion can be summarised as $SC\sigma^2<1$, where $S$ is the number of species in the ecological community, $C$ is the so-called `connectance' (the probability that any given pair of species interact with one another), and $\sigma^2$ is the variance of the distribution for the elements of the community matrix. 

May's finding -- complexity tends to make ecological equilibria unstable -- seemed inconsistent with empirical findings, and so May's model immediately drew criticism. May himself asked what `devious strategies' nature might use to sustain ecosystems that are both complex and stable \cite{may2}. This sparked the so-called `diversity-stability debate' (or `complexity-stability debate') \cite{allesina2015}. Critics argued that the model is too stylised and that the bound on complexity required for stability could be removed if only more realistic features of natural ecosystems were taken into account. 

The complexity-stability debate is very much present in the current scientific literature, in no small part due to the work by Tang and Allesina~\cite{allesina2012}, who extended May's stability criteria to more general ensemble of community matrices. Recent studies include, among many other contributions, \cite{grilli,gibbs,allesina_et_al_2015, gravel, baron_nat_comms, stone, grilli2017,fyodorov2016, pigani}. The main outcome of this work is that there may be quantitative changes to May's rule for stability, but fundamentally the bound on complexity remains: too much complexity in an ecological system promotes instability.

In recent years, stylised models of complex ecological communities have been studied with tools from the theory of disordered systems \cite{mpv}. These methods were first developed in the context of spin-glass physics, and then later used to study neural networks (see for example \cite{coolendyn}). Much of this work focuses on the analysis of Lotka--Volterra systems with random interactions. In particular the stability diagram of such a model with pairwise interactions has been established in \cite{bunin2016,bunin2017,gallaepl2018}, and connections to the spectrum of the community matrix have been drawn \cite{baron_et_al_prl}. Recent work targets the unstable phase of the Lotka--Volterra model with random interactions \cite{BuninPRX}. The structure of the `energy landscape' of the model is also under intense study \cite{birolibunin, Ros1, Ros2}. Further studies of random Lotka--Volterra models with dynamic mean field theory include \cite{park,poley2,aguirre,azaele}.

The main difference between the random-matrix theory approach on the one hand, and the dynamical approach on the other is the starting point of the analysis. May's work focuses on the community matrix (the Jacobian at an equilibrium) but does not state how the equilibrium is arrived at. In contrast, work starting from an actual dynamics allows one to study the stability of feasible equilibria.

 In this paper we focus on Lotka--Volterra models with random higher-order interactions between species. This is motivated for example by the work in \cite{kleinhesselink, singh, bairey,Grilli_HOI, Mayfield, Letten, AlAdwani,Buche,Levine,GibbsPNAS}, see also \cite{GibbsEcoLett} and references therein. The spectra of interaction matrices with higher-order couplings have previously been studied for example in \cite{bairey}. Similar to \cite{GibbsPNAS} we use a complementary approach, and analyse a Lotka--Volterra model with higher-order interactions using dynamical mean-field theory. More specifically, we set up a (generalised) Lotka--Volterra model of species, allowing for interactions of multiple orders. That is to say, species can interact pairwise, but also in triplets, quadruplets etc. The tool we use, dynamical mean field theory (derived via the generating-functional approach, or the De-Dominici-Martin-Siggia-Rose-Janssen method \cite{msr,dominicis, opper, felixroy}), is rooted in the statistical physics of disordered systems. From this approach a typical (mean field) dynamics for a representative species is derived after averaging over the random interaction structure. Fixed points of this process can then be analysed, and criteria for their stability can be derived. A detailed account of the method can be found in \cite{GFA_notes}. Our work expands upon the analysis of \cite{GibbsPNAS} (conducted using the cavity approach to dynamic mean-field theory). The model in \cite{GibbsPNAS} focuses on interactions between pairs and triplets of species, whereas we include general orders of interaction. We also allow for more general correlations between interaction coefficients. Our aim is to establish the different types of instability the system can experience and to understand the resulting phase behaviour. In the language of spin glass physics our model is the $p$-spin analog of the Lotka-Volterra model with pairwise interaction \cite{cugliandolo,fontanari}.

The remainder of this paper is structured as follows: In Sec.~\ref{sec:model} we introduce the model, in particular we describe how the random higher-order interaction coefficients are constructed. Sec.~\ref{sec:analysis} then contains the main dynamic mean-field analysis. We derive the single effective species process governing the system within dynamic mean field theory. Making a fixed-point ansatz we use this to obtain the properties of the surviving community in the stable phase. We also identify different types of instability, and obtain the phase diagram. In Sec.~\ref{sec:results} we evaluate this for specific orders of interaction. As a benchmark, we review the known results for the model with pairwise interactions. We then analyse the models with third-order and fourth-order interactions, respectively, and finally a model which combines second and third order interactions. In Sec.~\ref{sec:conclusions} we finally summarise our work and draw some conclusions.

\section{Model Definitions}\label{sec:model}
We consider ecological communities which evolve from a pool of $N$ species. The abundance of species $i$ ($i\in\{1,\dots,N\})$ at time $t$ is denoted by $x_i(t)$. The starting point of our analysis is a set of generalised Lotka--Volterra equations \cite{bunin2016,bunin2017,gallaepl2018}. These are conventionally written in the form 
\begin{equation}\label{eq:lv0}
    \dot x_i = r_ix_i \left(K_i - x_i + \sum_{j \neq i} \alpha_{ij}x_j\right).
\end{equation}
In this expression $K_i$ is the carrying capacity of species $i$ in `mono-culture', i.e. the abundance of species $i$ in the long run if no other species is present in the community. The per capita growth rate of species $i$ at low abundance is given by $r_iK_i$.
The coefficients $\alpha_{ij}$ in Eq.~(\ref{eq:lv0}) characterise the interaction between species. In the form of Eq.~(\ref{eq:lv0}) this interaction is taken to be pairwise. 

We now consider generalisations to higher-order interactions. For example, third-order interactions can be introduced as follows
\begin{equation}\label{eq:lv1}
    \dot x_i = r_ix_i \left(K_i - x_i + \sum_{j} \alpha_{ij}x_j+\sum_{j<k}\beta_{ijk}x_jx_k\right),
\end{equation}
where the additional coefficients $\{\beta_{ijk}$\} describe interactions between three species. Specifically, $\beta_{ijk}$ quantifies the effect of the combination of species $j$ and $k$ on the growth of species $i$, where $j$ and $k$ are assumed to be different species to $i$.
In a more general form, allowing for simultaneous interactions of multiple orders, we can write 
\begin{align}\label{lv}
    \dot{x}_i(t) &= x_i(t)\bigg[k_i - x_i(t)  \nonumber \\
    &+ \sum_{p = 2}^N\sum_{\{i_2, \dots, i_p\}} \alpha^{(p)}_{i, i_2, \dots, i_p}x_{i_2}(t)\dots x_{i_p}(t)\bigg].
\end{align}
The notation $\sum_{\{i_2, \dots, i_p\}}$ indicates a sum over all possible sets $\{i_2, \dots, i_p\}$ of size $p-1$ that do not contain the species $i$. Note that if two sets contain the same elements, they are considered to be the same set, regardless of the order of the elements, so each possible set is only counted once.
We have labelled the coefficients describing the different orders of interaction $p$ by a superscript $(p)$. More precisely, $\alpha^{(p)}_{i, i_2, \dots, i_p}$ denotes the reproductive benefit or detriment to species $i$ from interacting with the $p-1$ species $i_2, \dots, i_p$.

In order to keep the analysis compact, we have set the growth rates $r_iK_i = k_i$ and will assume that the $k_i\equiv k$ are the same for all species $i$. Specifically, we set $k = 1$ for all simulations, but we keep $k$ general in the analysis. This simplifying restriction has previously been made for example in \cite{gallaepl2018,bunin2018}.

In order to model {\em complex} ecological communities, we will assume that the interaction coefficients are drawn from a probability distribution. They are fixed at the beginning and then remain constant throughout the time-evolution of the Lotka--Volterra system. We will carry out the analysis for Gaussian distributions of the interactions, but universality found in the spectra of random matrices \cite{mehta1991,tao_vu} suggests that many of the results are applicable to more general distributions, provided some relatively mild conditions hold on the higher-order moments of the interaction coefficients. 

The equilibria of the ecosystem and their stability are then determined by the first two moments of the distribution of the interaction coefficients. Using an overbar to denote averages over the distribution of interaction coefficients, we write these as follows
\begin{align}
    \overline{\alpha^{(p)}_{i_1,i_2,\dots,i_p}} &= \frac{\mu_p(p-1)!(N-p)!}{(N-1)!}\nonumber \\
    &=\frac{\mu_p}{\binom{N-1}{p-1}}, \nonumber \\
    \overline{\left(\alpha^{(p)}_{i_1,i_2,\dots,i_p}\right)^2} - \left(\overline{\alpha^{(p)}_{i_1,i_2,\dots,i_p}}\right)^2 &= \frac{\sigma_p^2p!(N-p)!}{2(N-1)!} \nonumber \\
    &= \frac{p}{2}\frac{\sigma_p^2}{\binom{N-1}{p-1}}, \label{eq:alpha}
\end{align}
where the $\mu_p$ and $\sigma_p$ are assumed not to depend on the system size $N$. 
The scaling of the mean and variance of the interactions of order $p$ with $N^{p-1}$ is standard in the context of disordered systems \cite{cugliandolo, fontanari} and guarantees a well-defined thermodynamic limit. We have instead used $(N-1)!/(N-p)!=(N-1)(N-2)\cdots (N-p+1)$ to reflect the fact that none of the indices $i, i_2,\dots,i_p$ can take the same value as any other index. As indicated in Eqs.~(\ref{eq:alpha}) these factorials create a denominator of $\binom{N-1}{p-1}$, which is the number of interactions each species has with groups of $p-1$ other species. This choice has no effect on the leading-order scaling with $N$ of the mean and variance, and does not affect the resulting dynamic mean field theory, but turns out to reduce finite-size effect in simulations. The significance of finite-size effects, especially for models with higher-order interactions, has also been highlighted in \cite{GibbsPNAS}. Some further discussion can be found in Sec.~\ref{smsims} of the Supplemental Material. The factors $(p-1)!$ and $p!$ in the mean and variance respectively also follow existing conventions, where $(p-1)!$ is the number of re-orderings of the $p-1$ other indices besides $i$. The extra factor of $p$ in the variance cancels with the factor of $2$ in the denominator when $p=2$, but has no necessary role. 

Thus, $\mu_p$ sets the average of the coefficients describing $p$-th order interaction, and $\sigma_p^2$ their variance.
Positive coefficients describe beneficial interactions, and negative coefficients detrimental interactions. Therefore $\mu_p$ characterises the average amount of co-operation (if $\mu_p>0$) or competition (if $\mu_p<0$). We note that opposite sign conventions have been used in some other literature, including \cite{GibbsPNAS, GibbsEcoLett, bunin2016}. The parameter $\sigma_p$ controls the heterogeneity between $p$-th order interactions. 

We also allow for correlations between different interaction coefficients within a given order. More precisely, we assume
\begin{align}\label{eq:gamma}
    \overline{\alpha^{(p)}_{i, i_2, \dots,i_p}\alpha^{(p)}_{j, j_2, \dots, j_p}} - \left(\overline{\alpha^p_{i, i_2, \dots, i_p}}\right)^2 =  \gamma_p\frac{\sigma_p^2p!(N-p)!}{2(N-1)!},
\end{align}
where the sets $\{i, i_2, \dots, i_p\}$ and $\{j, j_2, \dots, j_p\}$ contain the same species, but where $i \neq j$. I.e., both coefficients, $\alpha^{(p)}_{i, i_2, \dots,i_p}$ and $\alpha^{(p)}_{j, j_2, \dots, j_p}$, relate to the same combination of $p$ species, but describe the effect of this interaction on the growth of two different species $i$ and $j$. The parameter $\gamma_p$ controls the symmetry of the interactions, and can take values $-1/(p-1) \leq \gamma_p \leq 1$ \cite{james}. If $\gamma_p = 1$ then $\alpha^{(p)}_{i, i_2, \dots, i_p}=\alpha^{(p)}_{j, j_2, \dots, j_p}$ with probability one, all $p$ species $i,i_2,\dots,i_p$ receive the same payoff in an interaction. For $\gamma_p = -1/(p-1)$ on the other hand, we have $\sum_{i \in \{i_1, i_2, \dots, i_p\}} \alpha^{(p)}_{i, \{i_1, i_2, \dots, i_p\}\setminus \{i\}} = \frac{\mu_pp!(N-p)!}{(N-1)!}$. This sum extends over the payoffs received by each of the species in the group $\{i,i_2,\dots,i_p\}$, where we recall that $\alpha^{(p)}_{i,i_2,\dots,i_p}$ describes the effect of the interaction of the $p$ species on species $i$, with first index $i$. For the case of $\mu_p = 0$ this is akin to zero-sum game.
 
The objective of this paper is to investigate how higher-order interactions, quantified by the parameters $\mu_p$, $\sigma_p$ and $\gamma_p$, affect the behaviour of the generalised Lotka--Volterra dynamics.

\section{Mathematical analysis}\label{sec:analysis}

\subsection{Generating functional analysis and dynamical mean-field theory}

To analyse the behaviour of the generalised Lotka--Volterra system we use a so-called generating-functional approach, going back to (among others) Martin, Siggia, Rose, De Dominicis and Janssen \cite{msr}. This results in what is known as `dynamic mean field theory'. The method has been used for a number of similar systems for example in \cite{opper,gallaasym,gallaepl2018}.
Central to the calculation is the generating functional,
\begin{align}\label{eq:gf}
Z[\pmb{\psi}]=\left<e^{i\sum_i\int dt~ x_i(t)\psi_i(t)}\right>.
\end{align}
The angle brackets in this expression describe an average over trajectories of the system (including an average over possible random initial conditions). The variables $\{\psi_i(t)\}$ constitute a source field. In essence, $Z[\pmb{\psi}]$ is the Fourier transform of the probability measure generated by the system in the space of trajectories.

The key first step of the statistical physics analysis is to average the generating functional over the disorder (i.e., over realisations of the interaction coefficients). This calculation follows standard steps \cite{coolendyn, cugliandolo,gallaepl2018,felixroy}, and we only report the final result here. Further details can be found in the Supplemental Material. We also refer to \cite{GFA_notes} for a step-by-step guide to this general type of calculation.

The outcome of the disorder average is a generating functional describing an `effective' single-species process. There are no species indices in this dynamic mean field description. The effective process involves a retarded interaction kernel, and correlated dynamic noise. The realisations generated by this process capture the statistics of the dynamic variables $\{x_i(t)\}$ in the original problem. 

For the Lotka--Volterra system in Eq.~(\ref{eq:lv1}) and with interaction coefficients as in Eq.~(\ref{eq:alpha}) (and setting $k_i\equiv k$) the effective process is found to be  
\begin{align}\label{effpro_main}
    \dot{x}(t) &= x(t)\bigg[k - x(t) + \sum_{p=2}^\infty\big\{ \mu_p M(t)^{p-1}  \nonumber \\
    &+\gamma_p\sigma_p^2\frac{p(p-1)}{2} \int_0^t G(t,t')C(t,t')^{p-2}x(t')dt'\big\} \nonumber \\
    & + h(t) + \eta(t)\bigg],
\end{align}
with the self-consistency relations
\begin{eqnarray}\label{eq:sc}
    \label{effpro1}M(t) &=& \left<x(t)\right>_*
    , \nonumber \\
    \label{effpro2}\left<\eta(t)\eta(t')\right>_* &=& \sum_{p=2}^\infty \sigma_p^2\frac{p}{2}\left<x(t)x(t')\right>_*^{p-1}\nonumber \\
    &=& \sum_{p=2}^\infty \sigma_p^2\frac{p}{2}C(t,t')^{p-1}, \nonumber \\
    \label{effpro3}G(t,t') &=& \left<\pdv{x(t)}{h(t')}\right>_*.
\end{eqnarray}
The second equation entails a formal definition of $C(t,t')\equiv\langle x(t)x(t)'\rangle_*$. The notation $\left<\cdots\right>_*$ represents an average over realisations of the effective process in Eq.~(\ref{effpro_main}) . The field $h(t)$ in (\ref{effpro_main}) describes external perturbations, and will be set to zero at the end of the calculation.
We highlight that the effective dynamics and self-consistency relations can also be obtained using the so-called `cavity method' \cite{bunin2016,bunin2017,barbier,GibbsPNAS}.

Based on the analysis of related random generalised Lotka--Volterra systems and similar models with random interaction coefficients, one can expect several different possible types of behaviour for the model. Which one of these is realised will depend on the values chosen for the model parameters. For particularly co-operative systems (high values of $\{\mu_p$\}), species abundances can grow indefinitely. The value of the order parameter $M(t)$ then diverges as $t\to\infty$. Other behaviours include reaching a fixed point, which could be unique and globally stable. Alternatively the system could have multiple marginally stable fixed points in certain regions of parameter space \cite{bunin2016,bunin2017,birolibunin,laura,Ros1, Ros2}. It is also possible that the system remains bounded (no divergence of abundances), but that the dynamics never settles down. For example, persistent oscillations may ensue (i.e., a limit cycle), or the system may turn out to be persistently volatile and potentially chaotic. We show some examples of these behaviours in Sec.~\ref{smsims} of the SM.

One purpose of our work is to characterise where in parameter space these different types of behaviour occur. Our analysis proceeds partially mathematically, starting from the effective dynamics in Eqs.~(\ref{effpro_main}, \ref{eq:sc}), and partially numerically.

\subsection{Fixed points of the effective dynamics}
\subsubsection{Fixed point ansatz}

We first proceed analytically, and assume that the original system reaches a unique stable fixed point. We then find the limits to this assumption and derive mathematical divergence and instability conditions.

If the system reaches a fixed point then $x_i(t)\to x_i^*$ at long times for all $i$ in the original system. At the level of the effective process this means that all realisations converge to fixed-point values,  $x(t) \to x^*$ as $t \to \infty$. Due to the random nature of the effective process, $x^*$ will be a random variable, as explained further below. This also means $M(t) \to M^*=\left< x^*\right>_*$ and $C(t+\tau, t) \to q\equiv \left<(x^*)^2\right>_*$ $\forall \tau$. In the fixed-point phase the response function $G(t + \tau, t)$ is time translation invariant and becomes $G(\tau)$. We define
\begin{align}
    \chi = \int_0^\infty G(\tau) d\tau,
\end{align}
which is finite if the fixed point is stable (i.e., perturbations decay to zero). 
Given that $C(t,t')\equiv q$, the second relation in Eq.~(\ref{eq:sc}) dictates that each realisation of the noise $\eta(t)$ tends to a static Gaussian random variable $\eta^*$ with $\left<\eta^*\right>_* = 0$ and
\begin{align}
    \left<\eta^{*2}\right>_* = \sum_{p=2}^\infty \sigma_p^2\frac{p}{2}q^{p-1}.
\end{align}

For later convenience we now introduce the following quantities
\begin{eqnarray}\label{summs}
    \sigma_\Sigma^2 &=& \sum_{p=2}^\infty \sigma_p^2\frac{p}{2}q^{p-1} \quad\left(= \left<\eta^{*2}\right>_*\right), \nonumber \\
    \mu_\Sigma &=&\sum_{p=2}^\infty \mu_pM^{*p-1}, \nonumber \\
    \gamma_\Sigma &=& \sum_{p=2}^\infty \gamma_p\sigma_p^2\frac{p(p-1)}{2}q^{p-2}.
\end{eqnarray}
These quantities $\sigma_\Sigma, \mu_\Sigma$ and $\gamma_\Sigma$ (which can be pronounced as `sigma-sum', `mu-sum', and `gamma-sum' respectively), are recognised as weighted combinations of the $\sigma_p^2$, $\mu_p$ and $\gamma_p$ respectively. As we will see below they govern the overall shape (e.g. mean, variance) of the distribution of abundances at the fixed point.
The introduction of these quantities allow us to simplify Eq.~(\ref{effpro_main}) for a fixed point to
\begin{align}\label{fpequation}
    x^*\left(k - x^* + \mu_\Sigma +\gamma_\Sigma\chi x^* + z\sigma_\Sigma\right) = 0,
\end{align}
where we have replaced $\eta^*$ with $z\sigma_\Sigma$, where $z$ is a standard Gaussian random variable. We find two potential solutions to this equation. One is $x^* = 0$, and the other is
\begin{align}\label{xstar}
    x^*=\frac{k + \mu_\Sigma + z\sigma_\Sigma}{1 - \gamma_\Sigma\chi},
\end{align}
where the latter solution is only valid when it is non-negative ($x^*$ is a species abundance and can therefore not be negative).

We conduct a linear stability analysis on these two solutions in Sec.~\ref{suplinstab} in the Supplementary Material. As part of this analysis we find that the solution $x^* = 0$ is stable only when it is the unique valid solution (i.e., when $\frac{k + \mu_\Sigma + z\sigma_\Sigma}{1 - \gamma_\Sigma\chi}<0$). We can therefore write
\begin{align}\label{soll}
    x^* = \max \left(\frac{k + \mu_\Sigma + z\sigma_\Sigma}{1 - \gamma_\Sigma\chi}, \quad 0\right)
\end{align}
for the physically relevant solution. Eq.~(\ref{soll}) ensures that the fixed-point abundance $x^*$ is non-negative.  

\subsubsection{Species abundance distribution at the fixed point}
The solution for the fixed point in Eq.~(\ref{soll}) describes a probability distribution (we note the random variable $z$ in this expression). When translated back to the original system of species with quenched disorder, this corresponds to the distribution of species abundances at a unique stable fixed point. Therefore, we would expect a histogram of species abundances, numerically determined at a fixed point of the original $N$-species system, to match the distribution in Eq.~(\ref{soll}), for sufficiently large values of $N$.
\begin{figure*}[t!!!]
    \centering
    \includegraphics[width = 0.8\textwidth]{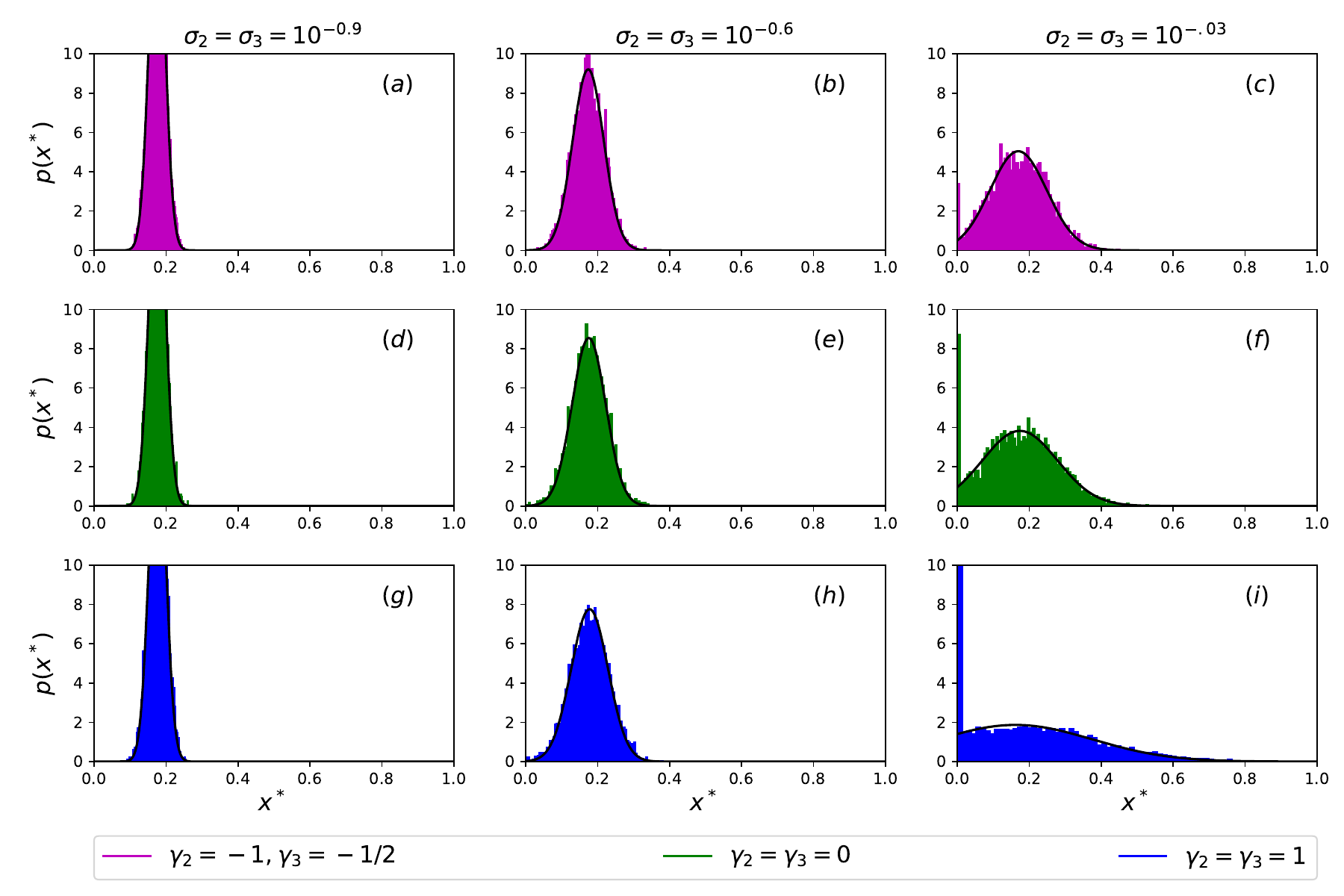}
    \caption{Species abundance distribution of the model with combined second and third order interactions. Examples of histograms from $20$ simulations of a setup with a combination of second and third order interactions with $N = 200$ species, run for 1000 units of time. Various values for $\gamma_{2,3}$ and $\sigma_{2,3}$ are used as indicated, $\mu_2 = \mu_3 = -4$. Simulation results are compared to analytical solutions of the $x^*$ distributions in black. 
    \label{hist}}
\end{figure*}

This is confirmed in Fig.~\ref{hist} where we show species abundance distributions from simulations, and compare these with the probability distribution in Eq.~(\ref{soll}).
The results in Fig.~\ref{hist} are for a combination of pairwise and third-order interactions ($p = 2$, $p = 3$).

The expression in Eq.~(\ref{soll}) describes a Gaussian distribution which is `clipped' at zero (see also \cite{bunin2016, bunin2017,gallaepl2018,GFA_notes}), where all of the mass that would have been in the range of negative $x^*$ is concentrated into a delta-function at $x^*=0$, describing extinct species. The pre-clipped Gaussian distribution has its mean at $(k+\mu_\Sigma)/(1-\gamma_\Sigma\chi)$, and a standard deviation of $\sigma_\Sigma/(1-\gamma_\Sigma\chi)$. For the case of $\gamma_\Sigma = 0$, the denominator in these expressions is one, and for other $\gamma_\Sigma$ (and assuming sufficiently low $\sigma_p$), we have found this denominator stays close to one. This means that we can understand the parameter $\mu_\Sigma$ as controlling the mean, and $\sigma_\Sigma$ as controlling the variance of the pre-clipped Gaussian distribution. For negative values of the $\gamma_p$ (i.e., antisymmetric interactions) the denominator, $(1-\gamma_\Sigma\chi)$ [which is found by using the expression in Eq.~(\ref{help}) see below] is higher than one, causing the standard deviation of the pre-clipped distribution to be smaller than $\sigma_\Sigma$. For positive values of $\gamma_p$ (positively correlated interactions) the denominator is smaller than one, causing the standard deviation to be higher than $\sigma_\Sigma$. This can be seen in the width of the distributions in Fig.~\ref{hist}. As the variance of interactions ($\sigma_p$) is increased, this denominator becomes further away from one, so it increases for $\gamma_p < 0$ and decreases for $\gamma_p > 0$. This causes the mean of the pre-clipped distribution to decrease as the distribution spreads out for antisymmetric interactions $\gamma_p < 0$, and increase as it spreads out for symmetric interactions $\gamma_p > 0$ (this change is too small to be seen in Fig.~\ref{hist}). We also find (see again Fig.~\ref{hist}) that increasing the correlation coefficient $\gamma_p$ leads to an increased fraction of extinct species. As we will see below, this effect relates to the stability of the system.

\subsubsection{Introduction of `clipping threshold' $z_1$}

As the distribution of $x^*$ can be expressed in terms of the standard Gaussian random variable $z$ [see Eq.~(\ref{soll})], we can find the value of $z$ which corresponds to $x^* = 0$, where the distribution is clipped. We call this value $z_1$. Equating the first expression inside the maximum in Eq.~(\ref{soll}) to zero leads to
\begin{align}\label{z11}
    z_1 = \frac{-\left(k + \mu_\Sigma\right)}{\sigma_\Sigma}.
\end{align}
This allows us to re-write the fixed-point solution as
\begin{align}\label{eq:clipping}
    x^* =
    \begin{cases}
      \frac{k + \mu_\Sigma + z\sigma_\Sigma}{1 - \gamma_\Sigma\chi} & z \geq z_1 \\
      0 & z \leq z_1.
 \end{cases}
\end{align}
Using $M^*=\left<x^*\right>_*$, $q=\left< (x^*)^2\right>_*$ and $\chi = \left<\pdv{x(\eta^*)}{\eta^*}\right>_*$ (where we recall that averages $\left<\cdots\right>_*$ over the effective process reduce to averages over $z$ at the fixed point), we can next find expressions for the macroscopic order parameters $M^*, q$ and $\chi$. These relations are
\begin{eqnarray}\label{orderparams}
    M^* &=& \frac{\sigma_\Sigma}{1 - \gamma_\Sigma\chi}\int_{z_1}^\infty(z - z_1) Dz, \nonumber \\
    q &=&\frac{\sigma_\Sigma^2}{\left(1 - \gamma_\Sigma\chi\right)^2}\int_{z_1}^\infty(z - z_1)^2 Dz, \nonumber \\
    \chi &= &\frac{1}{1 - \gamma_\Sigma\chi}\int_{z_1}^\infty Dz,
\end{eqnarray}
where $Dz = \frac{dz}{\sqrt{2\pi}}e^{-z^2/2}$. Further explanation of the derivation of the expression for $\chi$ can be found in Section~\ref{suppfp} of the Supplemental Material. For later convenience we define the integrals
\begin{equation}\label{intn}
    I_n(z_1)=\int_{z_1}^\infty (z-z_1)^n Dz,
\end{equation}
for $n=0,1,2$. We note that $\phi\equiv I_0(z_1)$ is the fraction of surviving species at a stable fixed point.

\subsubsection{Limits of stability of the fixed point}
The assumption of a fixed point made previously, does not hold for all combinations of system parameters $\{\sigma_p, \mu_p, \gamma_p\}$. We now consider where this assumption is valid, and the consequences for the system when it is not.

One type of instability comes from the lack of solutions to the macroscopic order parameters in Eq.~(\ref{orderparams}), for certain combinations of system parameters. A transition point can be found in parameter space, beyond which solutions cease to exist. The region without solutions mostly occurs for system parameters above the transition point (as opposed to below), but in some cases there can also be a lower limit of the region with solutions. We find simulations to display unbounded growth of population abundances for system parameters outside the range with solutions. We call these transition points the `divergence points' because of the divergence of $M^*$ beyond them.

Increasing the value of one of the $\mu_p$ promotes growth in general, with a bias towards the more abundant species as the effect of raising $\mu_p$ on growth is proportional to species abundance $x_i(t)$. This is why divergent behaviour is found for values above the transition point for $\mu_p$ (keeping all other parameters fixed). Conversely, lowering the value of $\mu_p$ can promote stability with the detriment to growth having a greater impact on the more dominant species. Further, lower values of a single $\gamma_p$ cause different species to keep each other in check, which is why we find divergent behaviour above the transition point for $\gamma_p$ (again keeping all other parameters fixed). Higher values of $\sigma_p$ lead to more variance in the population abundances which can lead to divergence, but sometimes too little variance can also cause divergence. Therefore we always find an upper limit, but can sometimes also find a lower limit to the non-divergent range of $\sigma_p$.

Another instability can occur when solutions for the order parameters continue to exist, but the fixed points become unstable against small perturbations. The solutions for the order parameters are then only approximations to results from simulations.
All other parameters fixed, we generally find that the fixed point solution is stable when the $\{\sigma_p\}$ are sufficiently low, and that instabilities set in at higher values of the variances of the interactions.
Where in parameter space this instability point occurs can be determined from linear stability analysis (see Sec.~\ref{suplinstab} in the Supplemental Material). We find the following condition for the onset of linear instability,
\begin{align}\label{instab}
    \left(1 - \gamma_\Sigma\chi\right)^2 = I_0\sum_{p=2}^\infty \sigma_p^2\frac{p(p-1)}{2}q^{p-2}.
\end{align}
Similar conditions for related generalised Lotka--Volterra equations were derived in \cite{bunin2016,bunin2017,gallaepl2018}, see also \cite{opper}.
For a system with a single order of interactions $p$, we find that Eq.~(\ref{instab}) can be simplified to
\begin{align}\label{instabp}
    I_2(z_1) = (p-1)I_0(z_1),
\end{align}
where the integrals $I_n(z_1)$ are defined in Eq.~(\ref{intn}). For a single value of $p$ we also find that
\begin{align}\label{Hgamintro}
    1 - \gamma_\Sigma\chi = \frac{1}{1 + \gamma_p}
\end{align}
from Eq.~(\ref{Hgam}) (derived in Section \ref{singlepinstab} of the SM) which is useful for simplifying expressions for order parameters at the instability point. For simulations with parameters beyond the instability point, the system displays either multiple marginally stable fixed points or persistent dynamics.

\subsection{The threshold $z_1$ and its relationship to the different types of instability}

In this section we briefly discuss some general elements of the behaviours of the order parameters. We also describe some mathematical details regarding the role of the threshold $z_1$ for the further analysis, and in particular its relation to the different types of instability.

\subsubsection{General relationship of order parameters as a function of $\sigma_p$}

\begin{figure*}[t!!]
    \centering
\includegraphics[width = 0.95\textwidth]{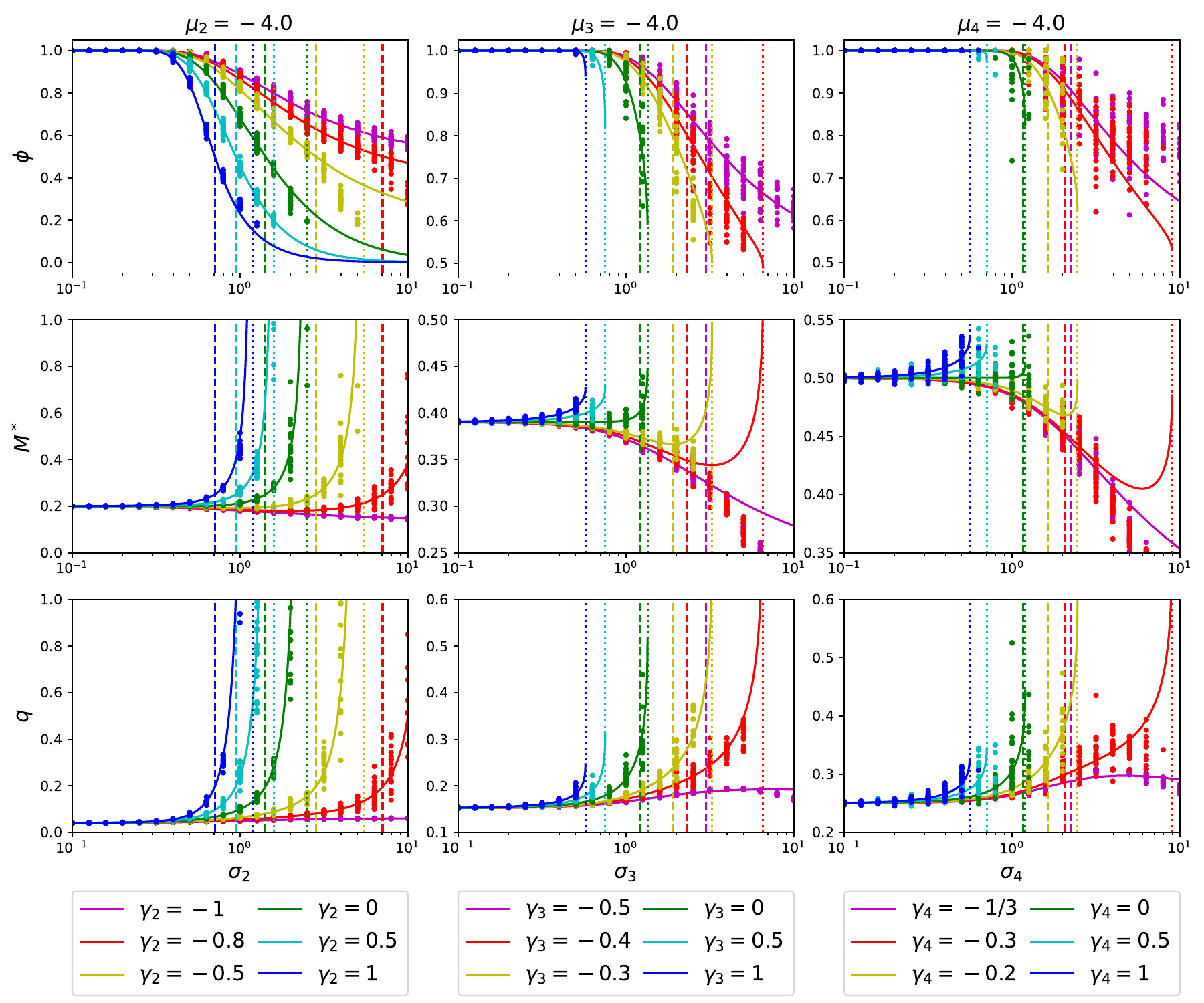}
    \caption{ 
Plots of analytical and simulation results for $\phi$, $M$, and $q$ for second order interactions ($p=2$) on the left, third order interactions ($p=3$) in the middle, and fourth order ($p=4$) on the right, with $\mu_p = -4$, and various choices of $\gamma_p$. The solid lines show the analytical predictions from Eqs.~(\ref{orderparams}) and the points show results measured from simulations. Each point is from one single realisation. Simulations are for $N = 500$ for $p = 2$, $N = 200$ for $p = 3$, and $N = 50$ for $p = 4$ species run for a maximum of 10000 units of time. For each value of $\gamma_p$, the relevant dashed line represents the instability point which satisfies Eq.~(\ref{instab}), and the dotted line represents the point where the system diverges. We find the simulation results to match the predictions up to $\sigma_c$, and roughly from $\sigma_c$ to $\sigma_d$, they do not match at all after $\sigma_d$.}
    \label{fivep_0}
\end{figure*}

The relationship of the macroscopic order parameters $\phi$, $M^*$ and $q$ with $\sigma_p$ is shown in Fig.~\ref{fivep_0} for $\mu_p= -4$ and various choices of $p$ and $\gamma_p$. This is for the model with a single order of interactions. The parameter $\phi = \int_{z_1}^\infty Dz$ represents the fraction of species that have not gone extinct, this quantity decreases with $\sigma_p$. As discussed above this is a consequence of an increasing spread of the abundance distribution of surviving species, leading to more and more abundances being clipped at $x^*=0$; the corresponding species then have abundance $x_i=0$, i.e., they are extinct.

This means that the species abundance distribution does not spread out symmetrically; the less abundant species are limited to the range $x_i \geq 0$ whereas the more abundant species can increase indefinitely. For positively correlated interactions ($\gamma_p > 0$), the mean of the pre-clipped distribution increases as the distribution spreads out (due to the decreasing denominator $1 - \gamma_\Sigma\chi$) which causes both $M^*$ and $q$ to be continuously increasing functions of $\sigma_p$ (see Figs.~\ref{allMplots2}, \ref{allMplots3} and \ref{allMplots4}). For negatively correlated interactions however ($\gamma_p < 0$) the mean of the pre-clipped distribution decreases with increasing $\sigma_p$, so this can initially cause $M^*$ and $q$ to decrease until it spreads out enough and enough species have become extinct to cause it to eventually increase as it spreads out further. This is true unless the interactions are fully antisymmetric [$\gamma_p = -1/(p-1)$] in which case these two effects balance out so both $M^*$ and $q$ tend to finite constants dependent on $\mu_p$ for $\sigma_p \to \infty$, this is shown later.
The system shows the same general behaviour for all values of $\mu_p < 0$, however for $\mu_p > 0$ there is more potential for the divergence of the species abundances, and qualitatively different behaviour is seen.

\subsubsection{Parametric solution for the macroscopic order parameters}

The system parameters are the $\{\mu_p\}$, $\{\sigma_p\}$ and $\{\gamma_p\}$, and the macroscopic order parameters at the fixed point are $M^*$, $q$ and $\chi$. In simulations, we set values for $\{\mu_p\}$, $\{\sigma_p\}$ and $\{\gamma_p\}$, and the order parameters are then an outcome of the dynamics.

We can also find the order parameters corresponding to a set of system parameters mathematically from Eqs.~(\ref{orderparams}). As the parameter $z_1$ appears as the lower limit of integrals, prior knowledge of $z_1$ is required in order to proceed with the evaluation of these integrals. Therefore we set $z_1$ as an independent variable. To find analytical solutions, we set $z_1$, all values of $\mu_p$, all values of $\gamma_p$, and all but one of the values of $\sigma_p$ as independent variables. For these fixed values, we find solutions for the order parameters $M^*$, $q$ and $\chi$, and the one remaining value of $\sigma_p$ using Eqs.~(\ref{z11}) and (\ref{orderparams}). We call this the ``unfixed" $\sigma_p$ where the value of $p$ can be any order we choose. The details of the solution method are described in Secs.~\ref{singlep} and \ref{combinationpmethod} of the Supplement. We can then use these solutions to plot the order parameters parametrically as functions of the unfixed $\sigma_p$, as has been done in Fig.~\ref{fivep_0}.

\subsubsection{The range of values for $z_1$ and its relationship with the unfixed $\sigma_p$}\label{sec:z_1-sigma_p}

We recall that the species abundance distribution is a clipped Gaussian [Eq.~(\ref{eq:clipping})]. Effective species with $z \leq z_1$ go extinct ($x^*=0$), and species with $z>z_1$ have a positive abundance. It is useful to first consider the limiting case in which $z_1\to-\infty$. There is then no clipping of the species abundance distribution, and no species goes extinct. Eq.~(\ref{z11}) indicates that this only possible if $\sigma_\Sigma=0$, or if $M\to\infty$. The latter indicates a breakdown of the stable phase, and will be discussed below. For $\sigma_\Sigma=0$ the species abundance distribution is a delta-distribution at its mean. The quantity $\sigma_\Sigma^2$ is a weighted sum of the $\{\sigma_p^2\}$ with strictly positive weights, see Eq.~(\ref{summs}). In order for $\sigma_\Sigma$ to be zero all $\sigma_p$ must therefore also be zero. We find that this solution ($\sigma_p = 0$) does sometimes not exist in the limit $z_1\to-\infty$ (e.g. when the $\mu_p$ are high), as discussed later. In this case the abundances diverge.

As we increase $z_1$ an increasing finite fraction of species goes extinct. These are all those species corresponding to values $z < z_1$. This is a result of an increase in the variance of the (unclipped) species abundance distribution, with more species crossing the $x = 0$ line as it spreads out. Fixing the $\{\mu_p, \gamma_p\}$ and all but one $\sigma_p$, we observe in simulations that increasing the unfixed $\sigma_p$ always corresponds to an increase in the fraction of extinct species for single orders of interactions, and therefore an increase in $z_1$. For a combination of orders, we usually find this to be true, but can find the opposite effect for some specific parameter combinations, shown later.

In the stable phase, where simulations reach a unique fixed point, a unique value of $\phi$ can be measured for any given set of $\{\sigma_p\}$, corresponding to a unique value of $z_1$. For orders higher than 2, we often find the same solutions for the unfixed $\sigma_p$ from two different values of $z_1$, whereas only one of these $z_1$ correspond to the physically attained value of $\phi$. This is due to the non-monotonous behaviour of the unfixed $\sigma_p$ as a function of $z_1$, displaying a maximum point, and nonphysical solutions thereafter.
As a consequence of this, not all values of the unfixed $\sigma_p$ are attainable in this way, some values do not have a corresponding $z_1$ [i.e., there are no solutions of Eqs.~(\ref{orderparams}) for those values of the unfixed $\sigma_p$ above the maximum point]. 

\subsubsection{Range of vales for $z_1$ for which $M^*$ is real and finite (introduction of $z_d$)}

As mentioned earlier, there are particular parameter combinations which result in the divergence of the mean species abundance $M(t)$ at long time $t$ in simulations. For a given combination of $\{\mu_p\}$, $\{\gamma_p\}$, and all but one of $\{\sigma_p\}$, there can be particular values of the unfixed $\sigma_p$, above (or below) which, the system is found to diverge in simulations. From a mathematical perspective, one possible cause of this divergent behaviour is that the solution for $M^*$ in Eq.~(\ref{orderparams}) becomes either infinite or complex. Again it is more meaningful to first find the value of $z_1$ where this occurs, rather than to directly operate in terms of the unfixed $\sigma_p$. The value of $z_1$ for which $M^*$ becomes infinite or complex is denoted by $z_d$, and we call the value of the unfixed $\sigma_p$ at that point $\sigma_d$. In some cases (see later), this value may not be unique, there may be two values of $\sigma_d$, where solutions of Eqs.~(\ref{z11},\ref{orderparams}) exist only for unfixed $\sigma_p$ between these values, and we find divergence outside this range.
  
To find the value of $z_d$ we first rearrange the expression for $M^*$ in Eq.~(\ref{orderparams}). This leads to
\begin{align}
    \sigma_\Sigma = \frac{(1-\gamma_\Sigma\chi)M^*}{I_1(z_1)}.
\end{align}
This is then substituted for $\sigma_\Sigma$ in Eq.~(\ref{z11}), along with the expression for $\mu_\Sigma$ in Eq.~(\ref{summs}) resulting in the following polynomial for $M^*$,
\begin{align}\label{eq:m_poly}
    \sum_p \mu_p (M^*)^{p-1} + \frac{z_1(1-\gamma_\Sigma\chi)}{I_1}M^* + k = 0.
\end{align}
We also have
\begin{align}\label{help}
    (1 - \gamma_\Sigma \chi) = \frac{I_2}{I_2 + \gamma_p(p-1)I_0}
\end{align}
for single values of $p$, as derived in Sec.~\ref{singlepmethod} in the Supplemental Material. Sec.~\ref{combinationpmethod} further describes for how this factor is found for a model combining second and third order interactions ($p = 2$ and $p=3$).
The value of $M^*$ is one of the $p-1$ roots of the polynomial in Eq.~(\ref{eq:m_poly}), and the value of $z_1$ where this root becomes either infinite or complex leads to the ``divergence point", after which, we find simulations to diverge. 

This is illustrated in Fig.~\ref{fivep_0}, where the solution for $M^*$ tends to infinity at $\sigma_d$ (dotted line) in the model with pairwise interactions ($p=2$). For other orders $M^*$ becomes complex beyond $\sigma_d$, but there is no divergence of the mathematical solution. Instead, $M^*$ assumes a finite value at the dotted line ($\sigma_p=\sigma_d$), which is where the discriminant of the polynomial is zero.

\subsubsection{Range of values for $z_1$ for which the unfixed $\sigma_p$ is strictly increasing with $z_1$ (introduction of $z_m$)}\label{sec:z_m_intro}

For orders higher than $p = 2$, we find another mathematical condition which results in a lack of solutions for Eq.~(\ref{orderparams}), and subsequently system divergence. For given $\{\mu_p\}$, $\{\gamma_p\}$, and all but one of $\{\sigma_p\}$, we can vary $z_1$ to find solutions for the unfixed $\sigma_p$ as described above. We find there can exist some values of the unfixed $\sigma_p$ which do not correspond to any value of $z_1$. In most cases, there is a maximum possible value of the unfixed $\sigma_p$ that can be found while varying $z_1$, we call the value at which this point occurs $z_m$. As there exist no solutions to Eq.~(\ref{orderparams}) for values of the unfixed $\sigma_p$ greater than $\sigma_p(z_m) \equiv \sigma_m$, we again find system divergence in simulations for such conditions. For values of $z_1$ past $z_m$ Eqs.~(\ref{orderparams}) have a solution, but these states are unattainable by the physical system, as this would mean two possible values for $z_1$ and $\phi$ for a given value of $\sigma_p$, and therefore two possible branches of solutions for $\phi$ (and other order parameters), however only one branch is observed, we confirm this later in Fig.~\ref{bendfive}. Further details of this can be found in Sec.~\ref{p=3sec} of the SM.

\subsubsection{Linear instability (introduction of $z_c$)}

As mentioned previously, there is a limit of the parameter regime in which a non-divergent system shows a unique stable fixed point. Eqs.~(\ref{orderparams}) then cease to hold. 
A breakdown of this phase can occur through an instability against small perturbations. The onset can be found via linear stability analysis as discussed in more detail in Sec.~\ref{suplinstab} of the Supplemental Material. The onset of the instability is signalled by the condition in Eq.~(\ref{instab}). The instability can also be identified in simulations. For a system with fixed $\{\mu_p\}$, $\{\gamma_p\}$, and all but one of $\{\sigma_p\}$, the instability sets in at a critical value of the unfixed $\sigma_p$. Within our analytical approach, this translates to a critical value for $z_1$, which we will denote by $z_c$, and the value of the unfixed $\sigma_p$ at this point is denoted $\sigma_c$. Values of $z_1 < z_c$ correspond to the phase with a unique stable fixed point. For a system with a single order of interactions $p$, $z_c$ is defined as the value of $z_1$ which satisfies equation Eq.~(\ref{instabp}). It is clear from this simplified form that $z_c$ is independent of $\mu_p$ and $\gamma_p$ and only a function of $p$. For example, we find $z_c = 0$ for $p = 2$ as previously known \cite{bunin2016,bunin2017,gallaepl2018}, $z_c\approx-0.840$ for $p = 3$ (see Sec.~\ref{p=3sec}) and $z_c\approx -1.326$ for $p=4$ (Sec.~\ref{p=4sec}). For systems with multiple orders of interaction, we find a range of values for $z_c$, this range is between the unique values of $z_c$ for each of the individual interactions.

As the parameter $z_1$ also determines the proportion of the initial species going extinct at the fixed point, we can characterise the onset of the linear instability as the point where a critical fraction of species has died out (the fraction of species corresponding to $z_1=z_c$). For a system with a single value of $p$, this critical fraction is the same regardless of the system parameters $\gamma_p$ and $\mu_p$. For example, less than one half of the initial species are extinct in a system with $p = 2$ in the phase with a unique stable fixed point, and at the point of linear instability exactly half of all initial species go extinct. In general the value of $\sigma_c$ at which this critical fraction is attained (and hence the instability sets in) depends on $\{\mu_p\}$, $\{\gamma_p\}$, and the rest of $\{\sigma_p\}$ for both models with single and multiple orders of interactions.

The linear instability point ($z_1=z_c$) can only be realised provided the system has not already diverged, i.e., we require $z_c < z_d$ (if divergence sets in both at an upper and at a lower bound, then $z_c$ must be between the two). We find instances in which there are formal solutions to Eqs.~(\ref{orderparams}) which also fulfil the condition in Eq.~(\ref{instab}), but where the solution of Eqs.~(\ref{orderparams}) is not physical (i.e., it cannot be attained by the system, as explained above). This can occur if the solution of Eqs.~(\ref{orderparams}) is such that the point $z_c$ at which the linear instability triggers is at a value $z_c>z_m$. In such cases the linear instability is not seen in simulations. This will be discussed in more detail below in Sec.~\ref{p=3sec}.

We note that the solution of Eqs.~(\ref{orderparams}) only describes the system in the regime of a unique stable fixed point.  This is the case for $z_1 < z_c$, but not when $z_1>z_c$ (the fixed point has then become linearly unstable).  Eqs.~(\ref{orderparams}) then only describe the system as an approximation. This can be seen in Fig.~\ref{fivep_0} where $\sigma_c$ is represented by the dashed line. Results from simulations match the analytical solutions well for $\sigma_p < \sigma_c$, but we find small discrepancies for values greater than $\sigma_c$. This also means that the calculated value of $\sigma_m$ beyond which the system diverges could potentially be an approximation, if this occurs at a greater value than $\sigma_c$.

\section{Results for different higher-order interactions}\label{sec:results}

The theory developed so far is general in the sense that arbitrary combinations of different orders of interactions are covered. In this section we now discuss the behaviour of specific examples in more detail. We start with the model with two-species interactions ($p=2$). This model has been studied intensively in existing literature (see e.g., \cite{bunin2016,bunin2017,gallaepl2018,felixroy,birolibunin}), but we include this for completeness and as a reference. We then move to the models with third-order and fourth-order interactions ($p=3$ and $p=4$ respectively). Finally we discuss a model that combines both second-order and third-order interactions.

\subsection{Model with second-order interactions}

\subsubsection{Mathematical solutions for $z_c$ and $z_d$}

\begin{figure}[t!!!]
    \centering
    \includegraphics[width=0.5\textwidth]{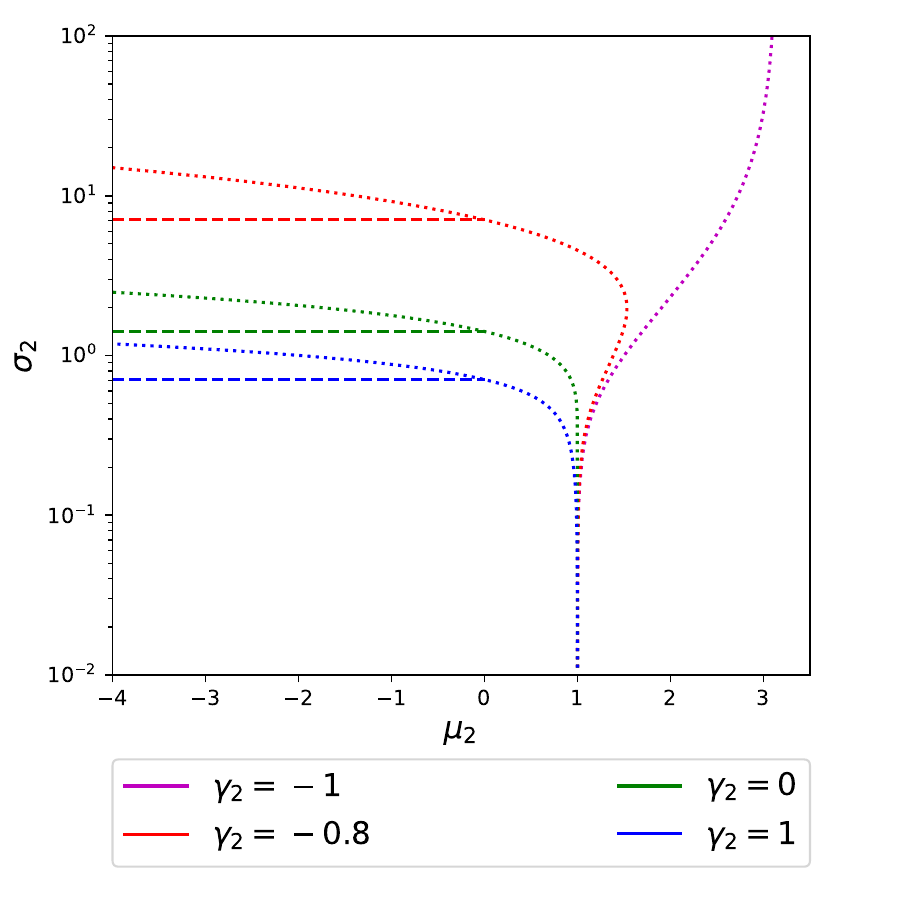}
    \caption{Phase diagram for the system with second order interactions only, for various values of $\gamma_2$. The horizontal dashed lines show the onset of the linear instability ($\sigma_c$) for each $\gamma_2$. The system reaches a unique fixed point below these lines. The curved dotted lines show $\sigma_d$, the system diverges above and to the right of these lines. Between the two transition points, the system displays either multiple fixed points or persistent dynamics.}
    \label{phase2}
\end{figure}

We here summarise existing results from the literature.
For the model with second-order interactions only we find from Eq.~(\ref{eq:m_poly}),
\begin{align}\label{eq:m_for_p=2}
    M^* = \frac{k}{\frac{-z_1(1 - \gamma_\Sigma\chi)}{I_1} - \mu_2},
\end{align}
where $\gamma_\Sigma=\gamma_2\sigma_2^2$ for the case of two-species interactions, and where $1 - \gamma_\Sigma \chi$ is found via Eq.~(\ref{help}). We also have
\begin{align}\label{sig2}
    \sigma_2 = \frac{\sqrt{I_2}}{I_2 + \gamma_2I_0}
\end{align}
from Eq.~(\ref{singlepsig}) in Sec. \ref{singlepmethod} of the Supplemental Material.
The linear instability is known to occur at $z_c = 0$, which means that half of the initial species have gone extinct \cite{opper,bunin2016,bunin2017,gallaepl2018}. The value of $\sigma_c$ where this occurs is given by $\sigma_c^2 = 2/(1 + \gamma_2)^2$, as shown in Sec.~\ref{singlepinstab} of the SM. The value of $\sigma_c$ is a function of $\gamma_2$ but independent of $\mu_2$, so the instability occurs along straight lines in the phase diagram in Fig.~\ref{phase2}. As $\sigma_2$ is a function of $z_1$ [Eq.~(\ref{sig2})] and independent of $\mu_2$, we find in simulations the value of $\phi$ (and $z_1$) depends only on $\gamma_2$ and $\sigma_2$, and is independent of $\mu_2$. The plot of $\phi$ for $\mu_2 = -4$ in Fig.~\ref{fivep_0} is the same for any value of $\mu_2$. This suggests that competition pressure has no effect on the diversity or stability of a community, unlike what has been found for real-world ecosystems in \cite{kat}. 

From Eq.~(\ref{eq:m_for_p=2}) we find $M^* \to \infty$ when, for given $\sigma_2$ and $\gamma_2$, $\mu_2$ and $z_1$ satisfy the following relation,
\begin{align}\label{div2}
    \mu_d = \frac{-z_1(1 - \gamma_\Sigma\chi)}{I_1}.
\end{align}
We define $z_d$ as the value(s) of $z_1$ which satisfies this for a given value of $\mu_2$.

For the case of $\sigma_2 = 0$ ($z_1 \to -\infty$), we find $\mu_d = 1$, this is true for all values of $\gamma_2$. For anti-symmetric interactions, where $\gamma_2 = -1$, $\mu_d \to \pi$ as $\sigma_2 \to \infty$ ($z_1 \to 0^-$). This is derived in Secs.~\ref{sec:singlepdiv} and \ref{sec:singlepanti} in the SM and can be seen in the phase diagram in Fig.~\ref{phase2}. As $z_1 = 0$ is also the condition for the linear-instability point, we find that this instability can only be reached in the limit of $\sigma_2 \to \infty$, and so systems with anti-symmetric interactions do not become unstable or diverge as long as the value of $\mu_2$ is below the divergence boundary.

\subsubsection{Phase diagram}

In Fig.~\ref{phase2} we have plotted $\sigma_c$ (linear instability) as a dashed line and $\sigma_d$ (divergent abundance) as a dotted line parametrically for various values of $\gamma_2$.
At $\mu_2 = 0$ we find $z_d = 0 = z_c$, and for $\mu_2 > 0$, $z_c > z_d$. This means that for $\mu_2 > 0$ the system will diverge before it can reach the point of linear instability. Therefore, we have not plotted the linear-instability line for $\mu_2 > 0$.
The expressions we have used to derive the divergence condition in Eq.~(\ref{div2}) are only valid for $z_d < z_c$ in the unique fixed point phase, therefore for $\mu_2 < 0$ the divergence line is only an approximation.

We expect the system to display three types of behaviour in Fig.~\ref{phase2}.
For values of $\sigma_2$ below the dashed lines and to the left of the dotted lines, the system reaches a globally stable unique fixed point, we call this the stable phase.
We find positive finite solutions for $M^*$ for values of $\mu_2$ to the left of the dotted lines only, to the right of these lines the abundance diverges.
Above the dashed lines but below the dotted lines the system remains finite but does not reach a globally stable unique fixed point.
It could either display multiple marginally stable fixed points, or the dynamics may continue indefinitely. We call this the unstable phase.
\begin{figure*}[t]
    \centering
    \includegraphics[width = \textwidth]{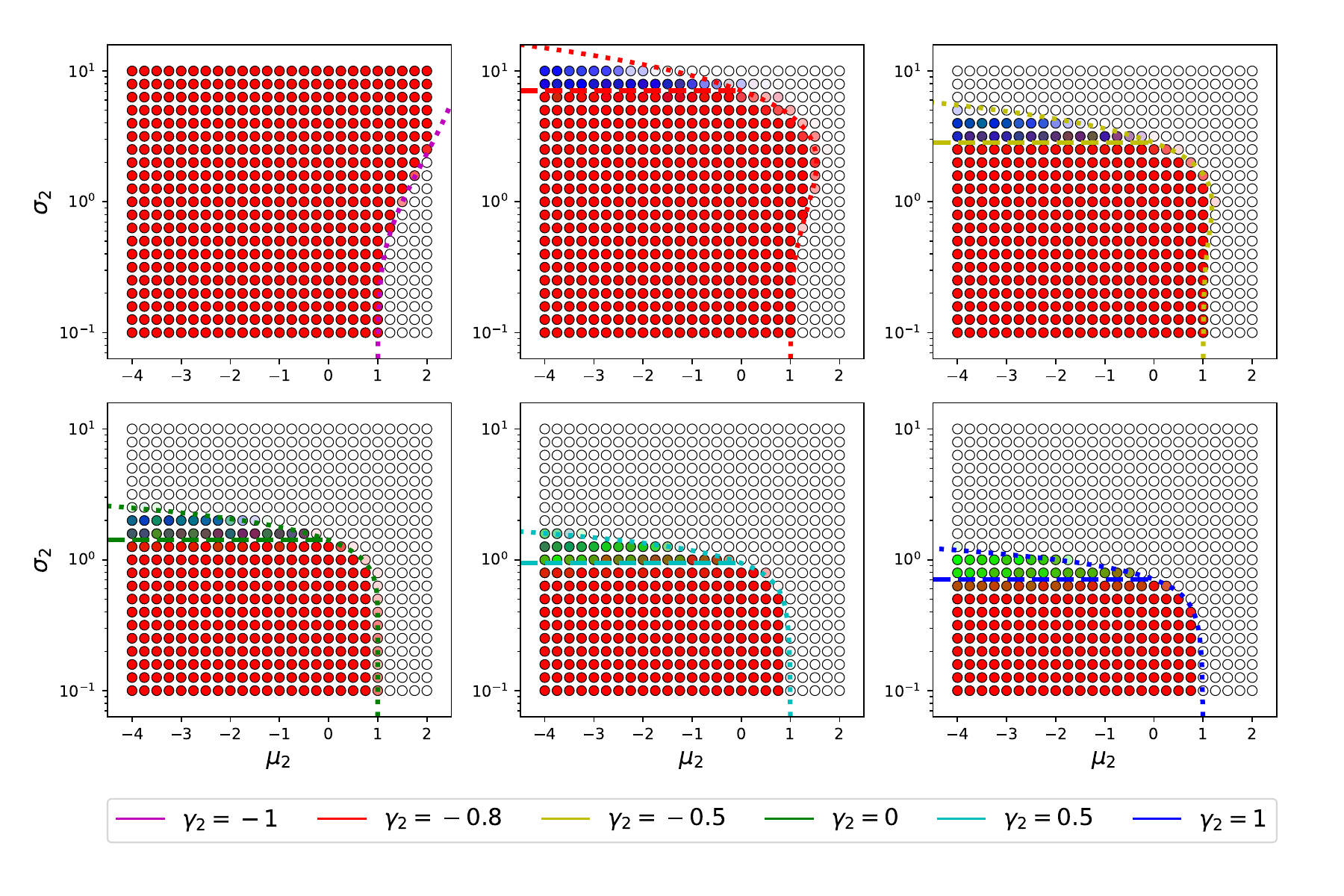}
    \caption{Phase diagrams for the model with  $p=2$ as in Fig.~\ref{phase2} with simulation results for each value of $\gamma_2$. Each point represents the average behaviour of 20 simulations for $N = 500$ species run for a maximum of 10000 units of time. Red indicates that the system reached a unique fixed point, green indicates multiple fixed points, blue persistent dynamics, and white divergence. The method for classifying the behaviour and converting the number of each type to a colour is described in Section \ref{smsims} in the SM.}
    \label{phasesim2}
\end{figure*}

\subsubsection{Simulations confirm phase diagram}

We show that the system displays the expected behaviour in the various phases in Fig.~\ref{phasesim2}. For each circle on the diagram, $20$ realisations of the system were run for $N = 500$ species for a maximum of 10000 units of time, the simulation was terminated at an earlier time if the system was found to have diverged or reached a fixed point. The behaviour of each run was classified as either: having a unique fixed point (red), having multiple fixed points (green), dynamically persisting after the allocated time (blue), or displaying divergent growth (white). Further details can be found in Sec.~\ref{sec:colour} in the SM. As expected, we find red colouring (unique fixed stable point) below the dashed lines, white colouring to the right of the dotted lines (system diverges), and both green and blue behaviour between the lines (multiple fixed points or persistent dynamics). In the instability phase we find multiple fixed points (green) more often for higher values of $\gamma_2$, and persistent dynamics more often for lower $\gamma_2$. 

\subsubsection{Discussion of the behaviour of the average abundance $M^*$}

\begin{figure*}
    \centering
    \includegraphics[width = 0.95\textwidth]{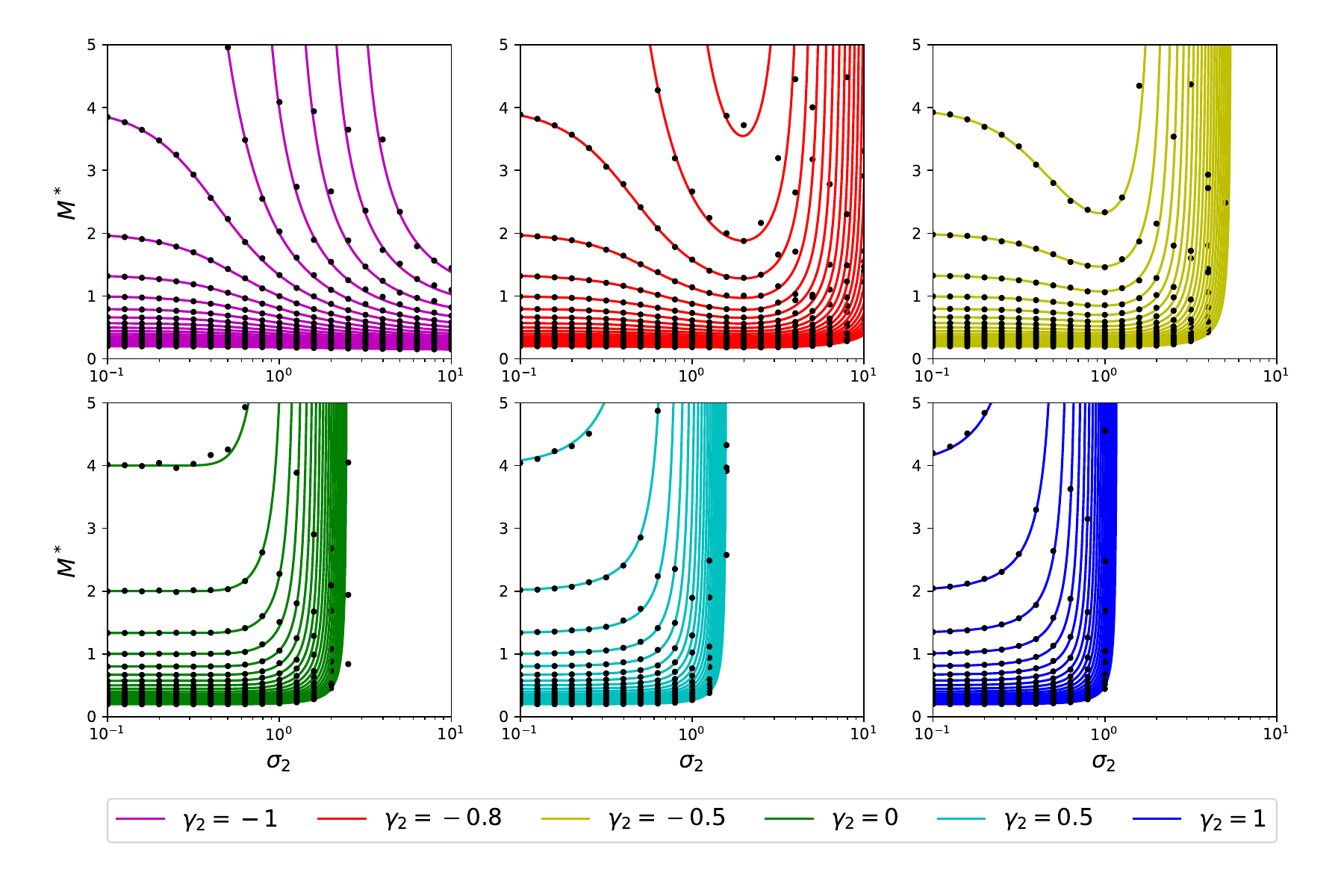}
    \caption{Plots of analytical predictions (lines) and simulation results (dots) for $M^*$ against $\sigma_2$ for all values of $\mu_2$ shown in Fig.~\ref{phasesim2}, between $-4$ and $2$ in steps of 0.25 (bottom to top in each panel). Each point shows the average results from 20 simulations of $N = 500$ species ran for a maximum of 10000 units of time.}
    \label{allMplots2}
\end{figure*}

Fig.~\ref{allMplots2} shows plots of $M^*$ against $\sigma_2$ for each value of $\mu_2$ between -4 and 2 in steps of 0.25. The results from simulations (points) are shown against the predictions from theory (lines). The lines with lower values of $M^*$ correspond to lower values of $\mu_2$. These plots show how the behaviour of $M^*$ changes with $\mu_2$ and $\gamma_2$.

Using Eq.~(\ref{div2}) we can rewrite Eq.~(\ref{eq:m_for_p=2}) as
\begin{align}\label{Mmud2}
    M^* = \frac{k}{\mu_d - \mu_2},
\end{align}
 where we note that $\mu_d$ depends on $\gamma_2$ and $\sigma_2$. The value of $M^*$ depends only on the distance of $\mu_2$ away from the divergence boundary at $\mu_d$, meaning that the shape of these graphs closely follow the shape of the divergence boundaries for each value of $\gamma_2$. For example, following along a red line in Fig.~\ref{allMplots2}, $\gamma_2 = -0.8$ is constant, $\mu_2$ is some constant value (depending on which line is followed) and $\sigma_2$ is increasing. This corresponds to following a vertical line of dots upwards from the $\gamma_2 = -0.8$ plot in Fig.~\ref{phasesim2}, as $\mu_2$ remains constant and $\sigma_2$ increases along the vertical line. The value of $M^*$ depends on $\mu_d - \mu_2$ for each value of $\sigma_2$, which is the horizontal distance between a vertical line of constant $\mu_2$, and the red divergence boundary in Fig.~\ref{phasesim2}. Starting from the bottom, as $\sigma_2$ is increased, moving upwards, the divergence boundary first moves further away from, and then back towards the chosen vertical line of constant $\mu_2$, until it eventually crosses it. This means the distance $\mu_d - \mu_2$ first increases, and then decreases, until it eventually becomes zero when they meet. As this distance is the denominator is the expression for $M^*$ in Eq.~(\ref{Mmud2}), this will cause the value of $M^*$ to first decrease, then increase and eventually diverge when the denominator becomes zero, and this is what we observe for each red line in Fig.~\ref{allMplots2}. The higher lines with higher values of $\mu_2$ display more pronounced changes as they would have a smaller denominator as they are closer to the divergence boundary. Therefore the shape of each red line in Fig.~\ref{allMplots2} looks similar to a $90^\circ$ clockwise rotation of the red divergence boundary in Fig.~\ref{phasesim2}. In fact, each plot of $M^*$ in Fig.~\ref{allMplots2} (except for $\gamma_2 = -1$ in magenta) resemble a $90^\circ$ clockwise rotation of the corresponding phase diagram in Fig.~\ref{phasesim2}.
 
\begin{figure*}
    \centering
    \includegraphics[width = 0.95\textwidth]{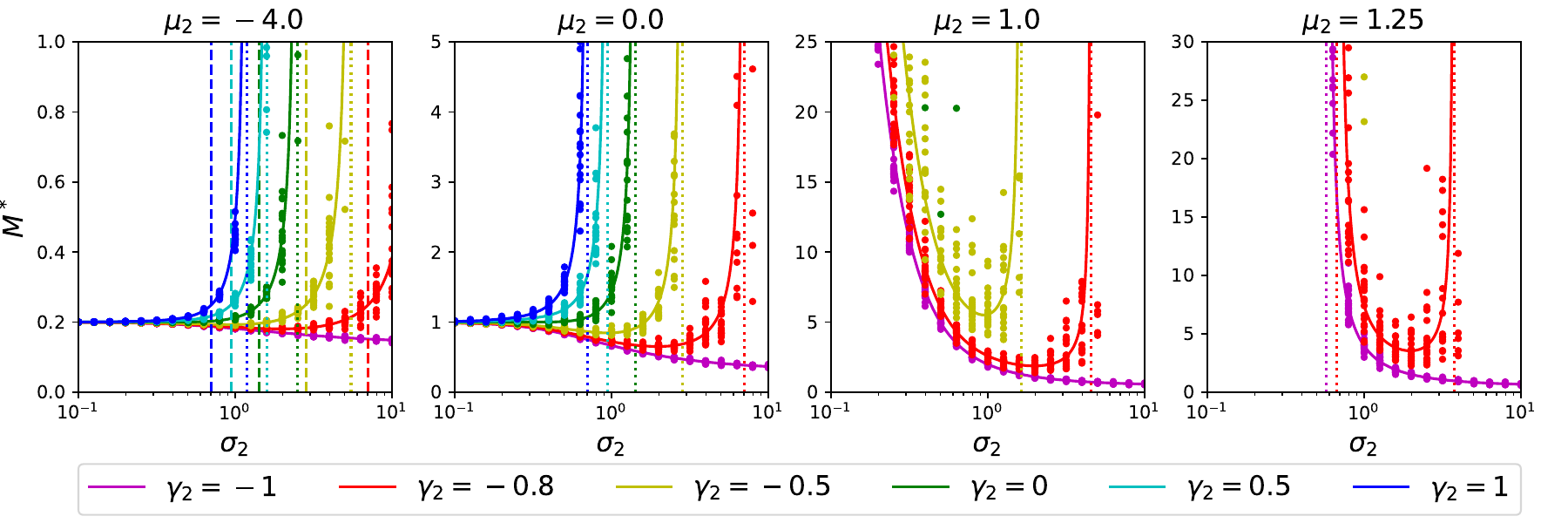}
    \caption{Plots of analytical predictions (lines) and simulation results (dots) for $M^*$ in the model with second-order interactions for various $\mu_2$ and $\gamma_2$, and varying $\sigma_2$. Each point shows the result of one of 20 simulations of $N = 500$ species run for a maximum of 10000 units of time. For each value of $\gamma_2$, the dashed line represents $\sigma_c$ and the dotted line(s) represent $\sigma_d$. }
    \label{Mplots2}
\end{figure*}

For antisymmetric interactions ($\gamma_2 = -1$), $\mu_d \to \pi$ as $\sigma_2 \to \infty$ ($z_1 \to 0^-$), so $M^* \to k/(\pi - \mu_2)$ and $q \to \pi M^{*2}$, both tend to finite constants with increasing $\sigma_2$. For $\sigma_2 = 0$ ($z_1 \to -\infty$), $\mu_d = 1$, $M^* = k/(1 - \mu_2)$ and $q = M^{*2}$ for all values of $\gamma_2$. This can be seen in Fig.~\ref{Mplots2} where lines for all $\gamma_2$ start at the same values for low $\sigma_2$.

In each panel of Fig.~\ref{allMplots2}, $M^*$ shows similar dependence on $\sigma_2$ for competitive interactions, where $\mu_2 < 0$. However, for higher $\mu_2$ closer to the divergence boundary, the behaviour changes and depends on the shape of the divergence boundary for the specific value of $\gamma_2$. As lower values of $\gamma_2$ generally have higher values of $\mu_d$, the system can tolerate more variance before divergence for lower $\gamma_2$.

In Fig.~\ref{Mplots2} we show how the average species abundance $M^*$ changes with increasing $\sigma_2$, for various fixed $\mu_2$ and $\gamma_2$, particularly across phase transitions. We show results from simulations (points) and analytical predictions (solid lines), along with the linear-instability point (dashed lines) and divergence point (dotted lines) for each value of $\gamma_2$. We note that there are necessarily some fluctuations in simulations for finite $N$. For example, some samples in the non-divergent region of the phase diagram can diverge in simulations, and some realisations in the divergent phase may result in a finite $M^*$. In the non-divergent  phase, if a realisation was classified as having divergent abundance, then the average abundance was not calculated, and an average was taken over the realisations that did not diverge.

For $\mu_2 = -4$ and $\mu_2 = 0$ (see the two left-most panels in Fig.~\ref{Mplots2}), the system is in the stable phase for low $\sigma_2$, in this phase the simulation results match the predictions well. We find $M^*$ increases with $\sigma_2$ and eventually diverges when it reaches the divergence point shown by the dotted lines. For $\mu_2 = -4$, the instability point is found at a lower value of $\sigma_2$ than the divergence point. Between these two points we find the unstable phase where analytical predictions are a good approximation for simulation results.

As shown in Fig.~\ref{phase2}, the system is in the divergent phase for low $\sigma_2$ if $\mu_2 \geq 1$. As a result we find divergent $M^*$ for $\mu_2 = 1$ and $\mu_2 = 1.5$ in Fig.~\ref{Mplots2} for low $\sigma_2$. For $\gamma_2 = -0.8$ (red) and $\gamma_2 = -0.5$ (yellow), we find that the system is divergent for small value of $\sigma_2$, enters the stable phase for intermediate $\sigma_2$ (after crossing the first dotted line in Fig.~\ref{phase2}), and returns to the divergent phase after at high $\sigma_2$ after crossing the second dotted line. This re-entry phenomenon can understood from the divergence transition boundary (dotted lines) in Fig.~\ref{phase2}, which curve back on themselves creating two divergence transition points for some values of $\mu_2$. The range of $\sigma_2$ between these two divergence points corresponds to the stable phase where simulation results match predictions. For $\gamma_2 = -1$ (magenta lines in Fig.~\ref{phase2}), we find one divergence point where the system diverges for values of $\sigma_2$ below this point and becomes stable after crossing it.
This demonstrates that low values of $\sigma_2$ do not always correspond to more stability, contrary to what one might expect.

In each panel of Fig.~\ref{Mplots2} the dependence of $M^*$ on $\sigma_2$ is similar across different values of $\gamma_2$. However, the change with $\sigma_2$ is more gradual for lower values of $\gamma_2$, this also means that the instability and divergence transitions in the first two panels occur at larger values of $\sigma_2$ for low values of $\gamma_2$. Generally, the stable region is larger the more anti-correlated the interactions.

\subsection{Third-order interactions}\label{p=3sec}

\subsubsection{Mathematical solutions for $z_c$ and $z_d$}

We now consider the system with third-order interactions only. We investigate how the system behaves for different values of the parameters $\mu_3$, $\gamma_3$ and $\sigma_3$, and find the transition points where the behavior changes.

To find $M^*$ we consider Eq.~(\ref{eq:m_poly}) for $p = 3$ only which becomes a quadratic equation for non-zero $\mu_3$. If $\mu_3 = 0$, we have 
\begin{align}\label{eq:M*_p=3mu0}
    M^* = \frac{-kI_1}{z_1(1 - \gamma_\Sigma\chi)},
\end{align}
and for non-zero $\mu_3$,
\begin{align}\label{eq:M*_p=3}
    M^* = \frac{-\frac{z_1(1-\gamma_\Sigma \chi)}{I_1} \pm \sqrt{\left(\frac{z_1(1 - \gamma_\Sigma \chi)}{I_1}\right)^2 - 4\mu_3 k}}{2\mu_3},
\end{align}
where $1 - \gamma_\Sigma \chi$ is found via Eq.~(\ref{help}). The solution for $M^*$ in Eq.~(\ref{eq:M*_p=3}) with a minus sign is positive for all combinations of $\mu_3$, $\gamma_3$ and $\sigma_3$ below the divergence points and matches simulation results, but the solution is nonphysical with the plus sign.
\begin{figure}[t]
    \centering
    \includegraphics[width = 0.5\textwidth]{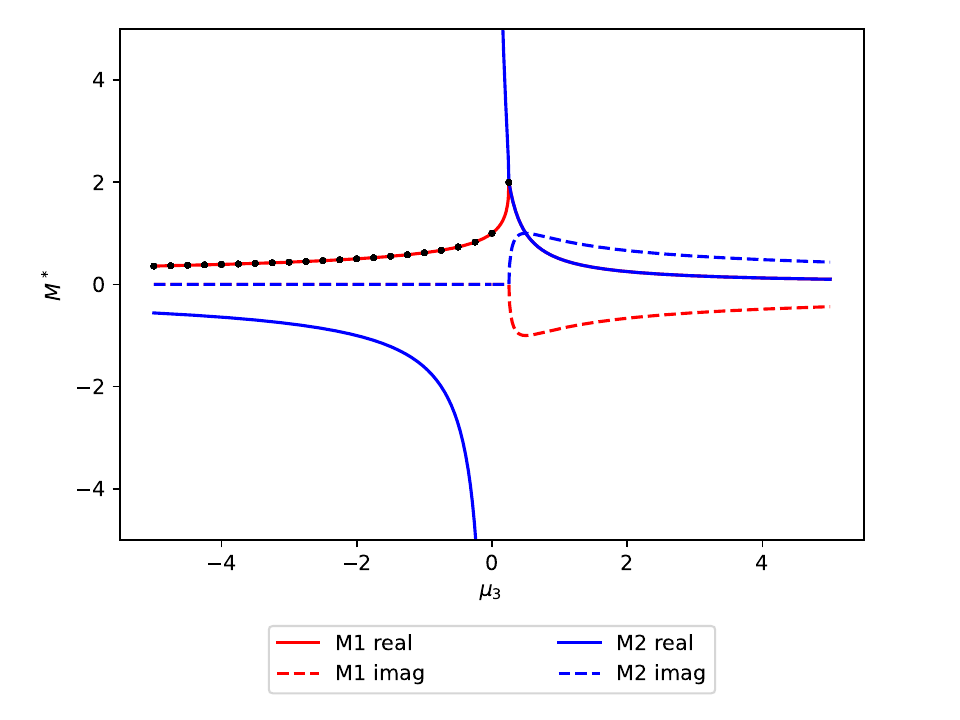}
    \caption{Plot of real and imaginary parts of the two solutions for $M^*$ from Eq.~(\ref{eq:M*_p=3}), with varying $\mu_3$ and $\sigma_3 = 0$. The solutions are denoted $M1$ and $M2$ in the figure. Red lines show the physically realised solution, matching the simulation results shown with black dots. The blue lines are the nonphysical solution. }
    \label{M3sig0}
\end{figure}

A plot of both solutions is shown in Fig.~\ref{M3sig0} for $\sigma_3 = 0$ ($\gamma_3$ is then irrelevant) with varying $\mu_3$. The other (plus sign) solution is negative and diverges across $\mu_3 = 0$, however the accepted (minus sign) solution is positive for all parameter combinations below the divergence point, where the two solutions become complex conjugates.
In simulations we find that the system becomes divergent (species abundances grow indefinitely) beyond this point, which occurs where the discriminant becomes zero ($z_1 = z_d$). The divergence point satisfies
\begin{align}\label{div3}
    \mu_d = \frac{1}{4k}\left(\frac{-z_1(1 - \gamma_\Sigma\chi)}{I_1}\right)^2,
\end{align}
where $z_d$ is the value(s) of $z_1$ which satisfied this for a given value of $\mu_3$, but can only exist for $\mu_3 \geq 0$. For lower values of $\mu_3$ the divergence point is found from a different mathematical condition, at $z_m$, found later.

Using the solution for $M^*$, we can also find
\begin{align}\label{sig3}
    \sigma_3 = \sqrt{\frac{2}{3}}\frac{(1 - \gamma_\Sigma\chi)I_1}{M^*I_2},
\end{align}
this is derived in Eq.~(\ref{singlepsig}) in Sec.~\ref{singlepmethod} in the Supplemental Material.
To find the onset of the linear instability, we numerically solve Eq.~(\ref{instabp}) with $p = 3$. We find $z_c \approx -0.84$, at this point $\phi \approx 0.8$ meaning the system becomes unstable when around $20\%$ of the initial $N$ species have become extinct.
Unlike the case for $p = 2$, the corresponding value of $\sigma_c$ depends on both $\gamma_3$ and $\mu_3$. Specifically, we find
\begin{align}\label{eq:sigma_c_p=3}
    \sigma_c = \sqrt{\frac{2}{3}}\frac{I_1}{M^*I_2(1 + \gamma_3)},
\end{align}
where $M^*$ depends on $\mu_3$.
Eq.~(\ref{eq:sigma_c_p=3}) is obtained by substituting the condition for the instability point from Eq.~(\ref{Hgamintro}) into Eq.~(\ref{sig3}).

\subsubsection{Mathematical solution for $z_m$}

\begin{figure*}
    \centering
    \includegraphics[width = 0.8\textwidth]{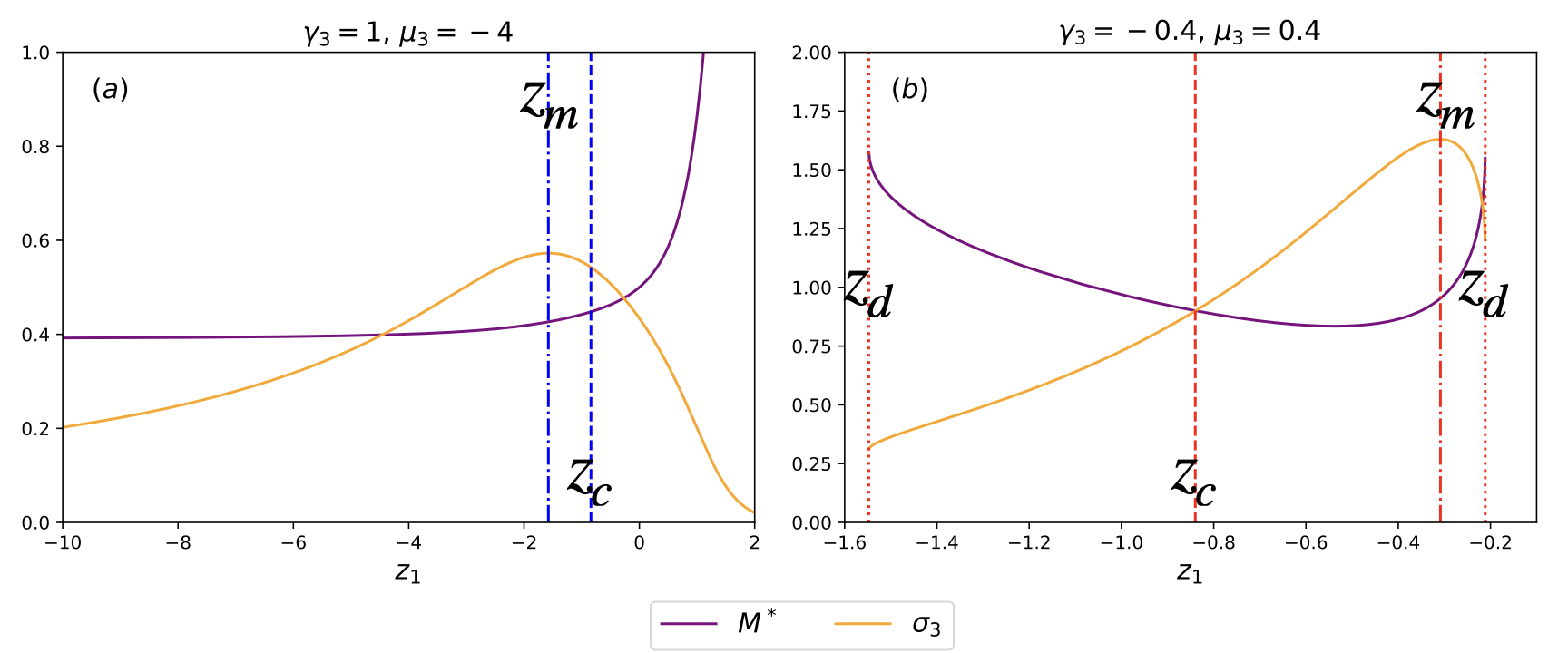}
    \caption{Dependence of $\sigma_3$ and $M^*$ on $z_1$ for the model with third-order interactions. The vertical dotted lines show $z_d$ (divergence point), the dashed lines show $z_c$ (linear instability), and the dash-dotted lines show $z_m$ (where $\sigma_3$ attains its maximum.}
    \label{z1plot}
\end{figure*}

The divergence condition above in Eq.~(\ref{div3}), can only be found for positive values of $\mu_3$, but for negative $\mu_3$ we still find divergence for large enough $\sigma_3$, this is due to another condition for divergence as discussed in Sec.~\ref{sec:z_m_intro}. For the case of third-order interactions, the existence of this other divergence condition is due to the non-monotonic dependence of $\sigma_3$ as a function of $z_1$, shown in Fig.~\ref{z1plot}. Specifically, $\sigma_3$ has a maximum at the point which we call $z_m$, this point is represented by the dashed-dotted line in the figure. The value of $\sigma_3$ at this point is denoted by $\sigma_m$. Although the maximum can in principle be found analytically by setting $d\sigma_3/dz_1 = 0$, this is very tedious in practice. Therefore, we use numerical methods to find $z_m$ and the corresponding $\sigma_m$.
As $\sigma_m$ is the maximum value of $\sigma_3$ that can be found, there exist no solutions to Eqs.~(\ref{orderparams}) for higher values of $\sigma_3$. Simulations with $\sigma_3$ above this value will result in divergence.

\begin{figure*}
    \centering
    \includegraphics[width = 0.8\textwidth]{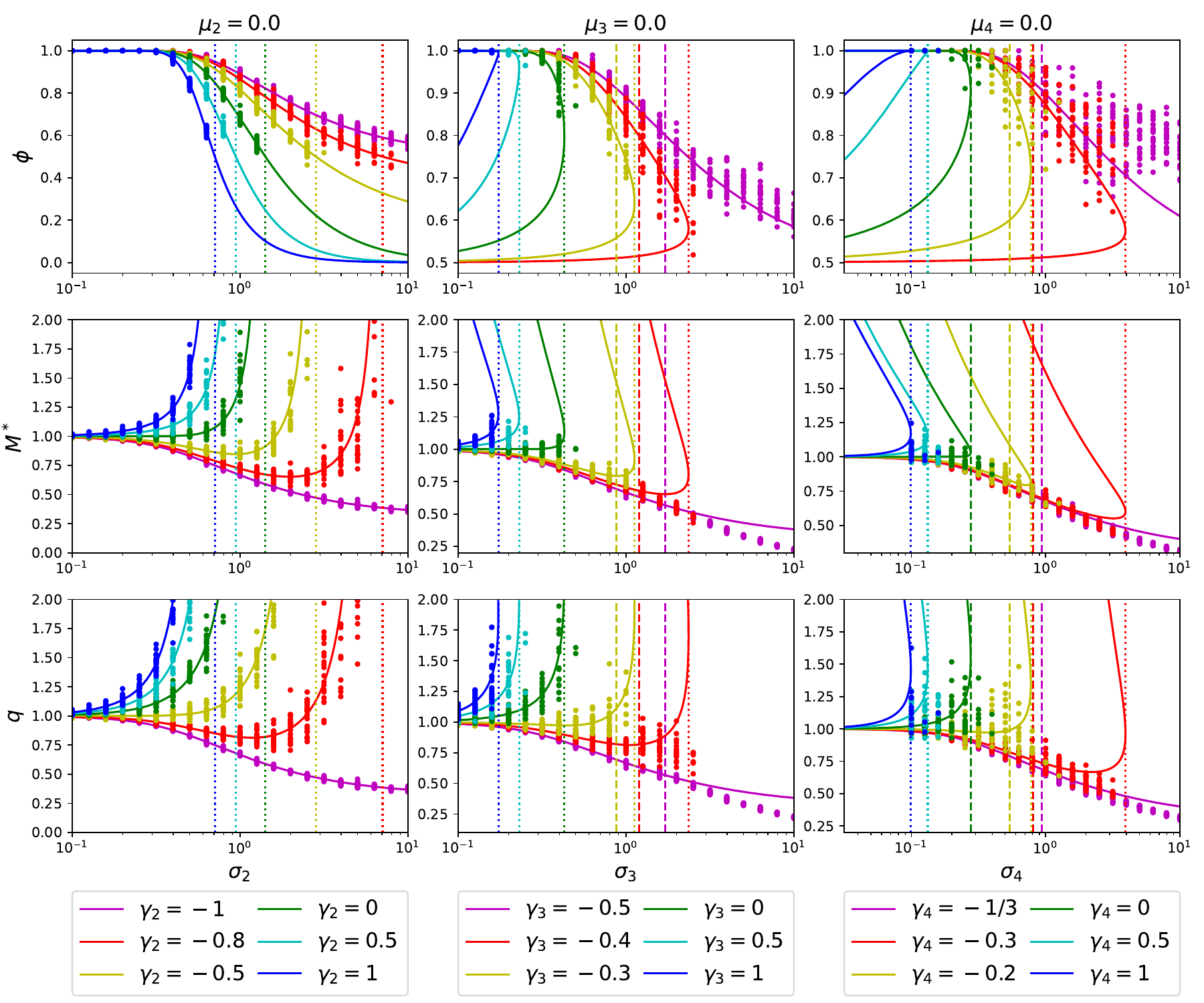}
    \caption{Plot of analytical results (lines) and simulation data (dots) for $\phi$ $M^*$ and $q$ for values of $z_m$ for interactions of orders higher than second order. Each point shows results from one of 20 simulations of $N = 500$ for $p = 2$, $N = 200$ for $p = 3$, and $N = 50$ for $p = 4$ species run for a maximum of 10000 units of time. For interactions of orders higher than 2, the analytical lines bend back after reaching $\sigma_m$ creating two branches of possible solutions for each value of $\sigma_p$.}
    \label{bendfive}
\end{figure*}

As mentioned in Sec.~\ref{sec:z_1-sigma_p}, for a single value of $p$, an increase of $z_1$ corresponds to an increase in $\sigma_p$ for a physically valid solution. We can therefore ignore the values of $\sigma_3$ obtained for values of $z_1$ past $z_m$, i.e., in the part of the curve in Fig.~\ref{z1plot} where $\sigma_3$ is decreasing in $z_1$. Any value of $z_1 > z_m$ does not correspond to a possible state of the physical system.
We confirm this in Fig.~\ref{bendfive} containing plots similar to Fig.~\ref{fivep_0}, however the analytical lines have been continued for $z_1 > z_m$ causing them to bend back on themselves for $p > 2$. In order to test whether any simulation results match the second branch, each of the 20 simulations is plotted as a separate point. Simulation results only match the lower branch, the other branch is nonphysical and can be disregarded. 

\subsubsection{Further discussion of the two examples shown in Fig.~\ref{z1plot}}

For the case of $\gamma_3 = 1$ and $\mu_3 = -4$ [Fig.~\ref{z1plot}(a)], $z_c$ (dashed line, linear instability) is higher than $z_m$ (dash-dot line). This means we cannot observe the linear instability transition, and the system should reach a unique fixed point up $\sigma_3(z_m)$ after which the system should diverge.

For the case of $\gamma_3 = -0.4$ and $\mu_3 = 0.4$ [Fig.~\ref{z1plot}(b)], we find two values for $z_d$ (dotted lines) as the divergence boundary in the $\mu_3,\sigma_3$-plane would in principle curve back on itself. However, one of these solutions for $z_d$ occurs in the nonphysical regime $z_d > z_m$, and is therefore not relevant. The other solution for $z_d$ is below $z_m$ (dash-dot line), therefore it corresponds to a physically observable transition point. At this point, $M^*$ transitions from divergent to finite (with increasing $\sigma_3$), and here $M^* = \sqrt{k/\mu_3}$. As $z_c$ (dashed line) is below $z_m$, the linear instability also occurs at a physically attainable point and can be observed in simulations.

We find that $\sigma_m$ exists for all values of $\gamma_3 \neq -1/2$ and therefore $z_m < z_d$ for the upper values of $z_d$ as shown for some cases in Fig.~\ref{z1plot}. Therefore the only physically attainable values of $z_d$ are the lower values, where no solutions exist for $z_1$ below $z_d$. The value of $\sigma_3$ above which simulations diverge is always $\sigma_m$ and never $\sigma_d$.

\subsubsection{Phase diagram}

\begin{figure}[t!!]
    \centering
    \includegraphics[width =0.45\textwidth]{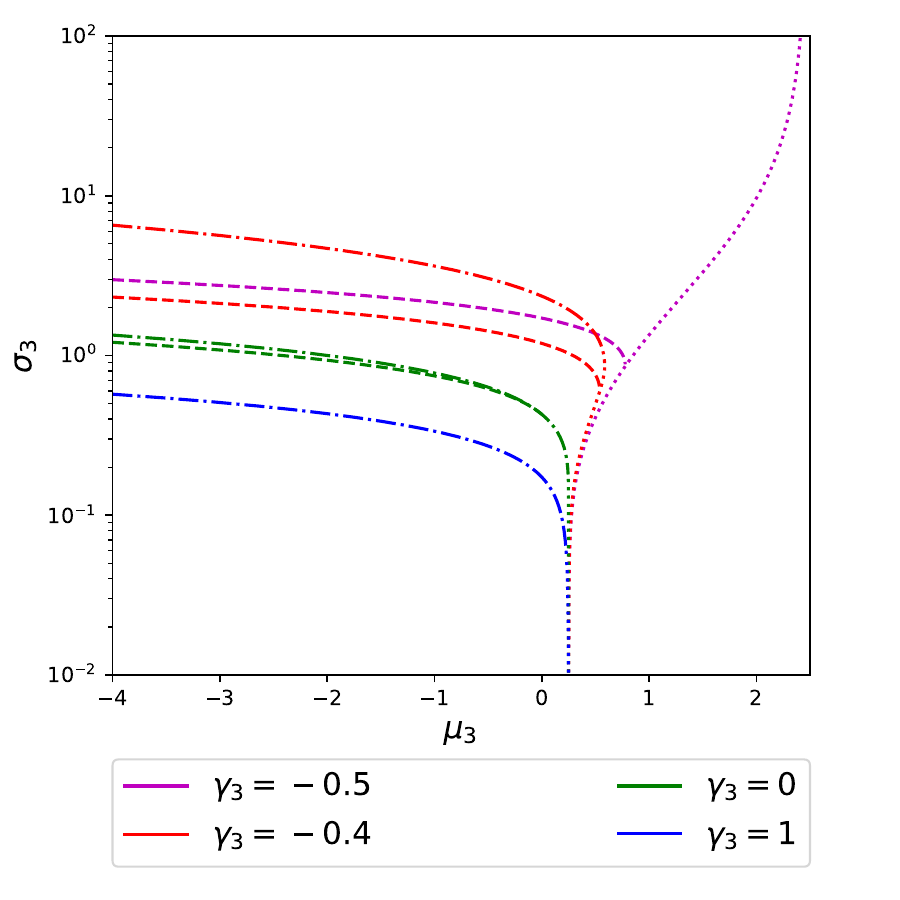}
    \caption{Phase diagram for the system with third-order interactions only, for various values of $\gamma_3$. The dash-dotted line represents $\sigma_m$, the dotted line represents $\sigma_d$, together these make up the divergence boundary, above and to the right of which the system diverges. The dashed line which can be seen under (some) divergence lines represents $\sigma_c$ and shows the linear instability boundary, below which the system reaches a unique fixed point. Between the two transition points, the system displays either multiple fixed points or persistent dynamics.}
    \label{phase3}
\end{figure}

We show the phase diagram for various values of $\gamma_3$ in Fig.~\ref{phase3}. For each value of $\gamma_3$, the dotted line represents $\sigma(z_d)$, the divergence point for analytical solutions of $M^*$. The system diverges to the right of these lines. The dotted lines have been plotted for $z_d \leq z_m$ only, as the $z_d$ transition is not observable otherwise. For $z_d > z_m$, the divergence transition occurs at $\sigma(z_m)$ which is represented by the dash-dotted line, above which the system diverges. Below this divergence point we also find the linear-instability point for some values of $\gamma_3$, shown by the dashed lines. Again, this transition is only observable for $z_c < z_m$, so the dashed lines have been ended at the point where they meet the dash-dotted line ($\sigma(z_m)$). 

\begin{figure*}[t!!]
    \centering
    \includegraphics[width = 0.95\textwidth]{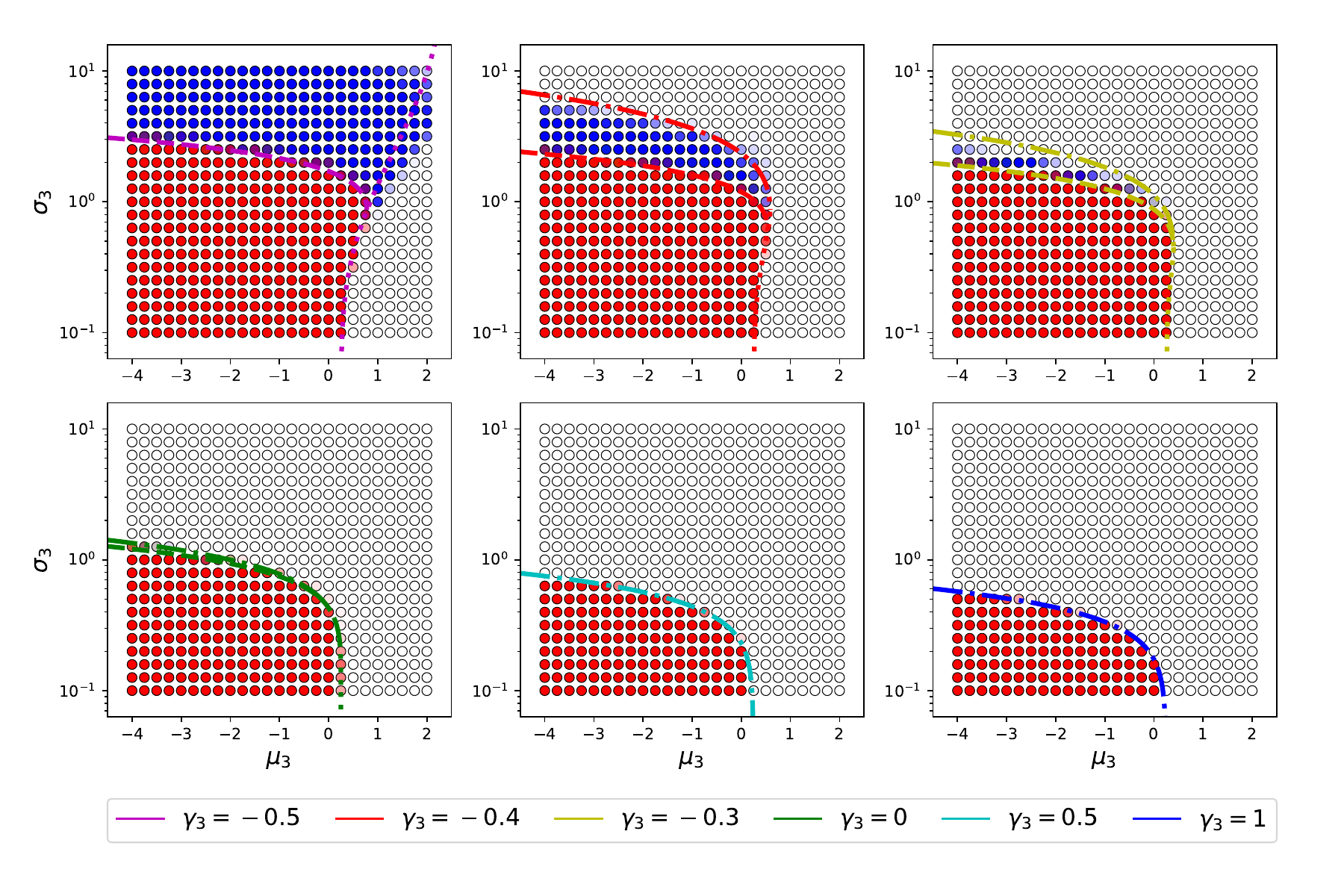}
    \caption{Phase diagrams as in Fig.~\ref{phase3} with simulation results for each value of $\gamma_3$. Each point represents the average behaviour of 20 simulations of $N = 200$ species ran for a maximum of 10000 units of time. Red indicates that the system reached a unique fixed point, green indicates multiple fixed points, blue persistent dynamics, and white divergence. The method for classifying behaviour and converting the number of each type to a colour is described in Section~\ref{smsims} in the SM.}
    \label{phasesim3}
\end{figure*}

In Fig.~\ref{phasesim3} we show the phase diagrams displayed in Fig.~\ref{phase3}, but with results from simulations,  similar to the case $p=2$ in Fig.~\ref{phasesim2}. Each circle represents average behaviour of 20 simulations of $N = 200$ species run for a maximum of 10000 units of time, but stopped before this if the system diverged or reached a fixed point. As expected, we find a unique fixed point (red) below the dashed lines, divergence (white) above the divergence lines, and instability (green and blue) between the two.

\subsubsection{Behaviour of fraction of surviving species}
\begin{figure*}
    \centering
    \includegraphics[width = 0.95\textwidth]{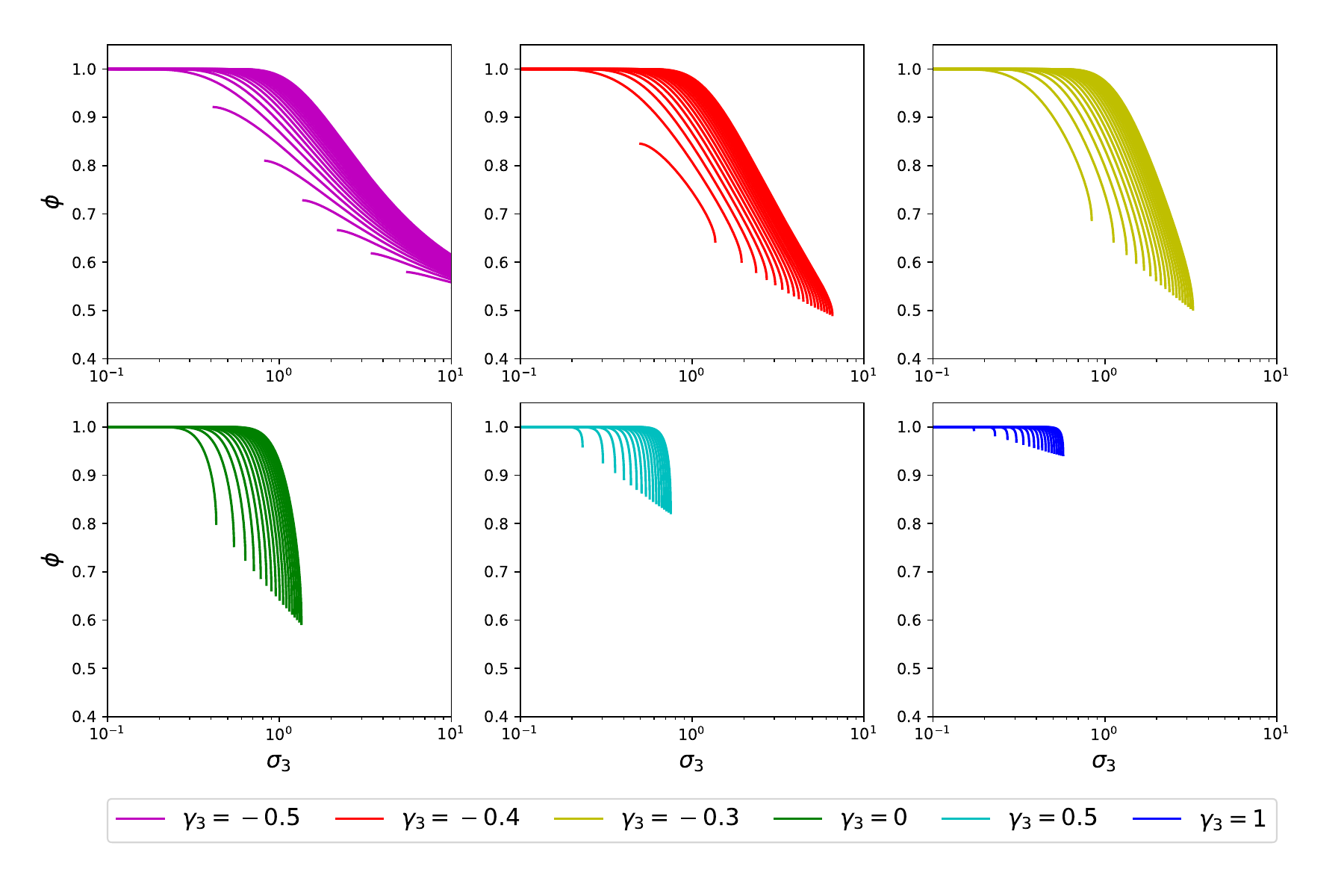}
    \caption{Fraction of surviving species in the model with third-order interactions. In each panel we show the analytical result for the fraction of surviving species, $\phi$. The different panels are for different choices of $\gamma_3$ as indicated. The different lines in each panel are for different choices of $\mu_3$, with $\mu_3$ increasing from top to bottom in each panel.}
    \label{allphiplots3}
\end{figure*}

In Fig.~\ref{allphiplots3} we show the fraction of surviving species $\phi$ as a function of $\sigma_3$ for each value of $\mu_3$ from -4 to 2 in steps of 0.25. (from top to bottom in the figure). This shows that lower values of $\mu_3$, $\sigma_3$ and $\gamma_3$ lead to more diversity. This can also be understood from the phase diagram (Fig.~\ref{phase3}). Unlike in the pairwise case, the linear instability line is not horizontal. Along this line we have a constant value of $z_1$ at $z_c \approx -0.84$ and therefore a constant value of $\phi$. For increasing values of $\mu_3$, this instability point occurs at lower values of $\sigma_3$, which means for a constant value of $\sigma_3$, $\phi$ decreases with increasing $\mu_3$.

\subsubsection{Behaviour of average species abundance $M^*$}
\begin{figure*}
    \centering
    \includegraphics[width = 0.95\textwidth]{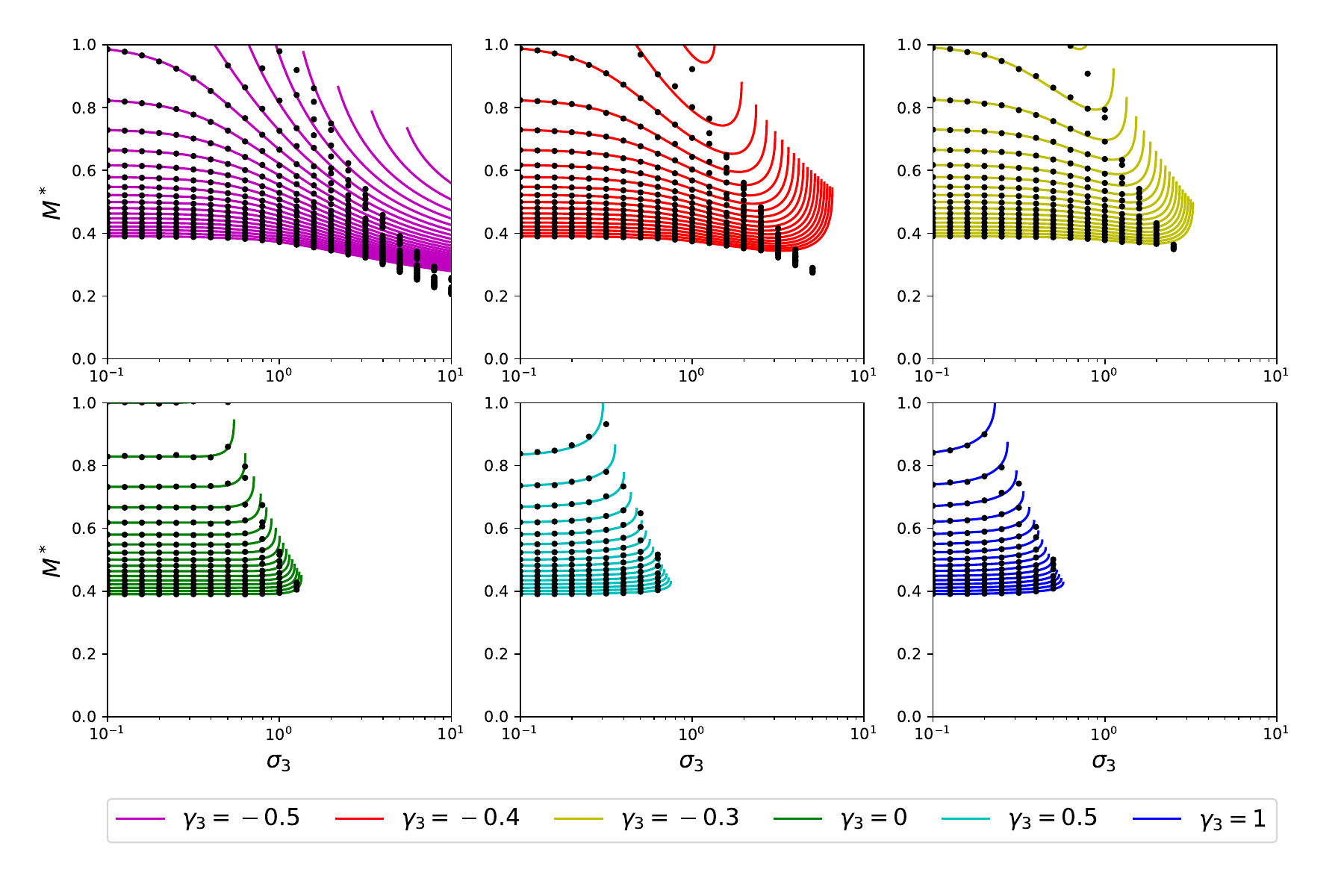}
    \caption{Plots of analytical and simulation results for $M^*$ against $\sigma_3$ for all values of $\mu_3$ shown in Fig.~\ref{phasesim3}, between -4 and 2 in increments of 0.25. The lines get higher with increasing values of $\mu_3$. Each point shows the average results from 20 simulations of $N = 200$ species ran for a maximum of 10000 units of time.}
    \label{allMplots3}
\end{figure*}

\begin{figure*}
    \centering
    \includegraphics[width = 0.95\textwidth]{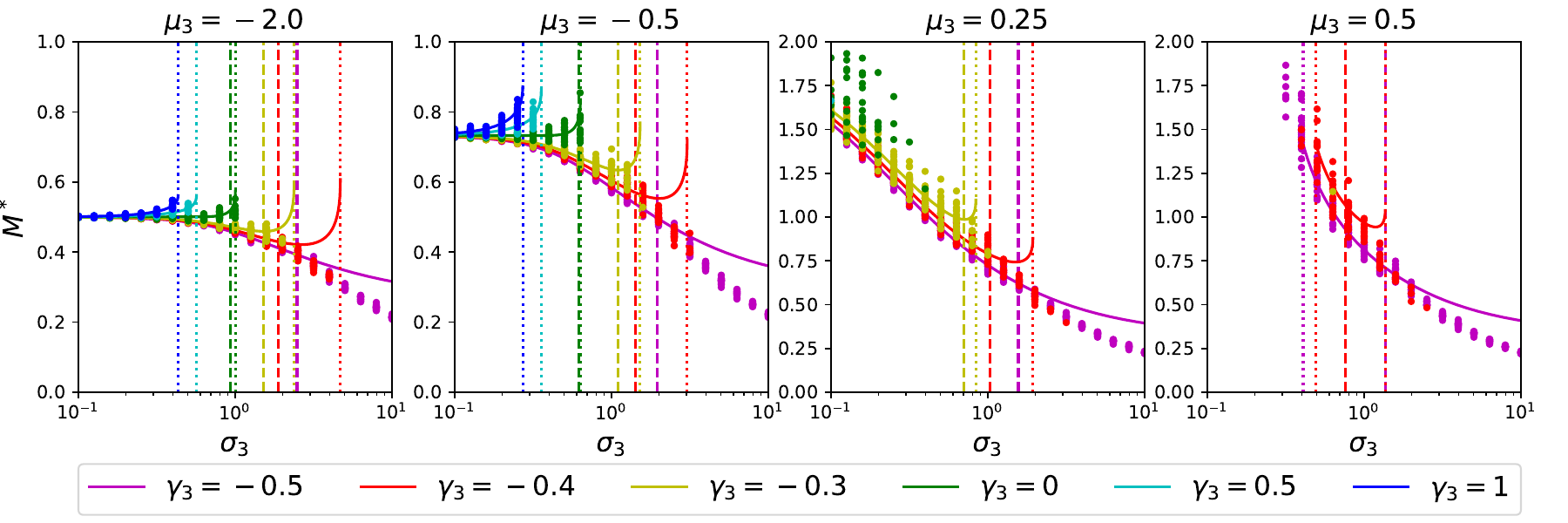}
    \caption{Plots of analytical and simulation results for $M^*$ in the model with third-order interactions for various $\mu_3$ and $\gamma_3$, and varying $\sigma_3$. Each point shows the average results from 20 simulations of $N = 200$ species ran for a maximum of 10000 units of time. For each value of $\gamma_3$, the dashed line represents $\sigma_c$ and the dotted line(s) represent the divergence point(s) at either $\sigma_d$ or $\sigma_m$.}
    \label{Mplots3}
\end{figure*}

Using Eq.~(\ref{div3}) we can rewrite Eq.~(\ref{eq:M*_p=3mu0}) as
\begin{align}
    M^* = \frac{1}{2}\sqrt{\frac{k}{\mu_d}}
\end{align}
for $\mu_3 = 0$, and Eq.~(\ref{eq:M*_p=3}) as
\begin{align}
    M^* = \frac{\sqrt{k}}{\mu_3}(\sqrt{\mu_d} - \sqrt{\mu_d -\mu_3})
\end{align}
for $\mu_3 \neq 0$. We can see that $M^*$ depends on the distance away from the divergence boundary ($\mu_d - \mu_3$), but this time not solely as before with $p = 2$. This means that the shape of the analytical lines in Fig.~\ref{allMplots3} loosely follow the shape of the divergence boundary in the phase diagram for each value of $\gamma_3$. At the lower divergence point ($z_1 = z_d$) we find $M^* = \sqrt{k/\mu_3}$, which is a finite value for $\mu_3 \neq 0$, and only depends on $\mu_3$. This can be seen in the last panel of Fig.~\ref{Mplots3} where both analytical lines begin at the same value of $M^*$. The value of $M^*$ is also finite at the upper divergence point ($z_1 = z_m$) as $\sigma_m$ is finite at this point. Unlike the case with $p = 2$, the value of $M^*$ does not diverge as it approaches the divergence boundary, so is limited to a finite value, unless the system is past the divergence point and in the divergence phase.

For $\sigma_3 = 0$ ($z_1 \to -\infty$), we have $\mu_d = 1/(4k)$ for all values of $\gamma_3$, this is derived in Sec.~\ref{sec:singlepdiv} of the SM. Eq.~(\ref{eq:M*_p=3}) becomes
\begin{align}
    M^* = \frac{1}{2\mu_3}(1 - \sqrt{1 - 4\mu_3k})
\end{align}
for $\mu_3 \neq 0$, and Eq.~(\ref{eq:M*_p=3mu0}) becomes $M^* = k$ for $\mu_3 = 0$. This function is plotted as the red line in Fig.~\ref{M3sig0}, which is continuous across $\mu_3 = 0$ and reaches $M^* = 2k$ at the divergence point after which the solutions become complex. The two left panels of Fig.~\ref{Mplots3} show $M^*$ begins at the same value for all values of $\gamma_3$ for low $\sigma_3$.

For fully antisymmetric interactions, where $\gamma_3 = -1/2$, there is no maximum value of $\sigma_3$ ($\sigma_m$), as we find $\sigma_3 \to \infty$ as $z_1 \to 0^-$, derived in Sec.~\ref{sec:singlepanti} of the SM. This is shown in Fig.~\ref{bendfive} where the magenta lines have only one branch and do not bend back on themselves, and so solutions exist for all values of $\sigma_3$. This is true as long as $\mu_3 \leq \mu_d$, where $\mu_d \to \pi^2/(4k)$, at which point $M^* = 2k/\pi$. In this limit Eq.~(\ref{eq:M*_p=3}) becomes
\begin{align}
    M^* = \frac{\pi - \sqrt{\pi^2 - 4\mu_3k}}{2\mu_3}
\end{align}
for $\mu_3 \neq 0$, and Eq.~(\ref{eq:M*_p=3mu0}) becomes $M^* = k/\pi$ for $\mu_3 = 0$. This demonstrates that both $M^*$ and $q = \pi M^{*2}$ remain finite for all values of $\sigma_3$ and no upper divergence point exists, only a lower point where simulations diverge for values of $\sigma_3$ below this point.
As the linear instability point occurs at $z_c \approx -0.84$, which is below $z_1 = 0$ where $\sigma_3 \to \infty$, antisymmetric interactions still can become linearly unstable for large enough $\sigma_3$.  This is very different from the system with $p=2$, where the linear instability cannot occur for fully antisymmetric interactions. Similar to the model with $p=2$ the mean abundance does not become divergent for any value of $\sigma_3$ as long as $\mu_3$ is below the divergence boundary.

\subsection{Fourth order interactions}\label{p=4sec}

We now consider the system with fourth-order interactions only. Again we investigate the behaviour of the system for parameters $\sigma_4$, $\mu_4$ and $\gamma_4$, and find the transition points. We find that the transition points follow similar conditions to the previous case with $p = 3$.

\subsubsection{Mathematical results}

We solve Eq.~(\ref{instab}) with $p = 4$ for find the instability point, $z_c \approx -1.326$. At this point we find $\phi \approx 0.9$, i.e., the system becomes unstable after around $10\%$ of the initial $N$ species have become extinct. Similarly to the case for $p = 3$, the value of $\sigma_4$ where this instability point occurs depends on both $\gamma_4$ and $\mu_4$,
\begin{align}\label{sig4c}
    \sigma_c = \frac{1}{(1 + \gamma_4)\sqrt{2I_2^3}}\left(\frac{I_1}{M^*}\right)^2.
\end{align}
This is obtained by substituting Eq.~(\ref{Hgamintro}) into Eq.~(\ref{singlepsig}) from Section \ref{singlepinstab} in the Supplemental Material.
The value of $M^*$ satisfies Eq.~(\ref{eq:m_poly}) which for $p = 4$ becomes the cubic equation
\begin{align}\label{Mcubic}
    \mu_4 M^{*3} + \frac{z_1(1 - \gamma_\Sigma\chi)}{I_1}M^* + k = 0,
\end{align}
where $1 - \gamma_\Sigma \chi$ is found via Eq.~(\ref{help}).
This can be solved using Cadano's cubic formula, resulting in three solutions for $M^*$,
\begin{align}\label{eq:M123}
    M_1 &= S + T, \nonumber\\
    M_2 &= - \frac{S + T}{2} + \frac{i\sqrt{3}(S - T)}{2}, \nonumber\\
    M_3 &= - \frac{S + T}{2} - \frac{i\sqrt{3}(S - T)}{2},
\end{align}
where
\begin{align}\label{p4ST}
    S &= \left(\frac{-k}{2\mu_4} + \sqrt{\left(\frac{z_1(1 - \gamma_\Sigma\chi)}{3\mu_4 I_1}\right)^3 + \left(\frac{k}{2\mu_4}\right)^2}\right)^{1/3}, \nonumber\\
    T &= \left(\frac{-k}{2\mu_4} - \sqrt{\left(\frac{z_1(1 - \gamma_\Sigma\chi)}{3\mu_4 I_1}\right)^3 + \left(\frac{k}{2\mu_4}\right)^2}\right)^{1/3}.
\end{align}

\begin{figure}[t!!]
    \centering
    \includegraphics[width = 0.5\textwidth]{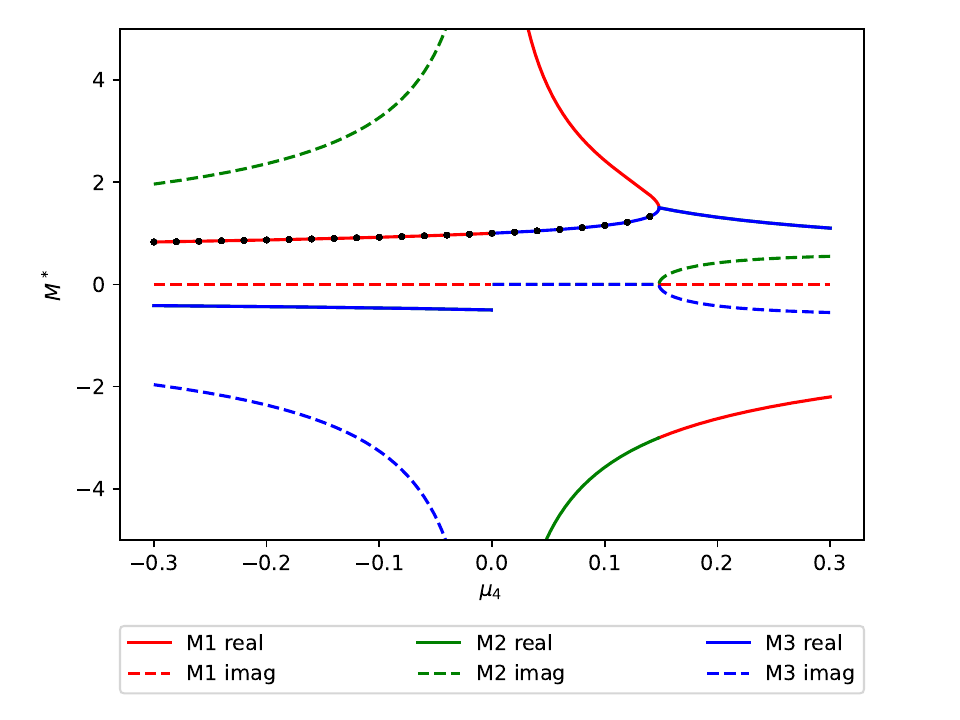}    \caption{Plot of real and imaginary parts of the three solutions for $M^*$ [Eq.~(\ref{eq:M123})] with varying $\mu_4$ and $\sigma_4 = 0$.  Black points are from simulations.}
    \label{M4sig0}
\end{figure}

A plot of both the real and imaginary parts of the three solutions is shown in Fig.~\ref{M4sig0} for $\sigma_4 = 0$ ($\gamma_4$ is then irrelevant) with varying $\mu_4$. From the requirement that the solution for $M^*$ is real, positive, and a continuous function of $\mu_4$, we find that for $\mu_4 < 0$, $M_1$ gives the correct solution, and for $\mu_4 > 0$, $M_3$ gives the correct solution. However, for $\mu_4 = 0$, Eq.~(\ref{Mcubic}) is no longer a cubic equation, and has the solution
\begin{align}
    M^* = \frac{-kI_1}{z_1(1 - \gamma_\Sigma\chi)}.
\end{align}
The solution $M_1$ (for negative $\mu_4$) is always real, but $M_3$ (for positive $\mu_4$) becomes complex when the discriminant,
\begin{align}
    D = \left(\frac{z_1(1 - \gamma_\Sigma\chi)}{3\mu_4 I_1}\right)^3 + \left(\frac{k}{2\mu_4}\right)^2,
\end{align}
is equal to zero. Setting this equal to zero gives an expression for the divergence point,
\begin{align}\label{eq:p4mud}
    \mu_d = -\frac{4}{k^2}\left(\frac{-z_1(1 - \gamma_\Sigma\chi)}{3I_1}\right)^3,
\end{align}
where $z_d$ is the value(s) of $z_1$ which satisfies this for a given value of $\mu_4$, and can be found for values of $\mu_4 \geq 0$.

Similarly to the case with $p = 3$, this $z_d$ determines the lower divergence point. Values of $\sigma_4$ below $\sigma_4(z_d)$ lead to divergence of species abundance. The upper divergence point is determined by $z_m$, the value of $z_1$ where $\sigma_4$ reaches its maximum value $\sigma_m$. For values of $\sigma_4$ above $\sigma_4(z_m)$ the mean abundance again diverges.
Fig.~\ref{bendfive} shows that values of $z_1$ beyond $z_d$, where the analytical lines bend back on themselves, do not correspond to physical states of the system, and can be disregarded as in the case with $p = 3$.

The right-hand side of Eq.~(\ref{eq:p4mud}) appears in Eq.~(\ref{p4ST}). Replacing this expression with $\mu_d$ in the solutions for $M^*$ shows that again the behaviour of $M^*$ depends on $\mu_4 - \mu_d$, the distance away for the divergence boundary, see Eq.~(\ref{STmud}) in the SM for the full expression. At the divergence point we find
\begin{align}
    M^* = \left(\frac{k}{2\mu_4}\right)^{1/3},
\end{align}
which becomes infinite for $\mu_4 = 0$, but remains finite for $\mu_4 \neq 0$, and depends on $\mu_4$ only. For $\sigma_4 = 0$ ($z_1 \to -\infty$), we find that divergence occurs at $\mu_d = 4/(27k^2)$ for all values of $\gamma_4$. The mean abundance is $M^* = k$ for $\mu_4 = 0$, and $M^* = 3k/2$ for $\mu_4 = \mu_d$. These values can be seen in Fig.~\ref{M4sig0} (we recall that $k=1$ in all figures). The derivation of these values along with the full expression for $M^*$ for each case can be found in Section \ref{sec:singlepdiv} of the SM.

For antisymmetric interactions, where $\gamma_4 = -1/3$,
\begin{align}
    \mu_d \to \frac{4}{k^2}\left(\frac{\pi}{3}\right)^3
\end{align}
as $\sigma_4 \to \infty$ ($z_1 \to 0^-$), at which point $M^* = 3k/(2\pi)$. This value of $\mu_d$ can be seen in Fig.~\ref{phase4}, and is derived in Section \ref{sec:singlepanti} of the SM. Similarly to the cases with lower values of $p$, both $M^*$ and $q$ tend to finite constants, with $q = \pi M^{*2}$ as $\sigma_4 \to \infty$ when interactions are fully antisymmetric.

\subsubsection{Phase diagram, fraction of surviving species and mean abundance}

\begin{figure}[t!!]
    \centering
    \includegraphics[width = 0.5\textwidth]{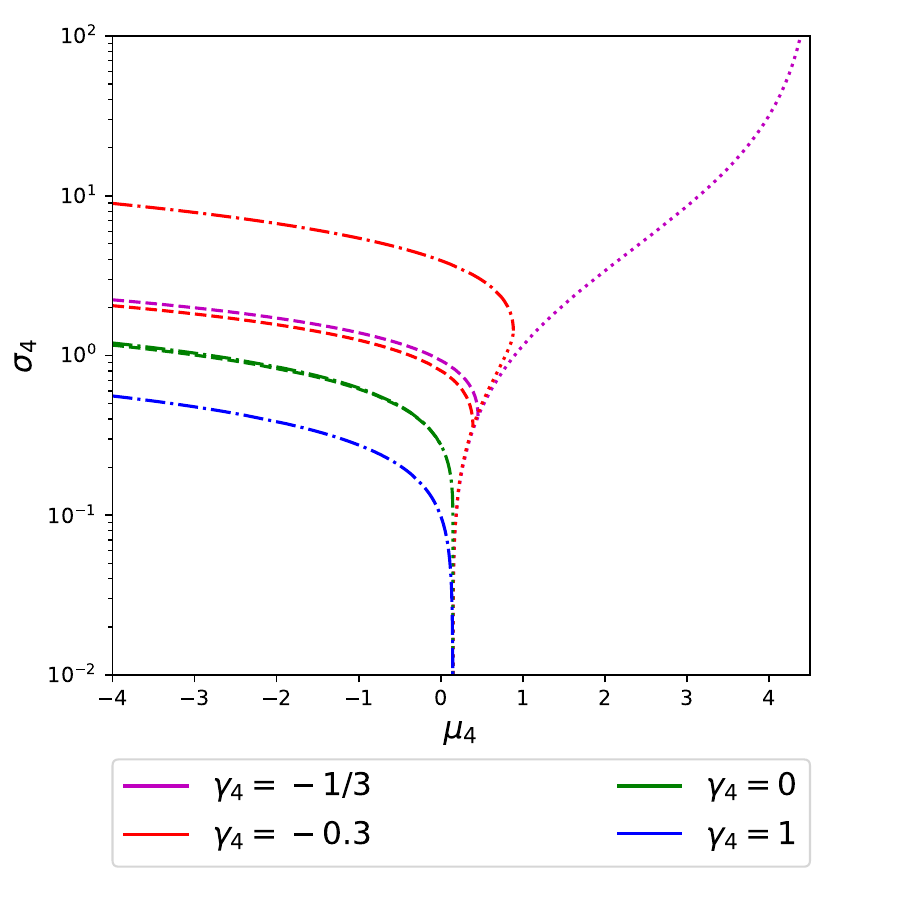}
    \caption{Phase diagram for the system with fourth order interactions only, for various values of $\gamma_4$. The dash-dot line represents $\sigma_m$, the dotted line represents $\sigma_d$, together these make up the divergence boundary, above and to the right of which the system diverges. The dashed line which can be seen under (some) divergence lines represents $\sigma_c$ and shows the instability boundary, below which the system reaches a unique fixed point. Between the two transition points, the system displays either multiple fixed points or persistent dynamics.}
    \label{phase4}
\end{figure}

\begin{figure*}[t!!]
    \centering
    \includegraphics[width = \textwidth]{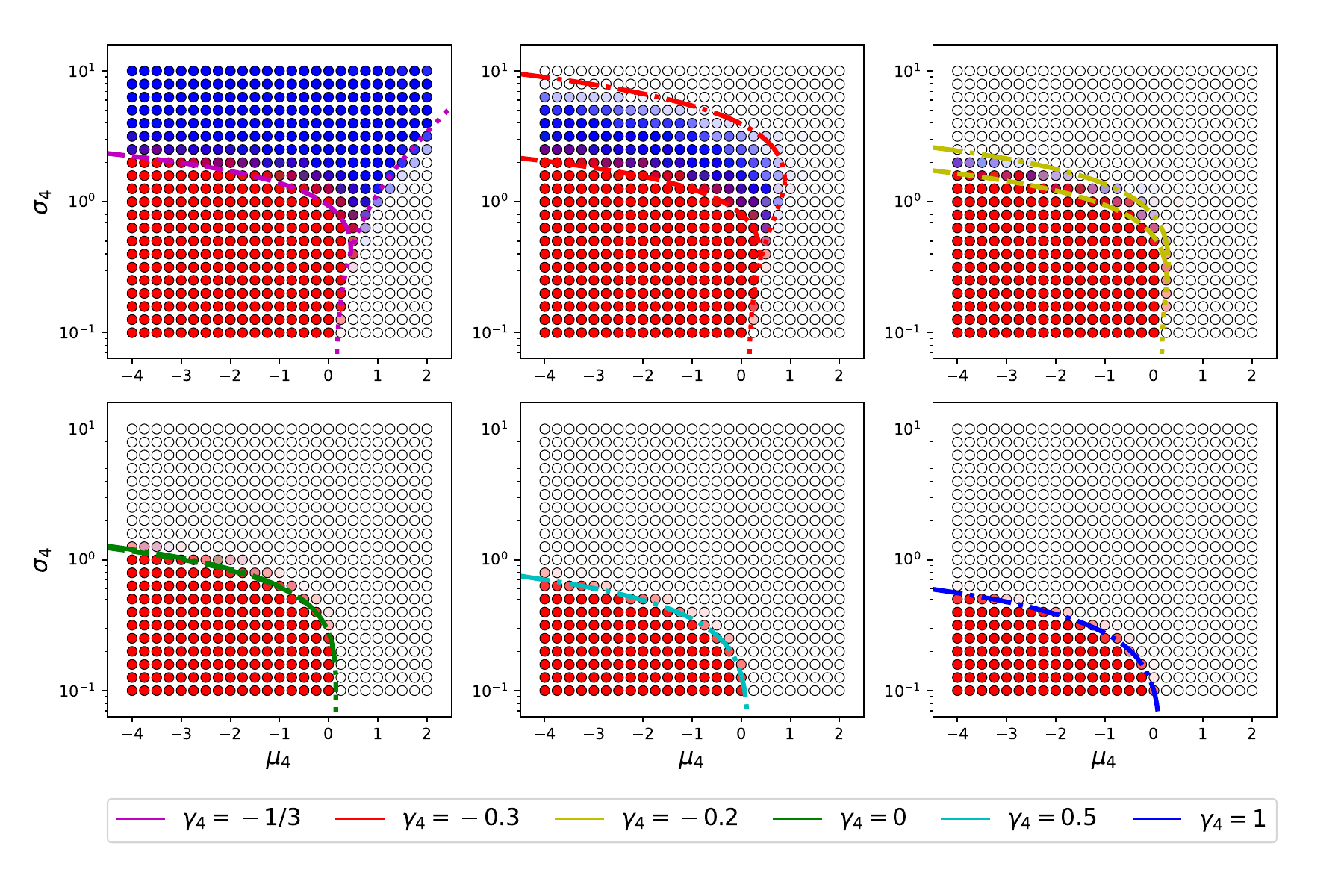}
    \caption{Phase diagrams as in Fig.~\ref{phase4} (model with fourth-order interactions) with simulation results for each value of $\gamma_4$. Each point represents the average behaviour of 20 simulations for $N = 50$ species run for 10000 units of time. Red indicates that the system reached a unique fixed point, green indicates multiple fixed points, blue persistent dynamics, and white divergence. The method for classifying behaviour and converting the number of each type to a colour is described in Section \ref{smsims} in the SM.}
    \label{phasesim4}
\end{figure*}

Figure~\ref{phase4} shows the analytical solutions for the transition points for various values of $\gamma_4$ in the phase diagram. These results are confirmed by simulations in Fig.~\ref{phasesim4}, where each circle represents the average behaviour of 20 runs with $N = 50$ species for a maximum of 10000 units of time. The simulation was stopped before this time if the system reached a fixed point or diverged.

\begin{figure*}
    \centering
    \includegraphics[width = 0.95\textwidth]{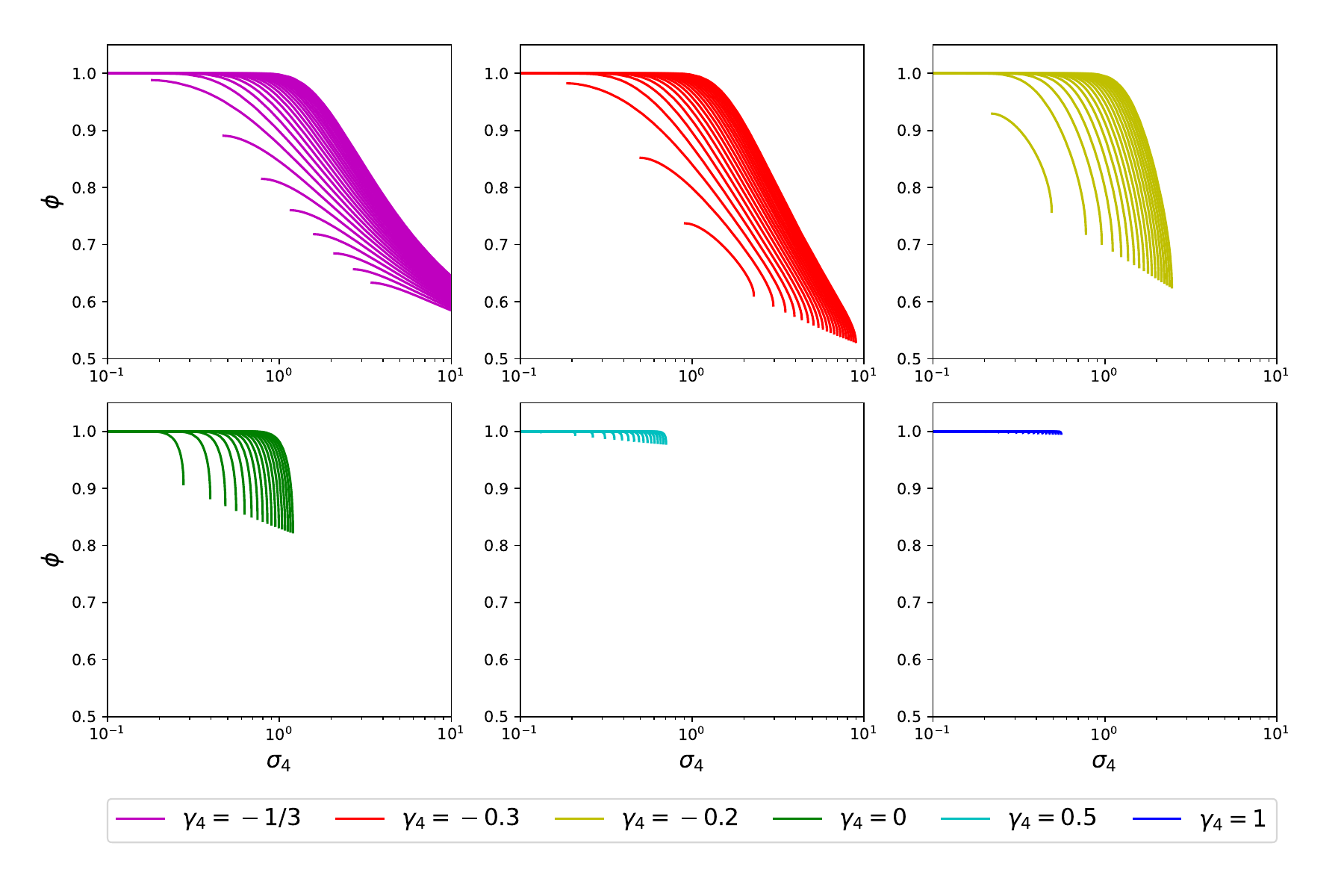}
    \caption{Fraction of surviving species in the model with fourth-order interactions. In each panel we show the analytical result for the fraction of surviving species, $\phi$. The different panels are for different choices of $\gamma_4$ as indicated. The different lines in each panel are for different choices of $\mu_4$, with $\mu_4$ increasing from top to bottom in each panel.}
    \label{allphiplots4}
\end{figure*}

The fraction of surviving species is shown in Fig.~\ref{allphiplots4} as a function of $\sigma_4$ for each value of $\mu_4$ between -4 and 2 in steps of 0.25 (top to bottom in the figure). Lower values of $\sigma_4$, $\gamma_4$ and $\mu_4$ lead to higher diversity as with third-order interactions.

\begin{figure*}
    \centering
    \includegraphics[width = 0.95\textwidth]{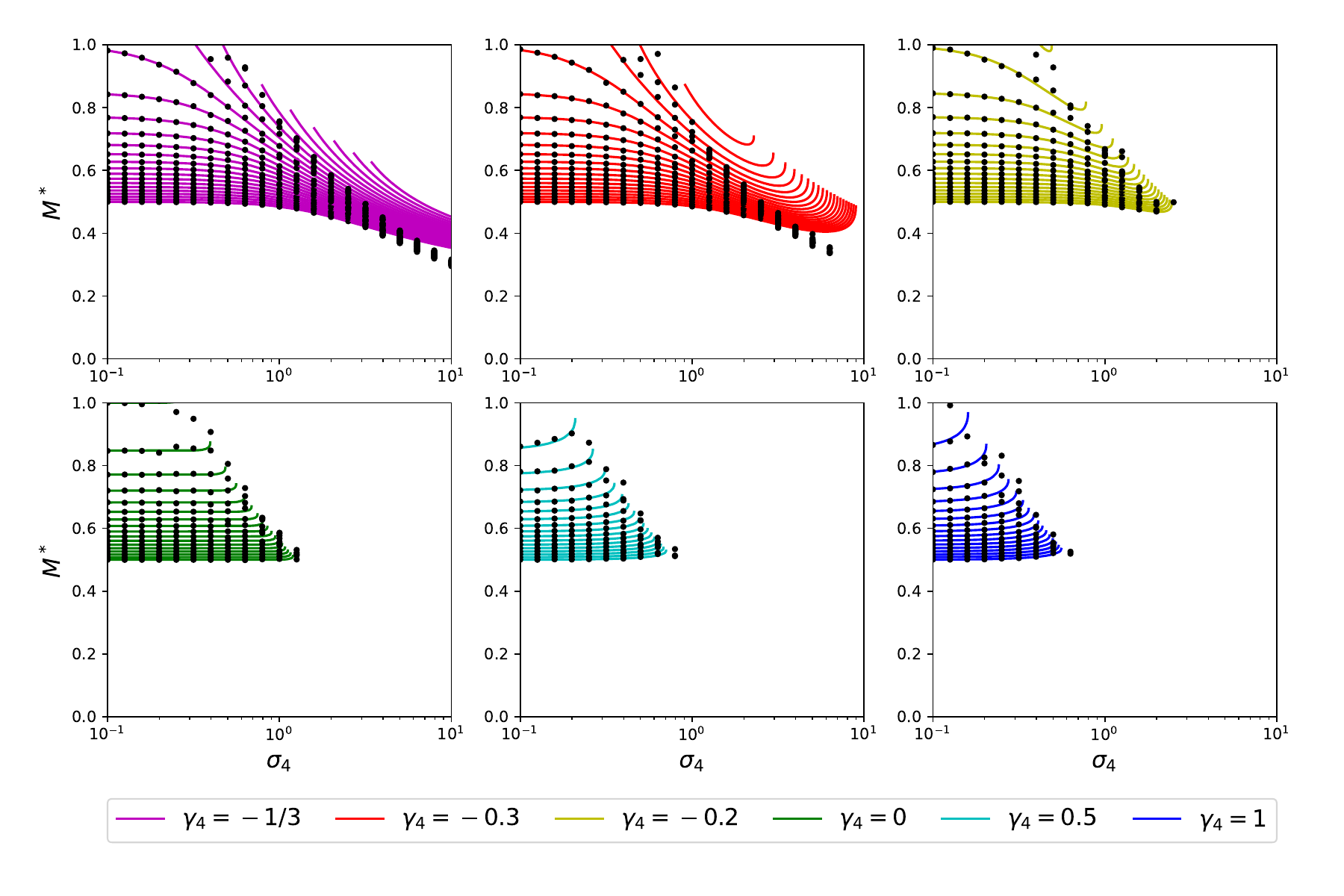}
    \caption{Plots of analytical and simulation results for $M^*$ against $\sigma_4$ for all values of $\mu_4$ shown in Fig.~\ref{phasesim4}, between -4 and 2 in increments of 0.25 (bottom to top). Lines are from the theory. Each point shows the average results from 20 simulations of $N = 50$ species run for a maximum of 10000 units of time.}\label{allMplots4}
\end{figure*}

\begin{figure*}[t!!]
    \centering
    \includegraphics[width = 0.95\textwidth]{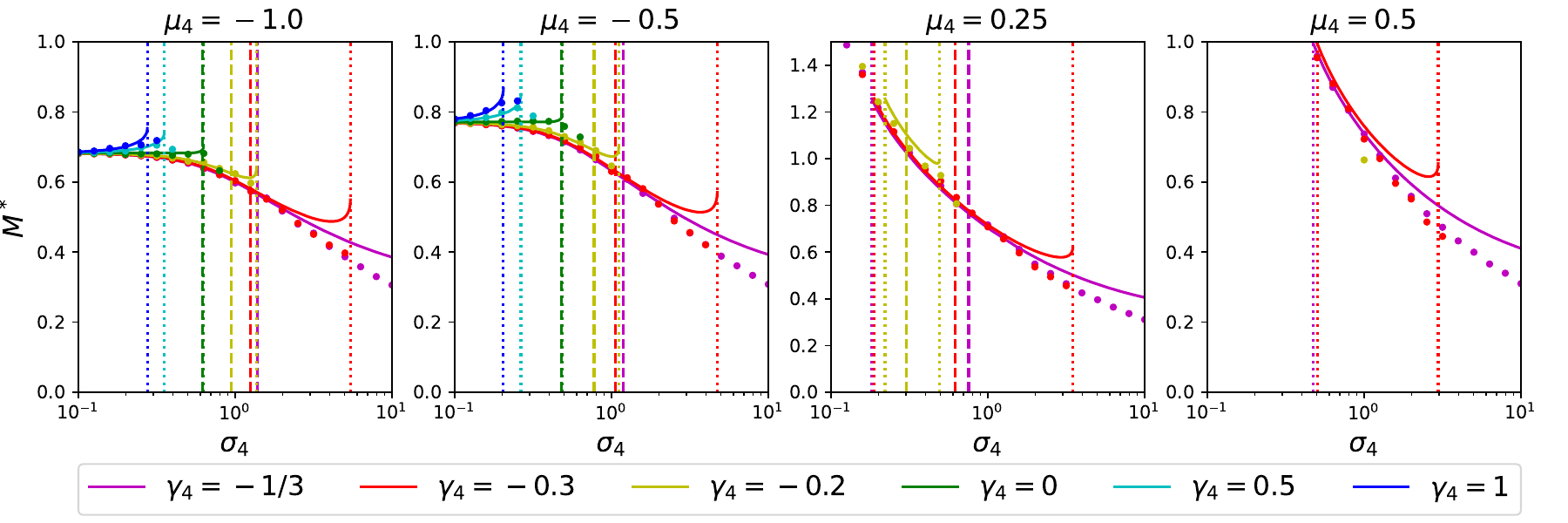}
    \caption{Plots of analytical and simulation results for $M^*$ in the model with fourth-order interactions for various $\mu_4$ and $\gamma_4$, and varying $\sigma_4$. Lines are from the theory. Each point shows the average results from 20 simulations of $N = 50$ species run for a maximum of 10000 units of time. For each value of $\gamma_4$, the dashed line represents $\sigma_c$ and the dotted line(s) represent the divergence point(s) at either $\sigma_d$ or $\sigma_m$.}
    \label{Mplots4}
\end{figure*}

We show how the behaviour of the average species abundance $M^*$ varies with $\sigma_4$ and $\mu_4$ for each value of $\gamma_4$ in Fig.~\ref{allMplots4}. The values of the system parameters are the same as in Fig.~\ref{phasesim4}, with $\mu_4$ between -4 and 2 in steps of 0.25, and the values of $\gamma_4$ as indicated in the figure. As the solution for $M^*$ depends on $\mu_4 - \mu_d$, but in a more complicated way than in previous values of $p$, the shape of the analytical lines very loosely follow the shape of the divergence boundary.
Figure~\ref{Mplots4} shows how the behaviour of $M^*$ varies with $\sigma_4$ and $\gamma_4$ for some chosen values of $\mu_4$. For $\sigma_4 = 0$, value of $M^*$ only depends on $\mu_4$, and the lines for all values of $\gamma_4$ begin at the same point for low $\sigma_4$. Comparing simulation results to analytical lines, we find they match well below the linear instability point $\sigma_c$, marked by the dashed line, but become different above this line.

\subsection{Higher orders of interactions}

Although we have not studied interactions with orders higher than $p = 4$, we can make speculations based on patterns. The polynomial for $M^*$ is of degree $p - 1$, so solutions for $M^*$ become increasingly complex for increasing $p$. For orders higher than $p = 2$, we found the solution for $\mu_d$ has to be a positive value. There is however another condition for divergence for other values of $\mu_p$. The existence of this other divergence condition was a result of $\sigma_p$ reaching a maximum point as a function of $z_1$. This maximum point may continue to exist for higher values of $p$, or divergence points may be due to other conditions.
The instability point, defined by  the condition in Eq.~(\ref{instabp}) was found to be satisfied by $z_c = 0$ for $p = 2$, $z_c \approx -0.84$ for $p = 3$, and $z_c \approx -1.33$ for $p = 4$. We find that as $p$ increases, $z_c$ continues to decrease, with $z_c \to -\infty$ as $p \to \infty$. This suggests that in general, the stability of the system decreases for interactions with higher orders.
For interactions without heterogeneity, $\sigma_p = 0$ we found the divergence point at the same value of $\mu_p$ for all values of $\gamma_p$. The divergence points were $\mu_d = 1$ for $p = 2$, $\mu_d = 1/(4k)$ for $p = 3$, and $\mu_d = 4/(27k^2)$ for $p = 4$. This point decreased as $p$ increased, and potentially $\mu_d \to 0$ as $p \to \infty$.
For each value of $p$, we consistently found that systems with fully antisymmetric interactions do not become divergent for increased variance, but for co-operative interactions, can have a point below which the abundance diverges. For the limit of $\sigma_p \to \infty$, we found $\mu_d \to \pi$ for $p = 2$, $\mu_d \to \pi^2/(4k)$ for $p = 3$, and $\mu_d \to \frac{4}{k^2}\left(\frac{\pi}{3}\right)^3$ for $p = 4$. For each case, this value is the same as $\mu_d$ for $\sigma_p = 0$, multiplied by $\pi^{p-1}$, so this may continue to increase as $p$ increases.

\subsection{Combination of second and third order interactions}
\subsubsection{Variance within only one of either second order or third order interactions}
\begin{figure*}
    \centering
    \vspace*{-3cm}
    \includegraphics[width = 0.85\textwidth]{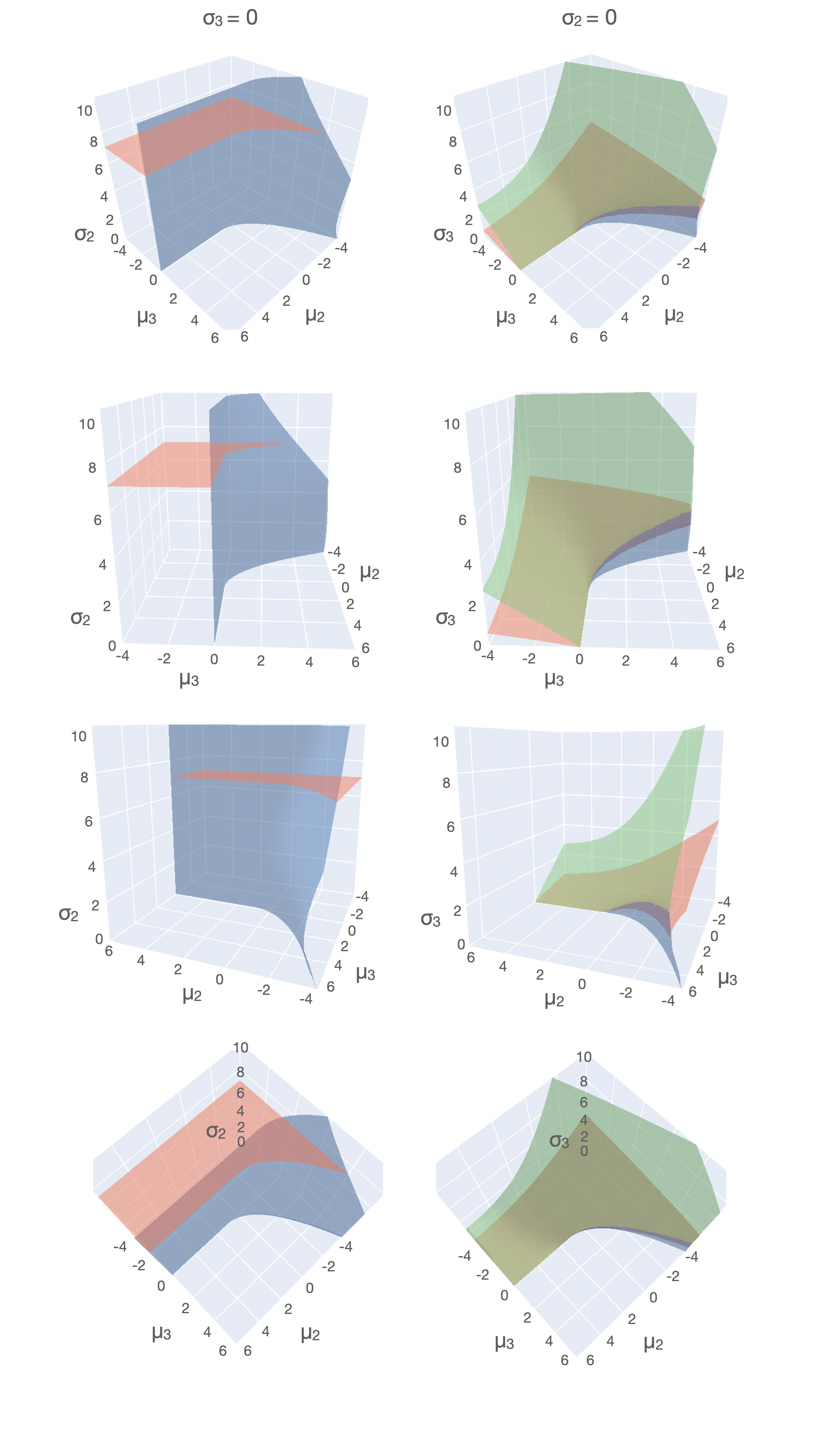}
    \vspace*{-1cm}
    \caption{Each column shows plots of four different perspectives of the same 3D phase diagram for $\gamma_2 = -0.8$ and $\gamma_3 = -0.4$ to help the reader understand the shapes in 3D. The left and right plots are three-dimensional  subspaces of the four-dimensional space spanned by $\mu_2$, $\mu_3$, $\sigma_2$, and $\sigma_3$. The plots on the left are for $\sigma_3 = 0$ with variation of second-order interactions only, and the right-hand plots are for $\sigma_2 = 0$, with variation of third-order interactions only. Each plot shows the transition point surfaces of the linear instability point $\sigma_c$ in red, and the divergence point $\sigma_d$ in blue and $\sigma_m$ in green against the $(\mu_2, \mu_3)$ plane.}
    \label{3dplots}
\end{figure*}

For a system with second- and third-order interactions, there are 6 different parameters ($\mu_2, \mu_3, \sigma_2, \sigma_3, \gamma_2, \gamma_3$). In 3 dimensions we can only show the phase diagram with at most 3 variables. In Fig.~\ref{3dplots} we show multiple perspectives of one individual phase diagram with fixed values of $\gamma_2 = -0.8$ and $\gamma_3 = -0.4$. The plane at the base of all diagrams show the divergence transition through $(\mu_2, \mu_3)$ space for both $\sigma_2$ and $\sigma_3$ set at zero. The four diagrams on the left show different perspectives of the 3-dimensional plot of the transition boundaries for increasing $\sigma_2$ (on the vertical axis) while keeping $\sigma_3$ set at zero, and the four diagrams on the right show different perspectives of instead increasing $\sigma_3$ with $\sigma_2$ set at zero.
Multiple perspectives are shown to help the reader visualise the shape in three dimensions, but interactive graphs are provided online at \cite{website}.

\begin{figure*}
    \centering
    \vspace*{-4cm}
    \includegraphics[width = 0.8\textwidth]{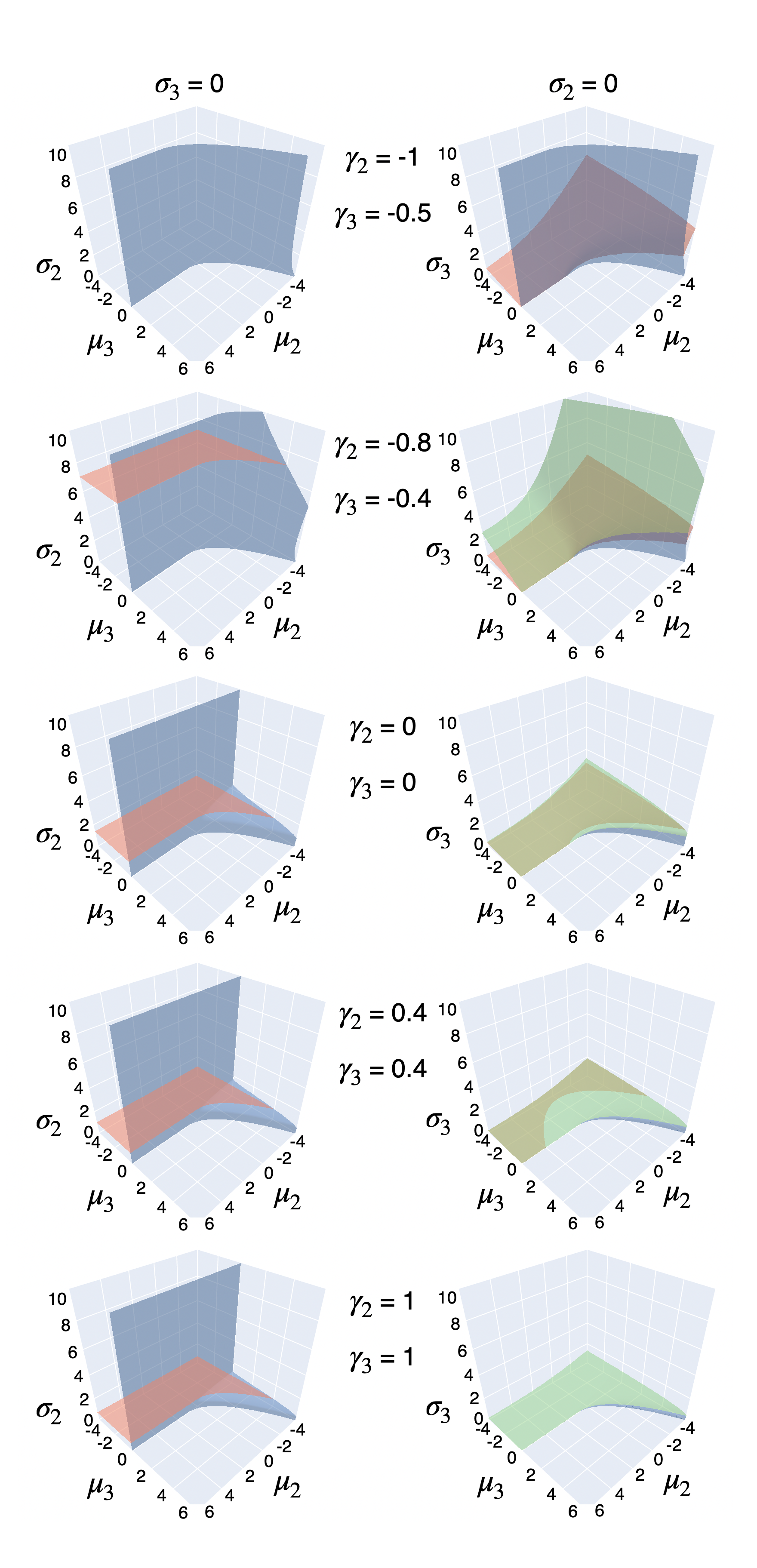}
    \vspace*{-1cm}
    \caption{Each plot shows a phase diagram of the model with second and third order interactions in 3D space. Plots on the left are for $\sigma_3=0$ and show how the transition boundaries vary with increasing $\gamma_2$ (top to bottom). The plots on the right are for $\sigma_2=0$, and show how the transition boundaries vary with increasing $\gamma_3$ (top to bottom). Each plot shows the transition point surfaces of the linear instability point $\sigma_c$ in red, and the divergence point $\sigma_d$ in blue and $\sigma_m$ in green against the $(\mu_2, \mu_3)$ plane.}
    \label{3dplots2}
\end{figure*}

In Fig.~\ref{3dplots2} we show one perspective for multiple examples of similar plots as in Fig.~\ref{3dplots} for different values of $\gamma_2$ and $\gamma_3$, again interactive versions of these are also available online at \cite{website}. The plots on the left-hand side of Fig.~\ref{3dplots2} all have $\sigma_3 = 0$, and therefore are independent of $\gamma_3$. The critical values of $\sigma_2$ at which the transition points occur depend only on $\mu_2$, $\mu_3$, i.e., the location above the base plane, and $\gamma_2$, where the stable region generally reduces as $\gamma_2$ is increased. Similarly the right-hand plots all have $\sigma_2 = 0$ meaning these are independent of $\gamma_2$, and the transition boundaries depend only on $\mu_2$, $\mu_3$ and $\gamma_3$, again reducing the size of the stable region as $\gamma_3$ is increased.
The 4-dimensional phase `diagram' with varying $\mu_2$, $\mu_3$, $\sigma_2$ and $\sigma_3$ can contain any combination of a left-hand plot and a right-hand plot as 3-dimensional subspaces for any chosen combination of $\gamma_2$ and $\gamma_3$, with a smooth transition between the two extremes existing in 4-dimensional space.

The red surface shows the critical point $\sigma_c$, below which the system should have a unique fixed point. For $\sigma_3 = 0$ (left) the condition is the same as for only second order interactions: $I_2(z_c) = I_0(z_c)$, which is satisfied for $z_c = 0$. For $\sigma_2 = 0$ (right) the condition is the same as for only third order interactions: $I_2(z_c) = 2I_0(z_c)$, which is satisfied for $z_c \approx -0.84$. However, if both $\sigma_2$ and $\sigma_3$ are non-zero, then the value of $z_c$ lies somewhere between these two values, depending on the chosen value of $\sigma_2$.

The solution for $M^*$ in Eq.~(\ref{eq:m_poly}) can be written as
\begin{align}\label{Msolp23}
    M^* = \frac{(\mu_{2d} - \mu_2) - \sqrt{(\mu_{2d} - \mu_2)^2 - 4\mu_3k}}{2\mu_3},
\end{align}
for the case of $\mu_3 \neq 0$, where
\begin{align}\label{zdp23}
    \mu_{2d} = \frac{-z_1(1 - \gamma_\Sigma \chi)}{I_1},
\end{align}
and the factor $1 - \gamma_\Sigma\chi$ can be found via Eq.~(\ref{p23Hnon0}) derived in Section \ref{p23method} in the Supplemental Material. As this factor plays a key role for mathematical results in this section, we define $H \equiv 1 - \gamma_\Sigma\chi$ for later convenience. The value of $\mu_3$ at the divergence point ($\mu_{3d}$) depends on whether $\mu_2$ is above or below $\mu_{2d}$.
By setting the discriminant in Eq.~(\ref{Msolp23}) to zero, we have
\begin{align}\label{divp23}
    \mu_{3d} = \frac{(\mu_{2d} - \mu_2)^2}{4k},
\end{align}
at which point the solution for $M^*$ would become complex.
However, for the case of $\mu_2 > \mu_{2d}$, the numerator in Eq.~(\ref{Msolp23}) would become negative, but as the solution for $M^*$ must be a positive value, the denominator must also be negative. This means that $\mu_3 < 0$ and is unable to attain the (positive) value in Eq.~(\ref{divp23}). Therefore, for $\mu_2 \geq \mu_{2d}$, the divergence point continues along the line of $\mu_{3d} = 0$. At $\mu_2 = \mu_{2d}$, this line meets the half-parabola defined by Eq.~(\ref{divp23}) for $\mu_2 \leq \mu_{2d}$ only. Substituting Eq.~(\ref{divp23}) into Eq.~(\ref{Msolp23}) leads to
\begin{align}
    M^* = \frac{\sqrt{k}}{\mu_3}\left(\sqrt{\mu_{3d}} - \sqrt{\mu_{3d} - \mu_3}\right)
\end{align}
for $\mu_2 \leq \mu_{2d}$ only.
For $\mu_2 = \mu_{2d}$, both of these cases lead to $M^* = \sqrt{k/(-\mu_3)}$, which is positive as $\mu_3 < 0$ below the divergence boundary.
For the case of $\mu_3=0$,
\begin{align}
    M^* = \frac{1}{2}\sqrt{\frac{k}{\mu_{3d}}},
\end{align}
which is positive and finite for $\mu_2 < \mu_{2d}$ ($\mu_{3d} > 0$), and diverges for $\mu_2 \geq \mu_{2d}$, where $\mu_{3d} = 0$ becomes the divergence condition.
The value of $M^*$ on the divergence boundary is $M^* = \sqrt{k/\mu_3}$, which is also positive and finite on the half-parabola, and infinite where it meets the line at $\mu_{3d} = 0$.
The divergence (blue) boundary can be seen to be the combination of a line (at $\mu_{3d} = 0$) and a half parabola (where $\mu_{3d} > 0$) in Fig.~\ref{3dplots}. This boundary continues for all $\sigma_2$ in the left ($\sigma_3 = 0$) plots, but can be seen only for low $\sigma_3$ in the right ($\sigma_2 = 0$) plots, where the condition of maximum $\sigma_3$ (green) takes over as the divergence boundary.

For both $\sigma_2 = \sigma_3 = 0$ ($z_1 \to -\infty$), $\mu_{2d} = 1$, and the divergence condition becomes
\begin{align}\label{p23divsig0}
    \mu_2 = 1 - 2\sqrt{\mu_{3d}k},
\end{align}
which is derived in Sec.~\ref{p23methodiv} of the SM. Again this only holds for $\mu_2 \leq 1$, for $\mu_2 > 1$ the boundary continues along the line $\mu_3 = 0$. This boundary can be seen at the base of every 3d plot shown in Fig.~\ref{3dplots} and Fig.~\ref{3dplots2} as a half-parabola as in Eq.~(\ref{p23divsig0}) for $\mu_2 \leq 1$, joined to the straight line $\mu_3 = 0$ for $\mu_2 > 1$. Along this divergence boundary, where $\sigma_2 = \sigma_3 = 0$, $M^* = \sqrt{k/\mu_3}$ and $q = M^{*2}$.
Taking a cross section through $\mu_3 = 0$ from the left plot would result in the phase diagram for $\gamma_2 = -0.8$ in Fig.~\ref{phase2}, and a cross section through $\mu_2 = 0$ from the right plot would result in the phase diagram for $\gamma_3 = -0.4$ in Fig.~\ref{phase3}. This confirms that if $\sigma_p = 0$ then $\gamma_p$ has no influence on the behaviour of the system, however $\mu_p$ still does.

The two plots in the top row of Fig.~\ref{3dplots2} show the transition boundaries for fully antisymmetric interactions with $\gamma_2 = -1$ and $\gamma_3 = -1/2$, they are also is shown online at \cite{website}. For the limit of $z_1 \to 0^-$, we have $\sigma_\Sigma \to \infty$, derived in Sec.~\ref{p23methodanti} of the SM, which means $\sigma_2 \to \infty$ for $\sigma_3 = 0$, or $\sigma_3 \to \infty$ for $\sigma_2 = 0$. For the case of $\sigma_3 = 0$, as the instability point occurs at $z_1 = 0$, this can only be attained in the limit of $\sigma_2 = \infty$, so there is no instability (red) surface in the left hand plot. For the right hand plot where $\sigma_2 = 0$, as $z_c < 0$ ($z_c \approx -0.84$), the linear instability point can be attained on the red surface in the diagram. However, as mathematical solutions exists for all values of $\sigma_3$, there is no maximum point (no $\sigma_m$) so there is no green surface in this diagram and the divergence boundary is the same blue surface in both plots.
For fully antisymmetric interactions the divergence boundary satisfies Eq.~(\ref{p23divsig0}) for $\sigma_p = 0$ and $\mu_2 \leq \mu_{2d}$, where $\mu_{2d} = 1$, and as $\sigma_p \to \infty$, $\mu_{2d} \to \pi$. The divergence boundary tends to
\begin{align}
    \mu_2 = \pi - 2\sqrt{\mu_{3d}k}
\end{align}
for $\mu_2 \leq \pi$ which connects to the line $\mu_{3d} = 0$ for $\mu_2 \geq \pi$. In the limit of $\sigma_p \to \infty$ ($z_1 \to 0^-$), the value of $M^*$ tends to a finite constant which depends on $\mu_2$ and $\mu_3$ only,
\begin{align}
    M^* = \frac{(\pi - \mu_2) - \sqrt{\left(\pi - \mu_2\right)^2 - 4\mu_3k}}{2\mu_3}
\end{align}
below the divergence boundary, unless $\mu_3 = 0$ in which case $M^* = k/(\pi - \mu_2)$. Along this boundary $M^* = \sqrt{k/\mu_{3d}}$. For this limit $z_1 \to 0^-$, $q$ also tends to a finite constant, $q \to \pi M^{*2}$.

\subsubsection{Variance within both second and third order of interactions}
Setting either $\sigma_2$ or $\sigma_3$ to zero as above causes the heterogeneity to come from one type of interactions only, so the results are similar to having a single order of interactions, but with two values of $\mu_p$ that can be varied instead of one. The 3-dimensional plots on the left and right  in Fig.~\ref{3dplots} and Fig.~\ref{3dplots2} represent three-dimensional subspaces of the four dimensional $\{\mu_2, \mu_3, \sigma_2, \sigma_3\}$ space, and there exists a smooth transition from any of the left-hand plots to any of the right-hand plots, depending on the chosen combination of $\gamma_2$ and $\gamma_3$. Given that we cannot plot a four-dimensional phase diagram, we have chosen some combinations of $\gamma_2$, $\gamma_3$, $\mu_2$, and $\mu_3$ and plotted phase diagrams in the $(\sigma_2,\sigma_3)$ plane in order to understand the transition between the two extremes.
\begin{figure*}
    \centering
    \includegraphics[width = \textwidth]{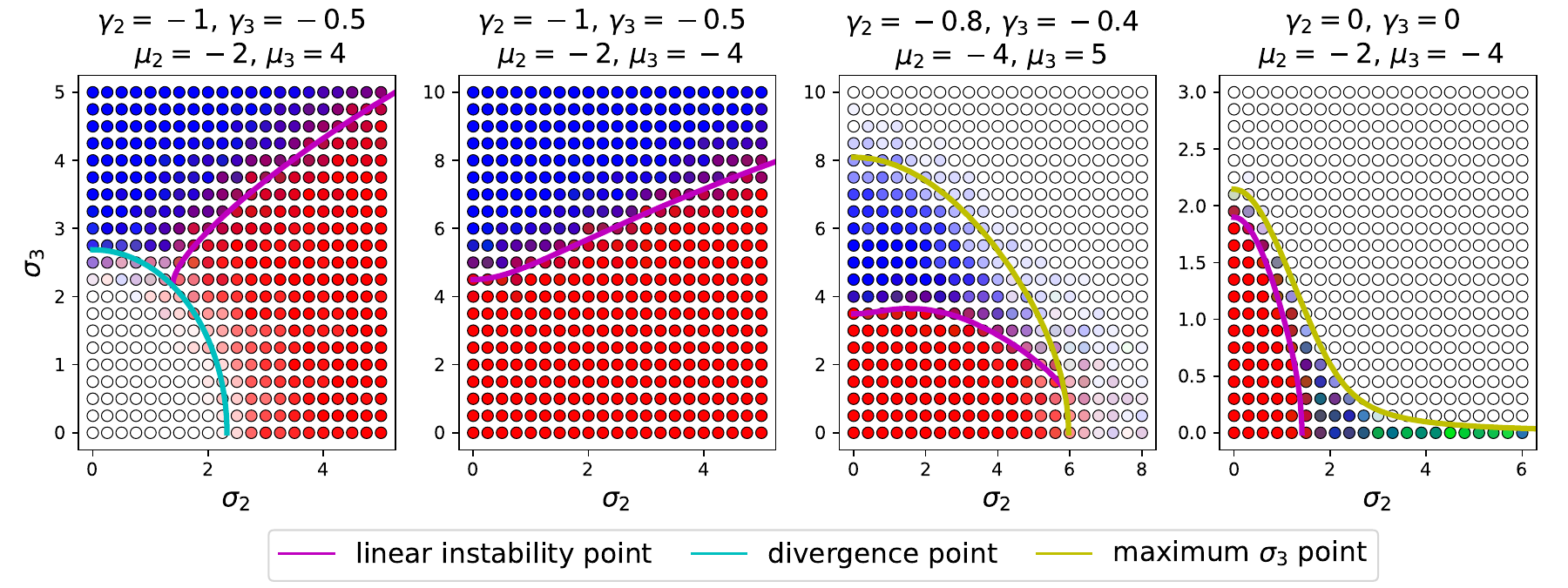}
    \caption{Phase diagrams in the $(\sigma_2, \sigma_3)$ plane for the model with second and third order interactions for various combinations of $\gamma_2$, $\gamma_3$, $\mu_2$ and $\mu_3$. The magenta lines represent the linear instability point $\sigma_c$ and both the cyan $\sigma_d$ and yellow $\sigma_m$ lines represent the divergence point. Each circle represents the average behaviour of 20 simulations for $N = 200$ species run for up to 10,000 units of time. Red indicates that the system reached a unique fixed point, green indicates multiple fixed points, blue persistent dynamics, and white divergence. The method for classifying the behaviour and converting the number of each type to a colour is described in Section \ref{smsims} of the SM.}
    \label{phasesim23}
\end{figure*}
Some examples are shown in Fig.~\ref{phasesim23}. Along the horizontal axis ($\sigma_3=0$) the system shows the same behaviours and transitions as along a line cut through the left-hand 3-dimensional plot on in Fig.~\ref{3dplots2} with the same value for $\gamma_2$, above the same point on the $(\mu_2, \mu_3)$ plane. Along the vertical axis in Fig.~\ref{phasesim23} ($\sigma_2=0$) the system shows the same behaviours and transitions as along the corresponding line in the right-hand 3-dimensional plot in Fig.~\ref{3dplots2} with the same value for $\gamma_3$. We see there is a smooth transition between the $\sigma_2 = 0$ and $\sigma_3 = 0$ axes.

The first two panels in Fig.~\ref{phasesim23} display results from simulations with fully antisymmetric interactions, $\gamma_2 = -1$ and $\gamma_3 = -0.5$. The locations on the $(\mu_2, \mu_3$) plane were chosen such that they begin on either side of the divergence boundary for $\sigma_2 = \sigma_3 = 0$. For asymmetric interactions, $\gamma_2 = -1$, $\gamma_3 = -0.5$, neither axis has the critical transition, but as either extreme is on different sides of this transition, the line runs through the middle of the plane. As the values for $\gamma_p$ are increased (third panel) the upper divergence boundary can be attained by finite values of $\sigma_p$ (yellow line). The last panel shows the case of uncorrelated interaction coefficients, where multiple fixed-point behaviour is possible (green colouring). 

The magenta lines in Fig.~\ref{phasesim23} show the linear instability transition point, where the critical condition becomes
\begin{align}
    H_c = \frac{I_2}{I_2 + \gamma_2 I_0 + (I_0 - I_2)(\gamma_2 - 2\gamma_3)}.
\end{align}
This is derived in Eq.~(\ref{Hc}) in Section~\ref{p23methodcrit} in the SM. This condition is satisfied at $z_c$, which occurs at values between $0$ (on the horizontal axis, only second-order interactions) and $-0.84$ (on the vertical axis, only third-order interactions). For each value of $z_c$ in this range, the corresponding $\sigma_2$ is found using
\begin{align}\label{sig2H}
    \sigma_2^2(H) = \frac{H(I_2 - I_2H - 2\gamma_3I_0H)}{I_0 I_2 (\gamma_2 - 2\gamma_3)}
\end{align}
from Eq.~(\ref{Hquad}), and $\sigma_3$ using
\begin{align}\label{sig3H}
    \sigma_3^2(H) = \frac{2I_1^2H(\gamma_2I_0H +I_2H - I_2)}{3I_0 I_2^2 M^2(\gamma_2 - 2\gamma_3)}
\end{align}
from Eq.~(\ref{p23sig3}).

The cyan lines in Fig.~\ref{phasesim23} show the lower divergence point, where solutions for $M^*$ become infinite or complex. This is given by the condition for $z_d$ obtained by combining Eqs.~(\ref{zdp23}) and (\ref{divp23}). This gives the following expression for $H$ at the divergence point
\begin{align}
    H_d = \frac{I_1}{z_1}\left(-2\sqrt{\mu_3k} - \mu_2\right).
\end{align}
As the value of $z_d$ is varied, this value for $H$ can be put into Eq.~(\ref{sig2H}) and Eq.~(\ref{sig3H}) to plot the lower divergence line in the same way as the linear instability line.

The yellow lines in Fig.~\ref{phasesim23} show the upper divergence point, the highest possible value of $\sigma_3$ for which solutions exist. For each value for $\sigma_2$, the maximum value for $\sigma_3$ was found varying $z_1$ and keeping all other parameters fixed.
For each combination of the variables, the quantity $H \equiv 1 - \gamma_\Sigma \chi$ satisfies the quadratic equation derived in Eq.~(\ref{Hquad}), and has two possible solutions
\begin{align}\label{H10}
    H_{1, 0} = \frac{I_2 \pm \sqrt{I_2^2 - 4\left(I_2 + 2\gamma_3I_0\right)(\gamma_2 - 2\gamma_3)\sigma_2^2I_0I_2}}{2(I_2 + 2\gamma_3I_0)},
\end{align}
($H_1$ is with the plus sign, and $H_0$ with the minus). This reduces to the single-order expression in Eq.~(\ref{help}) for the case of either $\sigma_p$ set to zero. In most cases, only the greater of these solutions $H_1$ leads to a valid $\sigma_3$, but there are some cases where both are valid, leading to two possible valid solutions for $\sigma_3$ for a given $z_1$,
\begin{align}
    \sigma_3 = \frac{I_1}{I_2M^*}\sqrt{\frac{2}{3}\left(H_{1,0}^2 - I_2\sigma_2^2\right)}.
\end{align}

\begin{figure*}
    \centering
    \includegraphics[width = \textwidth]{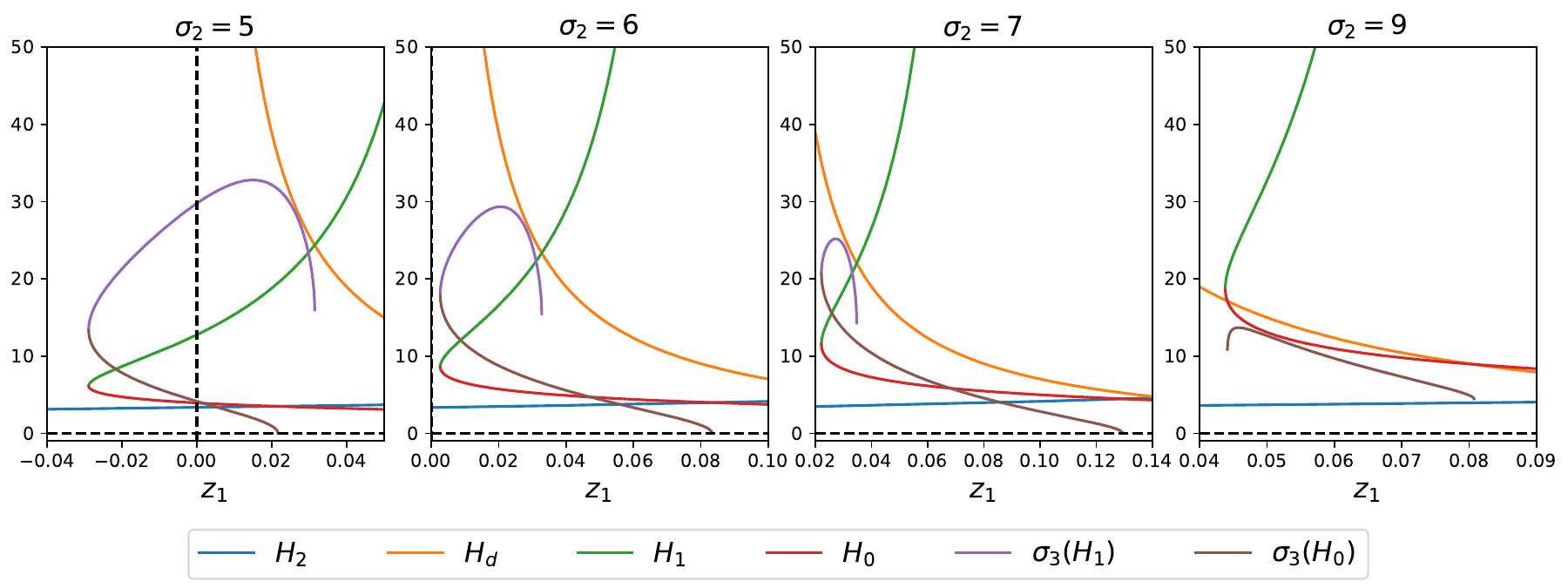}
    \caption{Plots of the quantities in the legend as function of $z_1$, for examples of fixed $\sigma_2$, with $\gamma_2=-0.7$, $\gamma_3=-0.47$ and $\mu_2=4$ and $\mu_3=1$.}
    \label{plotz23}
\end{figure*}

Fig.~\ref{plotz23} shows the dependence on $z_1$ of multiple expressions for $H$, and their corresponding solutions for $\sigma_3$, for various values of $\sigma_2$ (given in the figure), $\gamma_2 = -0.7$, $\gamma_3 = -0.47$, $\mu_2 = -4$ and $\mu_3 = 1$. The two solutions for $H$ in Eq.~(\ref{H10}) are shown as $H_1$ (the greater value) and $H_0$ (the lower value). These are shown along with $H_2$, the expression for $H$ that satisfies $\sigma_3 = 0$, and $H_d$ satisfying the divergence condition. The point where $\sigma_3 = 0$ and the divergence point can be identified where either $H_1$ or $H_0$ cross the lines for $H_2$ or $H_d$, respectively.
It can be seen that both branches of $H$ are required to attain the full range of $\sigma_3$ values. This is because only $H_0$ crosses $H_2$, and therefore only $\sigma_3(H_0)$ can attain zero. The value for $z_d$ corresponds to where $H_1$ intersects with $H_d$, but since this is beyond the maximum point, this point cannot physically be realised. However, in some cases (e.g., $\sigma_2 = 6, 7$) values of $z_1$ beyond this $z_d$ can still be attained, but by the lower branch at a much lower value of $\sigma_3$.
From these plots we can understand a new behaviour that has not been previously seen for systems with a single order of interactions, an increase in $z_1$ corresponding to a decrease in $\sigma_3$. As the value of $z_1$ directly corresponds to the fraction of surviving species, $\phi$, this suggests that during simulations for this set of parameters, and keeping $\sigma_2$ fixed, increasing $\sigma_3$ from zero would initially cause the fraction of surviving species to increase. This behaviour would continue until $z_1$ reaches its lowest possible value, at which point we move to the $H_1$ branch, and the usual behaviour of decreasing $\phi$ continues until $\sigma_3$ reaches its maximum value for which solutions exist, after which point the system diverges as usual.
In the last panel of Fig.~\ref{plotz23} we find that all solutions for $H_1$ (in green) are above $H_d$ (in orange), and therefore we are unable to find solutions for $\sigma_3$ using this expression. The only valid solutions are from $H_0$ (in red) which happens to cross $H_d$ (in green) in two places, which correspond to the greatest and lowest values of $z_1$ for which solution for $\sigma_3$ can be found. In this example, the lower divergence point, the lowest attainable value of $\sigma_3$ happens to occur at the highest attainable value for $z_1$, however the upper divergence point is still satisfied by the maximum point, instead of the other value for $z_d$.

\begin{figure}[t!!!]
    \centering
    \includegraphics[width = 0.5\textwidth]{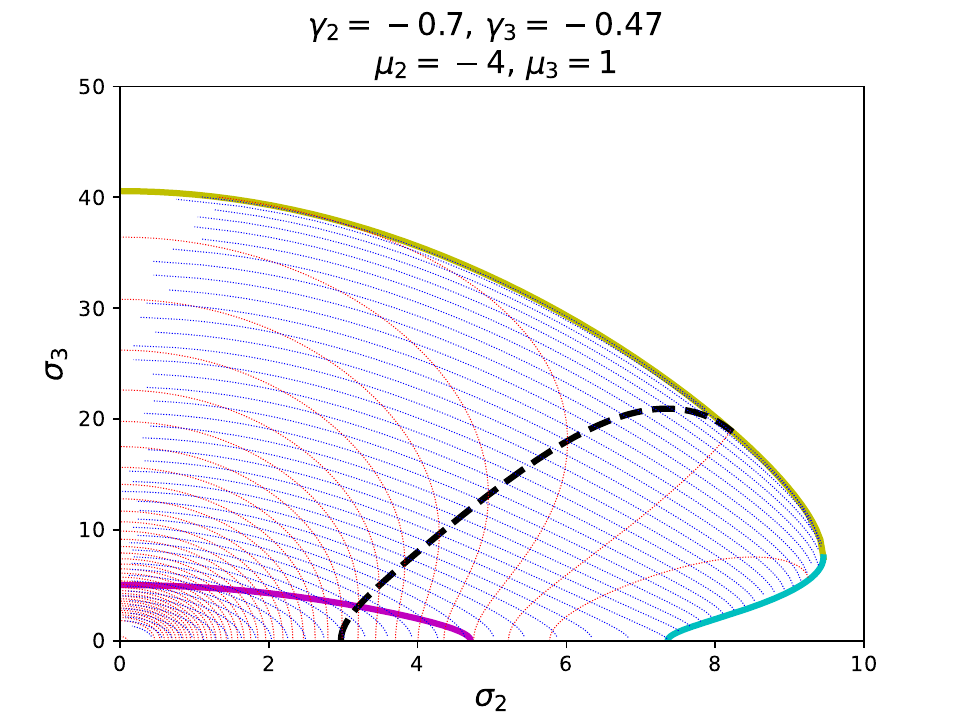}
    \caption{Phase diagram in $(\sigma_2,\sigma_3$) space for the model with second and third order interactions. The remaining model parameters are as given in the figure. The magenta line is the linear instability point ($\sigma_{3,c}$), the cyan line is the lower divergence point ($\sigma_{3,d}$), the yellow line is the upper divergence point ($\sigma_{3,m}$). The red lines show contours of constant $z_1$ (and hence constant $\phi$), and the blue lines are contours of constant $H$. The black dashed line indicates where the diversity $\phi$ is maximal as a function of $\sigma_3$ for fixed $\sigma_2$.
    \label{phase23}}
\end{figure}

Fig.~\ref{phase23} shows the phase diagram mapped out by observations made from Fig.~\ref{plotz23}. The magenta, cyan, and yellow lines show the instability, lower divergence, and upper divergence boundaries as before. The red contour lines trace out paths of constant $z_1$, and therefore paths of constant $\phi$. In the bottom left corner, where $\sigma_2 = \sigma_3 = 0$ we have $\phi = 1$ ($z_1 \to -\infty$). For most parameter combinations, $\phi$ decreases as either $\sigma_p$ is increased, but this is not the case for sufficiently large $\sigma_2$ where we find two branches for $\sigma_3$ as described above. The black dotted line shows the point where the two branches meet, at the lowest value of $z_1$ for each $\sigma_2$, therefore this line maps out $\sigma_3$ for which optimal diversity is achieved (greatest $\phi$) for each $\sigma_2$. The region below the black dotted line shows the region where diversity is an increasing function of $\sigma_3$, and above the dashed line it is decreasing (this is the more common behaviour). Along with the increase in diversity $\phi$, the increase of variance of third-order interactions $\sigma_3^2$ also reduces the spread of the (pre-clipped) species abundance distribution ($\sigma_\Sigma/H$) up to the black dotted line. This suggests there is an optimal (in terms of diversity) balance between the variance of both types of interactions for some parameter combinations.

\section{Conclusions}\label{sec:conclusions}
In summary, we have extended dynamic mean-field analyses of generalised Lotka--Volterra systems with pairwise couplings to higher-order interactions. In the language of spin-glass physics, if the pairwise model is an analog of the Sherrington-Kirkpatrick model, then  our model is the multi-spin (or $p$-spin) extension. More specifically, we allow for combinations of multiple orders of interaction. We have used generating functionals to derive the effective single-species process, for general symmetry or anti-symmetry of the coupling coefficients at any given order of interaction. This generalises the work of \cite{GibbsPNAS}, which focues on combinations of second and third order interactions, and includes a symmetry parameter only at the level of pairwise interaction.

Making a fixed-point ansatz we then studied the stable phase with unique attractors and calculated key order parameters such as the fraction of surviving species, the mean abundance and the abundance distribution. We find that the model can display the linear instability seen in the pairwise model. The location of this instability can depend on the mean of the higher-order interaction coefficients, which is not seen in the pairwise model (Fig.~\ref{phasesim2}). Here we again see that more competitive interactions typically have a stabilising effect in a model with higher-order interactions (Figs.~\ref{phasesim3} and \ref{phasesim4}). In the pairwise model with fully antisymmetric interactions the linear instability can never be attained \cite{bunin2016,gallaepl2018}. In the model with higher-order interactions instead we find that this instability can occur at finite $\sigma_p$. This is due to the lower critical fraction of species required to reach the instability.

Separately, the analytical solution for the mean abundance can be shown to diverge along certain manifolds in the multi-dimensional space spanned by the model parameters. Additionally we also find instances in which the self-consistent equations for the static order parameters in the fixed-point phase cease to have solutions. In simulations we find that abundances diverge beyond this point. However, within the analytical theory, this new type of divergence is distinct from the divergence point in the pairwise model, where the mean abundance increases without bound as the divergence point is approached. In the model with higher orders the divergence is instead of sudden nature. In fact, in simulations we find that in many cases $M^*$ actually decreases as one approaches the divergence point from below, see e.g. Fig.~\ref{Mplots3} and \ref{Mplots4}. This is very different from the behaviour in the model with $p=2$, see Fig.~\ref{Mplots2}. This suggests that higher-order interactions can have a limiting effect on the potential abundances of stable communities.

\medskip

Our analysis extends and generalises the work of \cite{GibbsPNAS} where the authors used the cavity method to derive the DMFT for a Lotka--Volterra system with combined second and third order interactions. We also note further work in \cite{GibbsEcoLett}. In-line with \cite{GibbsPNAS}, we find that increasing the variance $\sigma_3^2$ of third order interactions usually reduces the fraction of surviving species. Further, we also confirm the effect seen in \cite{GibbsPNAS}: making the third order couplings more harmful (i.e., reducing $\mu_3$ in our setup) can increase the fraction of survivors (see Fig.~\ref{allphiplots3}). We also find this in the system with fourth order interactions (Fig.~\ref{allphiplots4}), a case not covered by \cite{GibbsPNAS}. We note that the amount of competition pressure $\mu_2<0$ has no effect on the fraction of surviving species and the linear instability point for the model with only pairwise interactions. In light of \cite{kat} this suggests that the pairwise model may not be sufficient to describe natural ecosystems. Models with higher-order interactions on the other hand produce results that are more akin to what is reported in \cite{kat}.

We note that the setup in \cite{GibbsEcoLett} is somewhat different to ours, in that third-order interactions in \cite{GibbsEcoLett} are correlated with second-order couplings via a constraint ensuring desired fixed-point abundances. In our model, the interaction coefficients of different orders are not correlated with one another. Therefore, results cannot be compared directly. We note though that \cite{GibbsEcoLett} reports that an increased variance of the species abundance distribution is often associated with less stability. This is in-line with our findings. We find that the variance of species abundances usually increases with the variance in the interaction coefficients. Increasing the $\sigma_p$ in turn often brings the system closer to the linear instability or diverging abundances. This is valid, except for some cases with re-entrance behaviour and some very specific choices of parameters for combinations of interactions, see e.g. Fig.~\ref{phase23}).  

Further, the stability of equilibria in models with higher-order interactions has been studied via the spectra of random matrices in \cite{bairey}. As pointed our in \cite{GibbsPNAS} the setup is somewhat different though, as \cite{bairey} mostly considers replicator dynamics with an overall constraint on the sum of the degrees of freedom (the $x_i$). In our model there is no such constraint on the $x_i$, and therefore results are not directly comparable.
\medskip

Our analysis opens up several interesting avenues for future work. For example, one could ask what happens if self-interactions are not linear, i.e., if the starting point is a system of the form $ \dot{x}_i(t) = x_i\bigg[k - g(x_i) +\sum_p f_p(N^{-1}\sum_i x_i^p)+  \sum_{p = 2}^N\sum_{\{i_2, \dots, i_p\}} \alpha^{(p)}_{i, i_2, \dots, i_p}x_{i_2}\dots x_{i_p}(t)\bigg]$, with some nonlinear functions $f_p$ and $g$. The function $g$ could involve polynomial terms of different orders. Analysing such a model would likely be much more complicated as the 
expression in the bracket in Eq.~(\ref{fpequation}) would become nonlinear in $x^*$ for nonlinear $g$. The equivalent of Eq.~(\ref{xstar}) would then also be much more intricate. At the same time, introducing such nonlinear self-interaction may well enhance the phase behaviour even more and lead to additional types of instabilties or divergencies. Similarly, one could also extend the model to include nonlinear functional response for interactions across species, along the lines of \cite{laura}, or indeed some form of saturation of higher-order interaction coefficients with species abundances, as indicated in \cite{GibbsPNAS}.

Further, as suggested in \cite{GibbsEcoLett}, it would be interesting to allow for correlations between the interaction coefficients across different orders. This appears sensible in terms of ecological realism, and at the same time such correlations may again change the phase diagram. 

It would also be of interest to study the interactions between the surviving species in more detail. The community of survivors emerges from the Lotka--Volterra dynamics, and is therefore feasible by construction. In the pairwise model the statistics of the interactions, restricted to surviving species, is very different to that in the original pool of all species \cite{bunin2016,bunin2017,baron_et_al_prl}. In particular the `reduced' interaction matrix (the matrix of interactions, restricted to extant species) is non-Gaussian even if the original full interaction matrix is Gaussian. We would expect the see something similar in the model with higher-order interactions, but how exactly the extinction dynamics affects these interactions is unclear. In the pairwise model it is also known that the leading eigenvalue of the reduced interaction matrix  determines the stability of the community \cite{stone}. Further, it was recently demonstrated the the linear instability and the onset of divergent abundances in the pairwise model are signalled by the bulk spectrum or outlier eigenvalues to cross the imaginary axis respectively \cite{baron_et_al_prl}. If and how this generalises to higher-order interaction `tensors' is not known. As we have shown, the model with higher-order interactions shows the linear instability and two different types of onset of diverging abundances. It would be interesting to see how these latter two transitions are reflected in the spectrum of the reduced interaction tensor.

Similar to \cite{GibbsPNAS}, it would be interesting to vary the growth rates $k_i$, i.e., to give them a mean $\mu_R$ and a variance $\sigma_R^2$. One could then obtain phase diagrams such as those in Figs.~\ref{3dplots} and \ref{3dplots2}, combining the effects of varying growth rates ($\mu_R, \sigma_R$) and interaction coefficient parameters ($\mu_p, \sigma_p)$ for each order of interactions $p$. This would facilitate the comparison of the effect of $\mu_R$ on the stability and fraction of surviving species to that of the $\mu_p$ considered in this work.

Further study of Lotka--Volterra models with higher-order interactions is also of interest from the perspective of statistical mechanics in more general. The $p$-spin glass model can behave differently to the $2$-spin model for example in terms of replica symmetry breaking, ageing etc \cite{stariolo,cugliandolo,barrat,gillin}. Here, we have demonstrated that the Lotka--Volterra model with higher-order interactions can show divergences not seen in the model with pairwise interaction. It would be interesting to study this further, and to analyse the corresponding `energy' landscape, the number and stability of equilibria, and the relaxation dynamics, extending the work of \cite{birolibunin,Ros1,Ros2}. Related to this, the nature of the different unstable phases in the model with higher-order coupling may also be different to the pairwise model \cite{BuninPRX}. We note that \cite{GibbsPNAS} reported no evidence of instances of multiple fixed points in models with one type of higher-order interactions only. We make similar observations in our model, however, we do find the possibility of multiple fixed points in models which combine pairwise and higher-order interactions (Fig.~\ref{traj23_green}).

\clearpage
\onecolumngrid
\renewcommand{\theequation}{S\arabic{equation}}
\setcounter{equation}{0}
 \renewcommand{\thesection}{S\arabic{section}}
\setcounter{section}{0}
 
\renewcommand{\thefigure}{S\arabic{figure}}
\setcounter{figure}{0}
 
\renewcommand{\thepage}{S\arabic{page}}
\setcounter{page}{1}

\begin{center}
{\bf Higher-order interactions in random Lotka--Volterra communities} \\
~\\

{\bf Laura Sidhom and Tobias Galla}\\
\medskip
~\\
{\bf ---~Supplementary Material~---}
 
\end{center}
 
\label{supp}

\section{Generating-functional analysis}

\subsection{Expression for the generating functional}

The generalised Lotka--Volterra equations are written as
\begin{align}
    \frac{\dot{x}_i(t)}{x_i(t)} = k_i - x_i(t) + h_i(t) +\sum_{p=2}^N \sum_{\{i_2, \dots, i_p\}} \alpha_{i, i_2, \dots, i_p}^{(p)}  x_{i_2}(t) \dots x_{i_p}(t),
\end{align}
the second sum is a sum over the ${N-1 \choose p-1}$ possible sets of $p-1$ indices which are not equal to $i$. Here we have introduced $h_i(t)$ to generate response functions, these fields will be later set to zero. Using this expression we find the generating functional for the process,
\begin{align}
    Z[\pmb{\psi}] &= \int D[\pmb{x}] \exp\left(i\sum_i \int \psi_i(t)x_i(t) dt \right) \left[\prod_i \mathcal{P}(x_i(0))\right] \nonumber \\
    &\times \prod_{i,t} \delta \left\{ \frac{\dot{x}_i(t)}{x_i(t)} - \left[ k_i - x_i(t) + h_i(t) + \sum_{p=2}^N \sum_{\{i_2, \dots, i_p\}} \alpha_{i, i_2, \dots, i_p}^{(p)}  x_{i_2}(t) \dots x_{i_p}(t)  \right]\right\},
\end{align}
where $\mathcal{P}(x_i(0))$ represents the probability distribution from which the initial conditions are drawn, not to be confused with $p$, the order of an interaction.
The generating functional $Z[\pmb{\psi}]$ is the Fourier transform of the probability measure in the space of possible paths of the system, where the field $\pmb{\psi}$ is a source term.
The integral $\int D[\pmb{x}] = \prod_i\prod_t \int dx_i(t)$ is taken over all possible paths, and over uniformly distributed initial conditions $x_i(0)$, which we assume can be factorised for the different species $i$. We write the delta function as a Fourier transform to give
\begin{align}
    Z[\pmb{\psi}] &= \int D\hat{\pmb{x}} D\pmb{x} \left[\prod_i \mathcal{P}(x_i(0))\right] \exp\left(i\sum_i \int \psi_i(t)x_i(t) dt \right) \nonumber \\
    &\times \exp\left\{ i \int \sum_i \hat{x}_i(t) \left[\frac{\dot{x}_i(t)}{x_i(t)} - k_i + x_i(t) - \sum_{p=2}^N \sum_{\{i_2, \dots, i_p\}} \alpha_{i, i_2, \dots, i_p}^{(p)} x_{i_2}(t) \dots x_{i_p}(t) - h_i(t) \right]dt\right\},
\end{align}
where the factors of $2\pi$ have been absorbed into the measure via
\begin{align}
    D\hat{x}Dx\equiv \frac{D[\hat{x}]D[x]}{2\pi}.
\end{align}

\subsection{Calculation of moments from the generating functional}
Moments of the dynamical variables can be generated from the generating functional by taking derivatives. We have 
\begin{align}\label{diff1}
\left. \frac{\delta Z[\pmb{\psi}]}{\delta\psi_i(t)} \right|_{\pmb{\psi} = \mathbf{0}} = i\left<x_i(t)\right>,
\end{align}
and
\begin{align} \label{diff2}
\left.\frac{\delta^2Z[\pmb{\psi}]}{\delta\psi_i(t)\delta\psi_i(t')} \right|_{\pmb{\psi} = \mathbf{0}} = -\left<x_i(t)x_i(t')\right>.
\end{align}
Response functions can be found by
\begin{align}\label{diff3}
\left.\frac{\delta^2Z[\pmb{\psi}]}{\delta\psi_i(t)\delta h_i(t')}\right|_{\pmb{\psi} = \mathbf{0}} = i\pdv{\left<x_i(t)\right>}{h_i(t')} = \left<x_i(t)\hat{x}_i(t')\right>.
\end{align}
We note that $Z[\pmb{\psi} = \mathbf{0}] = 1$ for all choices of the perturbation $\{h_i(t)\}$, this is because $Z[\mathbf{0}]$ is a sum over the probabilities of possible paths of the system. This leads to
\begin{align}\label{diff4}
\frac{\delta Z[\pmb{\psi} = \mathbf{0}]}{\delta h_i(t)} = -i\left<\hat{x}_i(t)\right> = 0,
\end{align}
and
\begin{align}\label{diff5}
\frac{\delta^2Z[\pmb{\psi} = \mathbf{0}]}{\delta h_i(t)\delta h_i(t')} = -\left<\hat{x}_i(t)\hat{x}_i(t')\right> = 0.
\end{align}

\subsection{Disorder average}
The Gaussian random coefficients for interactions of order $p$ have moments
\begin{align}
    \overline{\alpha^{(p)}_{i, i_2, \dots, i_p}} &= \frac{\mu_p(p-1)!(N-p)!}{(N-1)!},\\
    \overline{\left(\alpha^{(p)}_{i, i_2, \dots, i_p}\right)^2} - \left(\overline{\alpha^{(p)}_{i, i_2, \dots, i_p}}\right)^2 &= \frac{\sigma_p^2p!(N-p)!}{2(N-1)!},\\
    \overline{\alpha^{(p)}_{i, i_2, \dots, i_p}\alpha^{(p)}_{j, j_2, \dots, j_p}} - \left(\overline{\alpha^{(p)}_{i, i_2, \dots, i_p}}\right)^2 &= \frac{\gamma_p\sigma_p^2p!(N-p)!}{2(N-1)!}.
\end{align}
These terms have been divided by $N-1 \choose p-1$ as this is the number of interactions each species has with the other species, and therefore the number of times it would receive each payoff. This is done so that the average payoffs are independent of $N$, the total number of species.

Focusing on the term in the generating functional containing the disorder, which we will call  $rt$ (`random term'),
\begin{align}
    rt = \overline{\exp\left(-i\int \sum_i \hat{x}_i(t)\sum_{p=2}^N \sum_{\{i_2, \dots, i_p\}} \alpha_{i, i_2, \dots, i_p}^{(p)}  x_{i_2}(t) \dots x_{i_p}(t) dt \right)},
\end{align}
we can rewrite this as
\begin{align}
    rt = \prod_{p=2}^N \prod_{s \in S(p)} \overline{\exp\left(-i\int\sum_{i \in s}\hat{x}_i(t)\alpha_{i, i_2, \dots, i_p}^{(p)} x_{i_2}(t) \dots x_{i_p}(t) dt \right)},
\end{align}
where we have introduced the new notation $S(p)$ which denotes the set of ${N \choose p}$ sets of $p$ species. Here we have taken the product over interaction orders $p$, the product over the sets of $p$ species, and the sum over the $p$ members of the set indexed by $i$ within the exponential.
Here we are able to factorise the expression in this way because there are no correlations between the interactions of separate orders or separate $p$-sets.
We can write the random interaction coefficients as
\begin{align}
    \alpha^{(p)}_{i, i_2, \dots, i_p}=\frac{\mu (p-1)! (N-p)!}{(N-1)!}+\sigma_p\sqrt{\frac{p!(N-p)!}{2(N-1)!}}z^{(p)}_{i, i_2, \dots, i_p}
\end{align}
where the $z^p_{i, i_2, \dots, i_p}$ are drawn from a standard $p$-variate Gaussian distribution with
\begin{align}
    \overline{z^{(p)}_{i, i_2, \dots, i_p}} = 0, \quad \quad \overline{\left(z^{(p)}_{i, i_2, \dots, i_p}\right)^2} = 1, \quad \quad \overline{z^{(p)}_{i, i_2, \dots, i_p}z^{(p)}_{j, j_2, \dots, j_p}} = \gamma_p
\end{align}
with $\{i, i_2, \dots, i_p\} = \{j, j_2, \dots, j_p\}$ but $i \neq j$.
This gives the random term (rt) as
\begin{align}
    rt &=\prod_{p=2}^N \prod_{s \in S(p)} \exp\left(-i\frac{\mu_p(p-1)!(N-p)!}{(N-1)!}\int\sum_{i \in s}\hat{x}_i(t) x_{i_2}(t) \dots x_{i_p}(t) dt \right) \nonumber \\ &\times\overline{\exp\left(-i\sqrt{\frac{\sigma_p^2p!(N-p)!}{2(N-1)!}}\int\sum_{i \in s} \hat{x}_i(t) z^{(p)}_{i,i_2,\dots,i_p} x_{i_2}(t) \dots x_{i_p}(t) dt \right)},
\end{align}
where we take the disorder average to obtain
\begin{align}
    rt &=\prod_{p=2}^N \prod_{s \in S(p)} \exp\left(-i\frac{\mu_p(p-1)!(N-p)!}{(N-1)!}\int\sum_{i \in s} \hat{x}_i(t) x_{i_2}(t) \dots x_{i_p}(t) dt \right) \nonumber \\
    &\times \exp\left(-\frac{\sigma^2_pp!(N-p)!}{4(N-1)!}\left(\int \sum_{i \in s} \hat{x}_i(t)\hat{x}_i(t')x_{i_2}(t)x_{i_2}(t') \dots x_{i_p}(t)x_{i_p}(t') dt dt'\right.\right. \nonumber \\
    &+ \left.\left. \gamma_p\int \sum_{i \in s, i \neq j} \hat{x}_i(t)x_i(t')\hat{x}_j(t')x_j(t) x_{i_3}(t)x_{i_3}(t') \dots x_{i_p}(t)x_{i_p}(t') dt dt \right)\right).
\end{align}
We now assume that the average over these $N-1 \choose p-1$ terms, that do not include self-interactions, is the same as the average of the $N^{p-1}$ terms that do contain self-interactions. In the limit of large $N$, the number of self-interaction terms is small compared to the total number of terms.
\begin{align}
    rt &=\prod_{p=2}^N \exp\left(-i\frac{\mu_p}{N^{p-1}}\int\left(\sum_i \hat{x}_i(t)\right) \left(\sum_{i_2}x_{i_2}(t)\right)^{p-1} dt \right) \nonumber \\
    &\times \exp\left(-\frac{\sigma^2_pp}{4N^{p-1}}\left(\int \left(\sum_i \hat{x}_i(t)\hat{x}_i(t')\right)\left(\sum_{i_2}x_{i_2}(t)x_{i_2}(t')\right)^{p-1} dt dt'\right.\right. \nonumber \\
    &+ \left.\left. \gamma_p(p-1)\int \left(\sum_i \hat{x}_i(t)x_i(t')\right)\left(\sum_jx_j(t)\hat{x}_j(t')\right) \left(\sum_{i_3}x_{i_3}(t)x_{i_3}(t')\right)^{p-2} dt dt \right)\right),
\end{align}
where the approximated expression includes coefficients with multiple instances of the same index.
We can now substitute this expression back into the generating functional to get $\overline{Z[\pmb{\psi}]}$:
\begin{align}
    \overline{Z[\pmb{\psi}]} &= \int D\hat{\pmb{x}} D\pmb{x} \left[\prod_i \mathcal{P}(x_i(0))\right] \exp\left(i\sum_i \int \psi_i(t)x_i(t) dt \right) \nonumber \\
    &\times \exp\left\{ i \int \sum_i \hat{x}_i(t) \left[\frac{\dot{x}_i(t)}{x_i(t)} - k_i + x_i(t) - h_i(t) \right]dt\right\} \nonumber \\
    &\times\prod_{p=2}^N \exp\left(-i\frac{\mu_p}{N^{p-1}}\int\left(\sum_i \hat{x}_i(t)\right) \left(\sum_{i_2}x_{i_2}(t)\right)^{p-1} dt \right) \nonumber \\
    &\times \prod_{p=2}^N\exp\left(-\frac{\sigma^2_pp}{4N^{p-1}}\left(\int \left(\sum_i \hat{x}_i(t)\hat{x}_i(t')\right)\left(\sum_{i_2}x_{i_2}(t)x_{i_2}(t')\right)^{p-1} dt dt'\right.\right. \nonumber \\
    &+ \left.\left. \gamma_p(p-1)\int \left(\sum_i \hat{x}_i(t)x_i(t')\right)\left(\sum_jx_j(t)\hat{x}_j(t')\right) \left(\sum_{i_3}x_{i_3}(t)x_{i_3}(t')\right)^{p-2} dt dt \right)\right).
\end{align}

\subsection{Introduce auxiliary variables}
We now introduce new variables
\begin{align}
C(t,t') &= \frac{1}{N} \sum_i x(t)x_i(t'),\\
L(t,t') &= \frac{1}{N} \sum_i \hat{x}_i(t)\hat{x}_i(t'),\\
K(t,t') &= \frac{1}{N} \sum_i x_i(t)\hat{x}_i(t'),\\
M(t) &= \frac{1}{N} \sum_i x_i(t),\\
P(t) &= \frac{i}{N} \sum_i \hat{x}_i(t).
\end{align}
We do this by inserting an expression for one in the following form
\begin{align}
\int \exp \left( i \int \hat{C}(t,t')\left(NC(t,t') - \sum_i x_i(t)x_i(t') \right) dtdt' \right) D\hat{C}DC = 1,
\end{align}
for each of the new variables. We now have the following expression for the generating functional,
\begin{align}
    \overline{Z[\pmb{\psi}]} &= \int \exp\left(\sum_i\ln\left[ \int \mathcal{P}(x_i(0)) \exp\left(i\int \psi_i(t)x_i(t) dt \right)\right.\right. \nonumber \\
    &\times \exp\left\{ i \int \hat{x}_i(t) \left[\frac{\dot{x}_i(t)}{x_i(t)} - k_i + x_i(t) - h_i(t) \right]dt\right\} \nonumber \\
    &\times \exp \left(- i \int \hat{C}(t,t') x_i(t)x_i(t') + \hat{L}(t,t') \hat{x}_i(t) \hat{x}_i(t') + \hat{K}(t,t') x_i(t)\hat{x}_i(t')dtdt' \right)\nonumber \\
    &\times \left.\left. \exp \left(-i \int \hat{M}(t) x_i(t) + i\hat{P}(t) \hat{x}_i(t) dt \right)D\hat{x}_iDx_i\right]\right) \nonumber \\
    &\times\prod_{p=2}^N \exp\left(-\mu_pN\int P(t) M(t)^{p-1} dt \right) \nonumber \\
    &\times \prod_{p=2}^N\exp\left(-\frac{\sigma^2_pNp}{4}\left(\int L(t,t')C(t,t')^{p-1} + \gamma_p(p-1) K(t',t)K(t,t') C(t,t')^{p-2} dt dt' \right)\right) \nonumber \\
    &\times \exp \left( iN \int \hat{C}(t,t')C(t,t') + \hat{L}(t,t')L(t,t') + \hat{K}(t,t')K(t,t') dtdt' \right) \nonumber \\
    &\times \exp \left(i N\int \hat{M}(t)M(t) + \hat{P}(t)P(t) dt \right) \nonumber \\
    &D\hat{C}DCD\hat{L}DLD\hat{K}DKD\hat{M}DMD\hat{P}DP.
\end{align}

\subsection{Saddle-point integration}
We now take the limit $N \to \infty$ and use the saddle-point approximation to evaluate the dynamical order parameters where the exponent becomes extremal.
We have
\begin{align}\label{saddlegf}
    \overline{Z[\pmb{\psi}]} = \int \exp\left(N\left(\Omega + \Phi + \Psi \right) \right) D\hat{C}DCD\hat{L}DLD\hat{K}DKD\hat{M}DMD\hat{P}DP,
\end{align}
with
\begin{align}\label{omega}
    \Omega &= \frac{1}{N} \sum_i\ln\left[ \int \mathcal{P}(x_i(0)) \exp\left(i\int \psi_i(t)x_i(t) dt \right)\right. \nonumber \\
    &\times \exp\left\{ i \int \hat{x}_i(t) \left[\frac{\dot{x}_i(t)}{x_i(t)} - k_i + x_i(t) - h_i(t) \right]dt\right\} \nonumber \\
    &\times \exp \left(- i \int \hat{C}(t,t') x_i(t)x_i(t') + \hat{L}(t,t') \hat{x}_i(t) \hat{x}_i(t') + \hat{K}(t,t') x_i(t)\hat{x}_i(t')dtdt' \right)\nonumber \\
    &\times \left. \exp \left(-i \int \hat{M}(t) x_i(t) + i\hat{P}(t) \hat{x}_i(t) dt \right)D\hat{x}_iDx_i\right]
\end{align}
describing the microscopic time evolution,
\begin{align}
    \Phi = \sum_{p=2}^N\left(-\mu_p\int P(t) M(t)^{p-1} dt -\frac{\sigma^2_pp}{4}\int L(t,t')C(t,t')^{p-1} + \gamma_p(p-1) K(t',t)K(t,t') C(t,t')^{p-2} dt dt'\right)
\end{align}
coming from the disorder average, and
\begin{align}
    \Psi &= i \int \hat{C}(t,t')C(t,t') + \hat{L}(t,t')L(t,t') + \hat{K}(t,t')K(t,t') dtdt' +i \int \hat{M}(t)M(t) + \hat{P}(t)P(t) dt
\end{align}
resulting from the introduction of macroscopic order parameters.
By extremising we obtain
\begin{align}
    i\hat{C}(t,t') &= \frac{1}{4}\sum_{p=2}^\infty \sigma_p^2p(p-1) \left[ L(t,t')C(t,t')^{p-2} + \gamma_p(p-2)K(t',t)K(t,t')C(t,t')^{p-3}\right],\\
    i\hat{K}(t,t') &= \frac{1}{2}\sum_{p=2}^N\sigma_p^2\gamma_pp(p-1) K(t',t)C(t,t')^{p-2}, \left(= \gamma_\Sigma(t,t')\chi(t,t')\right)\\
    i\hat{L}(t,t') &= \frac{1}{4} \sum_{p=2}^\infty \sigma_p^2pC(t,t')^{p-1}, \left(=\frac{1}{2}\sigma_\Sigma(t,t')\right)\\
    i\hat{M}(t) &= \sum_{p=2}^\infty \mu_p (p-1) P(t)M(t)^{p-2},\\
    i\hat{P}(t) &= \sum_{p=2}^\infty \mu_p M(t)^{p-1}, \left(= \mu_\Sigma(t)\right)\\
    \label{sad1}
   C(t,t') &= \lim_{N\to\infty} \frac{1}{N} \sum_i \left< x_i(t)x_i(t') \right>_\Omega,\\
   \label{sad2}
    K(t,t') &= \lim_{N\to\infty} \frac{1}{N} \sum_i \left< x_i(t)\hat{x}_i(t') \right>_\Omega,\\
    \label{sad3}
    L(t,t') &= \lim_{N\to\infty} \frac{1}{N} \sum_i \left< \hat{x}_i(t)\hat{x}_i(t') \right>_\Omega,\\
    \label{sad4}
    M(t) &= \lim_{N\to\infty}\frac{1}{N} \sum_i \left<x_i(t)\right>_\Omega,\\
    \label{sad5}
    P(t) &= \lim_{N\to\infty} \frac{i}{N} \sum_i \left< \hat{x}_i(t) \right>_\Omega.
\end{align}
In these expressions we have used the notation
\begin{align}
    \left<F\right>_\Omega =  \left( \frac{\int \mathcal{P}(x(0))\left( F\right)\exp\left(\omega\right)D\hat{x}Dx}{ \int \mathcal{P}(x(0)) \exp \left(\omega\right) D\hat{x}Dx} \right),
\end{align}
where $\omega$ is the argument of the exponential in $\Omega$ [Eq.~(\ref{omega})],
\begin{align}
    \Omega &= \ln \left[ \int \mathcal{P}(x(0)) \exp \left(\omega\right) D\hat{x}Dx\right].
\end{align}
We have assumed that any perturbation is identical across species, $h_i(t) = h(t)$, and that initial conditions are independent and identically distributed, $\mathcal{P}(x_i(0)) = \mathcal{P}(x(0))$ $\forall i$. The terms in the sum in Eq.~(\ref{omega}) are then all identical, so we drop the subscripts.

\subsection{Further simplification}
We differentiate the expression for the generating functional in Eq.~(\ref{saddlegf}), then take the limit $N \to \infty$, and compare with Eqs.~(\ref{diff1}) - (\ref{diff5}) to get
\begin{align}
    \overline{\left<x_i(t)\right>} = -i\left. \frac{\delta\overline{Z[\pmb{\psi}]}}{\delta\psi_i(t)}\right|_{\pmb{\psi} = \mathbf{0}} = \left<x_i(t)\right>_{\Omega[\pmb{\psi} = \mathbf{0}]}
\end{align}
and from Eq.~(\ref{sad4}),
\begin{align}
    M(t) = -i\left.\lim_{N \to \infty} \frac{1}{N}\sum_i \frac{\delta\overline{Z[\pmb{\psi}]}}{\delta\psi_i(t)}\right|_{\pmb{\psi} = \mathbf{0}}.
\end{align}
We also have
\begin{align}
    \overline{\left<x_i(t)x_i(t')\right>} = \left.-\frac{\delta^2\overline{Z[\pmb{\psi}]}}{\delta\psi_i(t)\delta\psi_i(t')}\right|_{\pmb{\psi} = \mathbf{0}} = \left<x_i(t)x_i(t')\right>_{\Omega[\pmb{\psi} = \mathbf{0}]}
\end{align}
and using Eq.~(\ref{sad1})
\begin{align}
    C(t,t') = - \left.\lim_{N \to \infty} \frac{1}{N}\sum_i \frac{\delta^2\overline{Z[\pmb{\psi}]}}{\delta\psi_i(t)\delta\psi_i(t')}\right|_{\pmb{\psi} = \mathbf{0}}.
\end{align}
Similarly we have
\begin{align}
    \overline{\left<x_i(t)\hat{x}_i(t')\right>} = i\pdv{\overline{\left<x_i(t)\right>}}{h_i(t')} = \left.\frac{\delta^2\overline{Z[\pmb{\psi}]}}{\delta\psi_i(t)\delta h_i(t')}\right|_{\pmb{\psi} = \mathbf{0}} =  \left<x_i(t)\hat{x}_i(t')\right>_{\Omega[\pmb{\psi} = \mathbf{0}]},
\end{align}
and from Eq.~(\ref{sad2})
\begin{align}
    K(t,t') = \lim_{N\to\infty} \frac{1}{N}\left. \sum_i \frac{\delta^2\overline{Z[\pmb{\psi}]}}{\delta\psi_i(t)\delta h_i(t')}\right|_{\pmb{\psi} = \mathbf{0}}.
\end{align}
In the same way, using Eq.~(\ref{sad5}) we have
\begin{align}
    \frac{\delta\overline{Z[\pmb{\psi} = \mathbf{0}]}}{\delta h_i(t)} = -i\left<\hat{x}_i(t)\right>_{\Omega[\pmb{\psi} = \mathbf{0}]} = 0,\\
    P(t) = 0,
\end{align}
and using Eq.~(\ref{sad3})
\begin{align}
    \frac{\delta^2\overline{Z[\pmb{\psi} = \mathbf{0}]}}{\delta h_i(t)\delta h_i(t')} = -\left<\hat{x}_i(t)\hat{x}_i(t')\right>_{\Omega[\pmb{\psi} = \mathbf{0}]} = 0,\\
    L(t,t') = 0.
\end{align}

\subsection{Effective single-species process}
We now insert our results from the saddle point method into our expression for $\Omega$ in Eq.~(\ref{omega}) to find
\begin{align}
    \left<F[x]\right>_{\Omega[\psi=0]} = \frac{\int F[x]\mathrm{P}[x] Dx}{\int \mathrm{P}[x]Dx},
\end{align}
where
\begin{align}\label{effmeasure}
    \mathrm{P}[x] &= \mathcal{P}(x(0))\exp\left\{ i \int \hat{x}(t) \left[\frac{\dot{x}(t)}{x(t)} - k + x(t) - h(t) \right]dt\right\} \nonumber \\
    &\times \exp \left(- \frac{1}{4} \sum_{p=2}^\infty \sigma_p^2 \int pC(t,t')^{p-1} \hat{x}(t) \hat{x}(t') + 2i\gamma_pp(p-1) G(t',t)C(t,t')^{p-2} x(t)\hat{x}(t')dtdt' \right)\nonumber \\
    &\times \exp \left(- i\sum_{p=2}^\infty \mu_p \int M(t)^{p-1} \hat{x}(t) dt \right)D\hat{x}.
\end{align}
We highlight that $\mathrm{P}[x]$ is not to be confused with $P(t)$.
$\mathrm{P}[x]$ is the probability of observing a path of the effective process given by the equation
\begin{align}\label{effpr}
    \dot{x}(t) = x(t)\left(k - x(t) + \sum_{p=2}^\infty\left( \mu_p M(t)^{p-1} + \gamma_p\sigma_p^2\frac{p(p-1)}{2} \int_0^t G(t,t')C(t,t')^{p-2}x(t')dt'\right) + h(t) + \eta(t)\right)
\end{align}
with
\begin{align}
    \label{effpr1}M(t) &= \left<x(t)\right>_*,\\
    \label{effpr2}\left<\eta(t)\eta(t')\right>_* &= \sum_{p=2}^\infty \sigma_p^2\frac{p}{2}\left<x(t)x(t')\right>_*^{p-1} = \sum_{p=2}^\infty \sigma_p^2\frac{p}{2}C(t,t')^{p-1}\\
    \label{effpr3}G(t,t') &= -iK(t,t') = \left<\pdv{x(t)}{h(t')}\right>_*
\end{align}
where $\left<\dots\right>_*$ is the average over realisations of the effective process, i.e., over random $\eta(t)$ and initial conditions $\mathcal{P}(x(0))$. If we take the generating functional for this process, average over the noise and use the saddle point approximation as before, we obtain the same measure $P[x]$.
This means that the statistics of realisations of this effective process are the same as those of the individual species trajectories in the original model. Therefore we can use the effective process to analyse the original system, its fixed points and their stability.

\section{Fixed-point solutions}\label{suppfp}
We have the equation for the effective process in Eq.~(\ref{effpr}) with the self-consistency relations in Eqs.~(\ref{effpr1} - \ref{effpr3}). We now make the following assumptions:
\begin{itemize}
    \item The system reaches a fixed point asymptotically, $x(t) \to x^*$ as $t \to \infty$. This means that both $M(t)$ and $C(t+\tau, t)$ tend to a constant, let $M(t) \to M^*$ and $C(t+\tau,t) \to q$ as $t \to \infty$ $\forall \tau$.
    \item Time-translation invariant response function $G(t+\tau,t) = G(\tau)$, we address only ergodic stationary stationary states which do not depend on initial conditions, where perturbations have no long term effects so that
    \begin{align}\label{Xinf}
        \chi = \int_0^\infty G(\tau) d\tau < \infty \quad \text{and} \quad \lim_{\tau\to\infty} G(\tau) = 0.
    \end{align}
    \item Dynamics started at $t = -\infty$.
    \item The fixed point assumption implies that each realisation of $\eta(t)$ must tend to a static Gaussian random variable $\eta^*$, with $\left<\eta^*\right> = 0$ and
    \begin{align}\label{sigsigdef}
        \left<\eta^{*2}\right> = \sum_{p=2}^\infty \sigma_p^2 \frac{p}{2}q^{p-1} \equiv \sigma_\Sigma^2,
    \end{align}
    where $\sigma_\Sigma$ is introduced here to simplify the expression.
\end{itemize}
This means that the fixed points will satisfy the condition
\begin{align}\label{fp1}
    x^*\left(k - x^* + \sum_{p=2}^\infty\left( \mu_p M^{*p-1} + \gamma_p\sigma_p^2\frac{p(p-1)}{2}\chi q^{p-2}x^*\right) + h(t) + \eta^*\right) = 0.
\end{align}
We can write $\eta$ as $z\sqrt{\sum_{p=2}^\infty \sigma_p^2\frac{p}{2}q^{p-1}} = z\sigma_\Sigma$ where $z$ is a standard Gaussian random number, and set $h(t) = 0$ as we no longer need it. Instead, we can generate response functions by differentiating with respect to $\eta(t)$,
\begin{align}
G(t,t') =\left<\pdv{x(t)}{\eta(t')}\right>_*.
\end{align}
As well as $\sigma_\Sigma^2$ defined in Eq.~(\ref{sigsigdef}), we also introduce quantities
\begin{align}
    \mu_\Sigma &= \sum_{p=2}^\infty \mu_p M^{*p-1},\label{musigdef}\\
    \gamma_\Sigma &= \sum_{p=2}^\infty \gamma_p\sigma_p^2 \frac{p(p-1)}{2}q^{p-2}.\label{gamsigdef}
\end{align}
This simplifies our condition for fixed points in Eq.~(\ref{fp1}) to
\begin{align}
    x^*\left(k - x^* + \mu_\Sigma + \gamma_\Sigma \chi x^* + z\sigma_\Sigma\right) = 0.
\end{align}
We have $x^*(z) = 0$ as a solution for all $z$, other solutions which satisfy
\begin{align}
k - x^* + \mu_\Sigma + \gamma_\Sigma \chi x^* + z\sigma_\Sigma = 0
\end{align}
are only valid for non-negative $x^*$ as this is the population concentration, giving
\begin{align}
    x^* = \max \left(\frac{k + \mu_\Sigma + z\sigma_\Sigma}{1 - \gamma_\Sigma\chi}, \quad 0\right).
\end{align}
The first argument becomes zero when
\begin{align}
    k + \mu_\Sigma + z\sigma_\Sigma = 0,
\end{align}
which happens at a specific value of $z$, namely $z_1$,
\begin{align}\label{z1}
    z_1 = \frac{- \left(k + \mu_\Sigma\right)}{\sigma_\Sigma},
\end{align}
giving the result for $x^*$ as
\begin{align}
    x^* =
    \begin{cases}
        \frac{k + \mu_\Sigma + z\sigma_\Sigma}{1 - \gamma_\Sigma\chi} & z \geq z_1 \\
        0 & z \leq z_1.
    \end{cases}
\end{align}
We can then simplify and solve the self-consistency relations in Eqs.~(\ref{effpr1}, \ref{effpr2}, \ref{effpr3}) to find the values of these parameters for fixed points,
\begin{align}\label{M1}
    M^* &= \int_{z_1}^\infty \frac{k + \mu_\Sigma + z\sigma_\Sigma}{1 - \gamma_\Sigma\chi} Dz,\\
    \label{q1}
    q &= \int_{z_1}^\infty \left(\frac{k + \mu_\Sigma + z\sigma_\Sigma}{1 - \gamma_\Sigma\chi}\right)^2 Dz,\\
    \chi &= \int_0^\infty G(\tau)d\tau
    \quad= \int_0^\infty \left<\pdv{x(t)}{\eta(t-\tau)}\right> d\tau \nonumber \\
    &= \left<\pdv{x(\eta^*)}{\eta^*}\right>
    \quad= \int_{z_1}^\infty \frac{1}{1-\gamma_\Sigma\chi} Dz
    \quad= \frac{\phi}{1 - \gamma_\Sigma\chi},\label{X1}
\end{align}
where $Dz = \frac{dz}{\sqrt{2\pi}}e^{-z^2/2}$ and $\phi \equiv \int Dz$ is the fraction of species alive at the fixed point. These equations, along with the definitions of $\mu_\Sigma, \sigma_\Sigma, \gamma_\Sigma$ can self-consistently be solved to find the values of the macroscopic order parameters at the fixed point.

\section{Linear stability analysis}\label{suplinstab}

To carry out the linear stability analysis, we use the effective process given in Eq.~(\ref{effpr}).
We follow \cite{diederich} (see also \cite{GFA_notes}) and write $x(t) = x^* + y(t)$ and $\eta(t) = z\sigma_\Sigma + v(t)$ with $\left<y(t)\right> = \left<v(t)\right> = 0$. The order parameter $M$ is not affected by the perturbation $y(t)$, as the ansatz assumes zero-average fluctuations from the fixed point. Self-consistency also requires
\begin{align}
    \left<\eta(t)\eta(t')\right> = \left<\left(z\sigma_\Sigma + v(t)\right)\left(z\sigma_\Sigma + v(t')\right)\right> = \sum_{p=2}^\infty\sigma_p^2\frac{p}{2}\left<\left(x^* + y(t)\right)\left(x^* + y(t')\right)\right>^{p-1},
\end{align}
which simplifies to
\begin{align}
    \sigma_\Sigma^2 + \left<v(t)v(t')\right> = \sum_{p=2}^\infty\sigma_p^2\frac{p}{2}\left(q + \left<y(t)y(t')\right>\right)^{p-1},
\end{align}
which we expand to first order of $\left<y(t)y(t')\right>$ to obtain
\begin{align}
    \sigma_\Sigma^2 + \left<v(t)v(t')\right> = \sum_{p=2}^\infty \sigma_p^2\frac{p}{2}\left(q^{p-1} + (p-1)q^{p-2}\left<y(t)y(t')\right>\right),
\end{align}
leading to
\begin{align}\label{v(t)}
    \left<v(t)v(t')\right> = \left<y(t)y(t')\right>\sum_{p=2}^\infty\sigma_p^2\frac{p(p-1)}{2}q^{p-2}.
\end{align}
Substituting this ansatz into the effective process leads to
\begin{align}
    \dot{y}(t) &= \left(x^* + y(t)\right)\left(k - x^* y(t) + \sum_{p=2}^\infty\left( \mu_p M^{*p-1} + \gamma_p\sigma_p^2\frac{p(p-1)}{2} \int_0^t G(t,t')C(t,t')^{p-2}\left(x^* + y(t')\right)dt'\right)\right.\nonumber\\
    &+ \left.z\sigma_\Sigma + v(t) \right)
\end{align}
We investigate the stability of solutions $x^*=0$ and $x^*>0$ separately:
\begin{enumerate}
\item[(i)]\label{x=0}
For the fixed point $x^*(z) = 0$, we have
\begin{align}
    \dot{y}(t) = y(t)\left(k - y(t) + \sum_{p=2}^\infty\left( \mu_p M^{*p-1} + \gamma_p\sigma_p^2\frac{p(p-1)}{2} \int_0^t G(t,t')C(t,t')^{p-2}y(t')dt'\right) + z\sigma_\Sigma + v(t) \right).
\end{align}
After linearising this becomes
\begin{align}
    \dot{y}(t) &= y(t)\left(k + \sum_{p=2}^\infty\mu_p M^{*p-1} + z\sigma_\Sigma \right) \nonumber\\
    &= y(t)\left(k + \mu_\Sigma + z\sigma_\Sigma \right)
\end{align}
It is now convenient to distinguish between the following two cases:
\begin{enumerate}
\item[(a)] For $z \geq z_1$, we find $\dot{y}(t) \geq 0$, hence $x^*=0$ is unstable. 
\item[(b)] For $z \leq z_1$, we find $\dot{y}(t) \leq 0$, hence $x^* = 0$ is stable.
\end{enumerate}
We note that $x^*=0$ is stable only for the cases where it is the unique fixed point, which is the case when $z \leq z_1$.
\item[(ii)]\label{otherfp}
The other fixed points ($x^*>0$) satisfy
\begin{align}
    k - x^* + \mu_\Sigma + \gamma_\Sigma\chi x^* + z\sigma_\Sigma = 0.
\end{align}
To study the stability of the fixed point for $z \geq z_1$,  we add a zero mean variance one noise variable $\xi(t)$ (following for example \cite{opper,GFA_notes}). We then have
\begin{align}
    \dot{y}(t) &= \left(x^* + y(t)\right)\left(k - x^* y(t) + \sum_{p=2}^\infty\left( \mu_p M^{*p-1} + \gamma_p\sigma_p^2\frac{p(p-1)}{2} \int_0^t G(t,t')C(t,t')^{p-2}\left(x^* + y(t')\right)dt'\right)\right.\nonumber\\
    &+ \left.z\sigma_\Sigma + v(t) + \xi(t) \right).
\end{align}
Linearising this we find,
\begin{align}
    \dot{y}(t) &= x^*\left(- y(t) + \sum_{p=2}^\infty\left( \gamma_p\sigma_p^2\frac{p(p-1)}{2} \int_0^t G(t,t')C(t,t')^{p-2} y(t')dt'\right) + v(t) + \xi(t) \right).
\end{align}
We perform a Fourier transform to obtain
\begin{align}
    \frac{i\omega\tilde{y}(\omega)}{x^*} = \left(\sum_{p=2}^\infty \gamma_p\sigma_p^2\frac{p(p-1)}{2}\tilde{G}(\omega)q^{p-2} - 1\right)\tilde{y}(\omega) + \tilde{v}(\omega) + \tilde{\xi}(\omega),
\end{align}
and from this we conclude
\begin{align}
    \tilde{y}(\omega) = \frac{\tilde{v}(\omega) + \tilde{\xi}(\omega)}{\frac{i\omega}{x^*} + \left(1 - \sum_{p=2}^\infty \gamma_p\sigma_p^2\frac{p(p-1)}{2}\tilde{G}(\omega)q^{p-2}\right)}.
\end{align}
We consider the case $\omega = 0$ (see also \cite{opper,sollich,GFA_notes}), using $\chi=\tilde{G}(0) = \int_0^\infty G(\tau) d\tau$, we find
\begin{align}
    \left<\tilde{y}(0)\overline{\tilde{y}(0)}\right>_\phi = \frac{\left<\tilde{v}(0)\overline{\tilde{v}(0)}\right>_\phi + \left<\tilde{\xi}(0)\overline{\tilde{\xi}(0)}\right>_\phi}{(1 - \gamma_\Sigma\chi)^2}.
\end{align}
The average is over the fraction $\phi$ of alive species only, indicated by the subscript `$\phi$', the species with $x^* = 0$ are stable as previously discussed and do not contribute to the the statistics of perturbations.
Using equation (\ref{v(t)}) we find
\begin{align}
    \left<|\tilde{y}(0)|^2\right> = \frac{\phi\left(1 + \left<|\tilde{y}(0)|^2\right>\sum_{p=2}^\infty \sigma_p^2\frac{p(p-1)}{2}q^{p-2}\right)}{(1 - \gamma_\Sigma\chi)^2},
\end{align}
where the factor of $\phi$ results because this only applies to the fraction $\phi$ of extant species with $z \geq z_1$.
Re-arranging we have
\begin{align}
\left<|\tilde{y}(0)|^2\right>  = \frac{\phi}{(1-\gamma_\Sigma\chi)^2 - \phi\sum_{p=2}^\infty \sigma_p^2\frac{p(p-1)}{2}q^{p-2}}.
\end{align}
This quantity needs to be finite in order for the fixed points to be stable, it must also be non-negative by construction. The onset of instability occurs for
\begin{align}\label{crit1}
    (1 - \gamma_\Sigma\chi)^2 = \phi\sum_{p=2}^\infty \sigma_p^2\frac{p(p-1)}{2}q^{p-2}.
\end{align}
\end{enumerate}

The expressions for the macroscopic order parameters $\chi$, $\phi$, $M^*$ and $q$ are derived on the assumption of a unique stable fixed point, hence they are only valid for model parameters in which such a unique fixed point exists, and is stable. 

\section{Further analysis of the model with a single order of interaction}\label{singlep}
Given values of $\mu_p$, $\sigma_p$ and $\gamma_p$ for a system with interactions of order $p$ only, we can find the fixed point values of $\phi$, $M^*$, $q$ and $\chi$. Given $\mu_p$ and $\gamma_p$ we can also find the value for $\sigma_p$ where the system changes its behaviour between stability, instability, and divergence.
\subsection{Summary of equations}
We can rewrite Eqs.~(\ref{M1}, \ref{q1}, \ref{X1}) as
\begin{align}\label{M2}
    M^* &= \frac{\sigma_\Sigma}{H}I_1,\\
    \label{q2}
    q &= \frac{\sigma_\Sigma^2}{H^2}I_2,
\end{align}
and
\begin{align}\label{X2}
    \chi = \frac{I_0}{H},
\end{align}
where
\begin{align}\label{H2}
    H \equiv 1 - \gamma_\Sigma\chi,
\end{align}
to simplify expressions as this factor $1 - \gamma_\Sigma\chi$ appears so frequently.
We define the relevant integrals,
\begin{align}
    \label{phi2}
    I_0(z_1) &= \int_{z_1}^\infty Dz \quad (\equiv \phi),\\
    \label{F2}
    I_1(z_1) &= \int_{z_1}^\infty (z - z_1) Dz,\\
    \label{S2}
    I_2(z_1) &= \int_{z_1}^\infty (z - z_1)^2 Dz
\end{align}
where $z_1$ is given by Eq.~(\ref{z1}) but also repeated here,
\begin{align}\label{singlepz1}
    z_1 = \frac{- \left(k + \mu_\Sigma\right)}{\sigma_\Sigma}
\end{align}
which is the value of $z$ where the species population reaches zero.
We also have the summed quantities from Eqs.~(\ref{sigsigdef}, \ref{musigdef}, \ref{gamsigdef}) reduced for a single value of $p$,
\begin{align}\label{sigtot2}
    \sigma_\Sigma^2 &= \sigma_p^2\frac{p}{2}q^{p-1},\\
    \label{mutot2}
    \mu_\Sigma &= \mu_p M^{*p-1},
\end{align}
and
\begin{align}\label{gamtot2}
    \gamma_\Sigma = \gamma_p\sigma_p^2\frac{p(p-1)}{2}q^{p-2}.
\end{align}
Using these equations, we can find all other quantities ($\sigma_p$, $I_0$, $I_1$, $I_2$, $M^*$, $q$, $\chi$, $H$, $\sigma_\Sigma$, $\gamma_\Sigma$, $\mu_\Sigma$) for specified parameters $\gamma_p$, $\mu_p$ and $z_1$, as we will explain next. The dependence of the macroscopic quantities on $\sigma_p$ is found parametrically from their dependence on $z_1$.

\subsection{Method for solving equations}\label{singlepmethod}

By combining Eq.~(\ref{X2}) with Eq.~(\ref{H2}) we obtain
\begin{align}\label{step1}
    \chi = I_0 + \gamma_\Sigma\chi^2.
\end{align}
Further, by substituting for $H$ from Eq.~(\ref{X2}) into Eq.~(\ref{q2}) we obtain an expression for $\chi^2$
\begin{align}
    \chi^2 = \frac{qI_0^2}{I_2\sigma_\Sigma^2},
\end{align}
which can be substituted into Eq.~(\ref{step1}) to obtain
\begin{align}\label{chii}
    \chi = I_0 + \frac{q\gamma_\Sigma}{\sigma_\Sigma^2}\frac{I_0^2}{I_2}.
\end{align}
The integrals $I_0, I_2$ on the right are functions of $z_1$, so they can easily be evaluated for given $z_1$. So in order to find $\chi$, we need to determine $\frac{q\gamma_\Sigma}{\sigma_\Sigma^2}$. For a single value of $p$ we find
\begin{align}\label{simp}
    \frac{q\gamma_\Sigma}{\sigma_\Sigma^2} = \gamma_p(p-1)
\end{align}
by combining Eqs.~(\ref{sigtot2},\ref{gamtot2}).
We note that $\frac{q\gamma_\Sigma}{\sigma_\Sigma^2}$ cannot be expressed in such simple terms in models with multiple order of interactions. 

By substituting Eq.~(\ref{simp}) into Eq.~(\ref{chii}), we are able to obtain $\chi$ via
\begin{align}\label{X(z1)}
    \chi = \gamma_p(p-1)\frac{I_0^2}{I_2} + I_0,
\end{align}
which is a function $z_1$ and $\gamma_p$ only, so this is the first macroscopic parameter we have solved for.
Using Eq.~(\ref{X2}) we can obtain $H$ via
\begin{align}\label{singlepH}
    H = \frac{I_2}{I_2 + \gamma_p(p-1)I_0}.
\end{align}
By substituting the expression for $\sigma_\Sigma$ from Eq.~(\ref{M2}), and $\mu_\Sigma$ from Eq.~(\ref{mutot2}) into Eq.~(\ref{singlepz1}) we obtain the following polynomial for $M^*$
\begin{align}\label{singlepM}
    \mu_p M^{*p-1} + \frac{z_1H}{I_1}M^* + k = 0.
\end{align}
For $p = 2$, this has the solution
\begin{align}\label{singlep2M}
    M^* = \frac{k}{\frac{-z_1H}{I_1} - \mu_2}.
\end{align}
For $p = 3$, we find two solutions to the quadratic equation, with
\begin{align}\label{singlep3M}
    M^* = \frac{-\frac{z_1H}{I_1} - \sqrt{\left(\frac{z_1H}{I_1}\right)^2 - 4\mu_3 k}}{2\mu_3}
\end{align}
as the only solution which is non-negative for all values of $\mu_3$ and agrees with results from simulations.
To see that the above solution for $M^*$ is positive for all combinations of $\mu_3, \gamma_3$ and $\sigma_3$, first consider $\mu_3 > 0$, this means the numerator should also be positive to ensure $M^*$ is positive. For positive $\mu_3$, the maximum possible value of $z_1$ is always negative (as it must be below the divergence boundary, discussed later), so $z_1 < 0$.
As $z_1 < 0$, $\frac{-z_1H}{I_1}$ is positive, and because the absolute value of the square root is less than this (because $\mu_3 > 0$), the numerator is still positive as a lower value has been subtracted. Conversely, if $\mu_3 < 0$, we require the numerator to also be negative to ensure $M^*$ is positive. This time, $\frac{-z_1H}{I_1}$ could either be positive or negative, but as the absolute value of the square root is more than $\frac{-z_1H}{I_1}$, the subtraction results in a negative value in either case. The other solution of the quadratic equation can change sign, and is not physical, in the sense that it does not agree with simulations.

The solution in Eq.~(\ref{singlep3M}) holds unless $\mu_3 = 0$, in which case
\begin{align}\label{singlep3mu0}
    M^* = \frac{-kI_1}{z_1H}.
\end{align}
For $p = 4$, the cubic equation (\ref{singlepM}) can be solved using Cadano's cubic formula, resulting in three solutions for $M^*$,
\begin{align}\label{singlep4M123}
    M_1 &= S + T, \nonumber\\
    M_2 &= - \frac{S + T}{2} + \frac{i\sqrt{3}(S - T)}{2}, \nonumber\\
    M_3 &= - \frac{S + T}{2} - \frac{i\sqrt{3}(S - T)}{2},
\end{align}
where
\begin{align}\label{singlep4ST}
    S &= \left(\frac{-k}{2\mu_4} + \sqrt{\left(\frac{z_1H}{3\mu_4 I_1}\right)^3 + \left(\frac{k}{2\mu_4}\right)^2}\right)^{1/3}, \nonumber\\
    T &= \left(\frac{-k}{2\mu_4} - \sqrt{\left(\frac{z_1H}{3\mu_4 I_1}\right)^3 + \left(\frac{k}{2\mu_4}\right)^2}\right)^{1/3}.
\end{align}

For $\mu_4 < 0$, the discriminant,
\begin{align}
    D = \left(\frac{z_1H}{3\mu_4 I_1}\right)^3 + \left(\frac{k}{2\mu_4}\right)^2,
\end{align}
is positive, resulting in real values of both $S$ and $T$, and $M_1$ as the only real solution with $M_2$ and $M_3$ complex conjugates. For $\mu_4 > 0$ but below the divergence point (found later), the discriminant is negative, resulting in complex conjugates for $S$ and $T$. Although all three solutions for $M^*$ are real, only $M_3$ is continuous with the previous solution as $\mu_4 \to 0$.
Simulations confirm that for $\mu_4 < 0$, $M_1$ gives the correct solution, and for $\mu_4 > 0$, $M_3$ gives the correct solution. For $\mu_4 = 0$, we get the same as before in Eq.~(\ref{singlep3mu0}), which is consistent with accepted solutions for non-zero $\mu_4$.

After we have found $M^*$, we can combine Eq.~(\ref{M2}) with Eq.~(\ref{q2}) to find $q$ via
\begin{align}\label{qq}
    q = I_2\left(\frac{M^*}{I_1}\right)^2,
\end{align}
and by inserting this expression for $q$, and the expression for $\sigma_\Sigma$ from Eq.~(\ref{M2}) into Eq.~(\ref{sigtot2}) we find
\begin{align}\label{singlepsig}
    \sigma_p = H\sqrt{\frac{2}{pI_2^{p-1}}}\left(\frac{I_1}{M^*}\right)^{p-2}.
\end{align}
Now we have found all other parameters as a function of $z_1$, $\gamma_p$ and $\mu_p$.

\subsection{Divergence point}\label{sec:singlepdiv}

For certain combinations of $z_1$, $\gamma_p$ and $\mu_p$, these equations do not produce a physically meaningful solution. This is a result of the solution(s) of $M^*$ becoming infinite or complex. The point where this occurs is the divergence point and the specific value(s) of $z_1$ at this point is called $z_d$, and the value of $\sigma_p$ at this point is called $\sigma_d$. As the equation for $M^*$ (\ref{singlepM}) has different solutions for each value of $p$, we look at each case separately. 
For a given value of $p$, we find that the divergence boundaries in the $\mu_p,\sigma_p$-plane for different value of $\gamma_p$ all intersect at a common value of $\mu_p$ when $\sigma_p=0$ (see Figs.~\ref{phase2}, \ref{phase3}, and \ref{phase4} in the main paper).
We will now determine this point of intersection for different orders $p$. 

When $\sigma_p=0$, we have $z_1 \to -\infty$. We also have

\begin{align}
    \lim_{z_1\to -\infty} I_0(z_1) &= 1 \nonumber\\
    \lim_{z_1\to -\infty} \frac{-z_1}{I_1(z_1)} &= 1 \nonumber \\
    \lim_{z_1\to -\infty} \frac{z_1^2}{I_2(z_1)}  &= 1,\label{sig0lim}
\end{align}
which also results in $\lim_{z_1\to -\infty} H = 1$ using Eq.~(\ref{singlepH}) and $q = M^{*2}$ from Eq.~(\ref{qq}) for $\sigma_p = 0$.

For $p = 2$, the solution
\begin{align}
    M^* = \frac{k}{\frac{-z_1H}{I_1} - \mu_2},
\end{align}
diverges when $\mu_2$ takes the value
\begin{align}
    \mu_d = \frac{-z_1H}{I_1}.
\end{align}
The solution for $M^*$ now becomes
\begin{align}
    M^* = \frac{k}{\mu_d - \mu_2}.
\end{align}
For the case of $\sigma_2 = 0$ and $z_1 \to -\infty$, $\mu_d = 1$, resulting in $M^* = k/(1 - \mu_2)$ and $q = M^{*2}$. Both $M^*$ and $q$ become divergent at the divergence point as well as beyond it.

For $p = 3$, the solution for $\mu_3 \neq 0$,
\begin{align}\label{singlep3Mdiv}
    M^* = \frac{-\frac{z_1H}{I_1} - \sqrt{\left(\frac{z_1H}{I_1}\right)^2 - 4\mu_3 k}}{2\mu_3},
\end{align}
becomes complex when the discriminate is zero, leading to
\begin{align}\label{singlep3div}
    \left(\frac{z_1H}{I_1}\right)^2 = 4\mu_d k.
\end{align}
For $\mu_3 = 0$, on the other hand, we have
\begin{align}\label{singlep3Mdivmu0}
    M^* = \frac{-kI_1}{z_1H},
\end{align}
which diverges when $z_1=0$. The combination $z_d=0$ and $\mu_3=0$  satisfies Eq.~(\ref{singlep3div}). We therefore see that Eq.~(\ref{singlep3div}) holds at the divergence point, in both cases, $\mu_3\neq 0$ and $\mu_3=0$.
However, Eq.~(\ref{singlep3div}) can only be satisfied for $\mu_d \geq 0$. We note that for $\mu_3 < 0$, there is still a point where simulations diverge, at $\sigma_m$, as there are no solutions for $\sigma_3 > \sigma_m$, this is described later. The divergence boundary defined by Eq.~(\ref{singlep3div}) defines the maximum possible value of $z_1$ ($\equiv z_d$) for a given value of $\mu_3$. As the point where $z_d$ becomes zero is where $\mu_3$ is also zero, while $\mu_3 > 0$, $z_1$ must be below zero.
Substituting Eq.~(\ref{singlep3div}) back into Eq.~(\ref{singlep3Mdiv}) leads to 
\begin{align}\label{singlep3Mmu}
    M^* = \frac{\sqrt{k}}{\mu_3}\left(\sqrt{\mu_d} - \sqrt{\mu_d - \mu_3}\right)
\end{align}
for $\mu_3 \neq 0$, and into Eq.~(\ref{singlep3Mdivmu0}) leads to
\begin{align}
    M^* = \frac{1}{2}\sqrt{\frac{k}{\mu_d}}
\end{align}
for $\mu_3 = 0$. Note that the value for $M^*$ in Eq.~(\ref{singlep3Mmu}) is necessarily positive for all non-zero values of $\mu_3$. In both cases of $\mu_3$, $M^* = \sqrt{k/\mu_3}$ at the divergence point (where $\mu_3 = \mu_d$), as for $\mu_3 = 0$ this value is infinite.
For the case of $\sigma_3 = 0$ and $z_1 \to -\infty$, Eq.~(\ref{singlep3div}) reduces to $\mu_d = 1/4k$, and
\begin{align}\label{singlep3Msig0}
    M^* = \frac{1}{2\mu_3}\left(1 - \sqrt{1 - 4\mu_3k}\right)
\end{align}
for $\mu_3 \neq 0$ and $M^* = k$ for $\mu_3 = 0$. At the divergence point ($\mu_3 = 1/4k$), $M^* = 2k$. These values are plotted in as the red line in Fig.~\ref{M3sig0}, which demonstrates these cases form a continuous function of $\mu_3$, i.e. Eq.~(\ref{singlep3Msig0}) tends to $k$ as $\mu_3 \to 0$.

For $p = 4$, we find the solution for $M^*$ for $\mu_4 > 0$ given by $M_3$ in Eq.~(\ref{singlep4M123}) becomes complex when the discriminant
\begin{align}
    D = \left(\frac{z_1H}{3\mu_4 I_1}\right)^3 + \left(\frac{k}{2\mu_4}\right)^2,
\end{align}
is equal to zero. Setting this equal to zero gives an expression for the divergence point,
\begin{align}\label{singlep4div}
    \mu_d = -\frac{4}{k^2}\left(\frac{z_1H}{3I_1}\right)^3,
\end{align}
where this expression would always give a positive value of $\mu_4$. For $\mu_4 = 0$, this condition also gives the divergence point, $z_d = 0$. Again for $\mu_4 > 0$, we can still find divergence in simulations, above $\sigma_m$.

Substituting Eq.~(\ref{singlep4div}) back into Eq.~(\ref{singlep4ST}) leads to
\begin{align}
    S = \left(\frac{k}{2\mu_4}\left(-1 + \sqrt{\frac{\mu_4 - \mu_d}{\mu_4}}\right)\right)^{\frac{1}{3}}, \nonumber\\
    T = \left(\frac{k}{2\mu_4}\left(-1 - \sqrt{\frac{\mu_4 - \mu_d}{\mu_4}}\right)\right)^{\frac{1}{3}}.
\end{align}
For $\mu_4 < 0$, both $S$ and $T$ are real, and $M^* = S + T$. For $\mu_4 = 0$,
\begin{align}\label{singlep4Mmu0}
    M^* = \frac{1}{3}\left(\frac{4k}{\mu_d}\right)^{\frac{1}{3}}.
\end{align}
For $0 < \mu_4 < \mu_d$, $S$ and $T$ are complex conjugates, with $S + T$ real and $S - T$ imaginary. We have
\begin{align}
    M^* = -\frac{1}{2}\left((S + T) + i\sqrt{3}(S - T)\right).
\end{align}
Finally, for $\mu_4 = \mu_d$ at the divergence point,
\begin{align}
    M^* = \left(\frac{k}{2\mu_4}\right)^{1/3},
\end{align}
which holds for $\mu_4 (= \mu_d) \geq 0$.
For the case of $\sigma_4 = 0$ and $z_1 \to -\infty$, Eq.~(\ref{singlep4div}) reduces to $\mu_d = 4/(27k^2)$. Using this value we find $M^* = k$ for $\mu_4 = 0$, and $M^* = 3k/2$ for $\mu_4 = \mu_d$. For the other values of $\mu_4$ we get the same expressions as above, but now with $S$ and $T$ being functions of $\mu_4$ only,
\begin{align}\label{STmud}
    S = \left(\frac{k}{2\mu_4}\left(-1 + \sqrt{1 - \frac{4}{27k^2\mu_4}}\right)\right)^{\frac{1}{3}}, \nonumber\\
    T = \left(\frac{k}{2\mu_4}\left(-1 - \sqrt{1 -\frac{4}{27k^2\mu_4}}\right)\right)^{\frac{1}{3}}.
\end{align}
These functions for $M^*$ at $\sigma_4 = 0$ are shown in Fig.~\ref{M4sig0}, where the different cases for $\mu_4$ (red and blue lines) join to form a continuous function of $\mu_4$.

\subsection{Anti-symmetric interactions}\label{sec:singlepanti}

For the case of $\gamma_p = -1/(p-1)$ (the lowest possible value for $\gamma_p$) , we find there is another condition for which we are unable to find mathematical solutions. This occurs for when the value of $H$ given in Eq.~(\ref{singlepH}) diverges, which happens when $I_0=I_2$, which in turns means $z_1 = 0$, and $I_0 = I_2 = 1/2$. As $z_1$ tends to zero from below, we find that $I_2 - I_0 \approx -z_1/\sqrt{2\pi}$, leading to
\begin{align}
    H \approx -\frac{1}{z_1}\sqrt{\frac{\pi}{2}}.
\end{align}
As $H$ diverges, $\sigma_p$ also diverges (shown later), which means solutions can be found for all values of $\sigma_p$ and there is no maximum value $\sigma_m$ beyond which no solutions would exist. This means that as we increase $\sigma_p$ we do not find a point where simulations diverge. As $\sigma_p$ tends to infinity, we find that both $M^*$ and $q$ tend to constant values.
As $z_1$ tends to zero from below, $I_1 = 1/\sqrt{2\pi}$ and therefore
\begin{align}\label{singlepz1H/F}
    \frac{-z_1H}{I_1} = \pi.
\end{align}
This common fraction appears within the solutions for $M^*$ and can now be substituted into the solutions for $M^*$ and $\sigma_p$ for each value of $p$. In each case for $p$, we find $q = \pi M^{*2}$ which also tends to a finite constant as $\sigma_p \to \infty$ for $\gamma_p = -1/(p-1)$.

For $p = 2$, substituting Eq.~(\ref{singlepz1H/F}) into Eq.~(\ref{singlep2M}) leads to
\begin{align}\label{singlep2antiM}
    M^* = \frac{k}{\pi - \mu_2},
\end{align}
and into Eq.~(\ref{singlepsig}) leads to
\begin{align}
    \sigma_2 = -\frac{\sqrt{\pi}}{z_1}.
\end{align}
Therefore as $z_1 \to 0-$, $\sigma_2 \to \infty$ and $M^*$ tends to a finite constant. However, if $\mu_2 = \pi$, the solution for $M^*$ in Eq.~(\ref{singlep2antiM}) diverges, so as $\sigma_2 \to \infty$, $\mu_d \to \pi$. In conclusion, the divergence boundary starts at $\mu_d = 1$ for $\sigma_2 = 0$, and tends to $\mu_d \to \pi$ for $\sigma_2 \to \infty$. For values of $\mu_2$ below this boundary, $M^*$ is always finite as solutions exist for all values of $\sigma_2$, and as $\sigma_2 \to \infty$, both $M^*$ and $q$ tend to a finite constants.

For $p = 3$, substituting Eq.~(\ref{singlepz1H/F}) into Eq.~(\ref{singlep3M}) leads to
\begin{align}\label{singlep3antiM}
    M^* = \frac{\pi - \sqrt{\pi^2 - 4\mu_3k}}{2\mu_3}
\end{align}
for $\mu_3 \neq 0$, and into Eq.~(\ref{singlep3mu0}) leads to $M^* = k/\pi$ for $\mu_3 = 0$. Eq.~(\ref{singlepsig}) becomes
\begin{align}
    \sigma_3 = -\frac{1}{z_1\sqrt{3}M^*},
\end{align}
which again tends to infinity as $z_1 \to 0-$. The solution for $M^*$ in Eq.~(\ref{singlep3antiM}) becomes complex when $\mu_3 = \pi^2/4k$, so for $\sigma_3 \to \infty$, $\mu_d \to \pi^2/4k$, at which point $M^* = 2k/\pi$.

For $p = 4$,
\begin{align}
    \sigma_4 = -\frac{1}{z_1\sqrt{2\pi}M^{*2}},
\end{align}
which again tends to infinity as $z_1 \to 0^-$, and $M^*$ tends to a finite constant which depends on $\mu_4$.
The values for $S$ and $T$ become
\begin{align}
    S = \left(\frac{k}{2\mu_4}\left(-1 + \sqrt{1 - \frac{4\pi^3}{27k^2\mu_4}}\right)\right)^{\frac{1}{3}}, \nonumber\\
    T = \left(\frac{k}{2\mu_4}\left(-1 - \sqrt{1 -\frac{4\pi^3}{27k^2\mu_4}}\right)\right)^{\frac{1}{3}}.
\end{align}
Again $M^* = S + T$ for $\mu_4 < 0$, and 
\begin{align}
    M^* = -\frac{1}{2}\left((S + T) + i\sqrt{3}(S - T)\right)
\end{align}
for $\mu_4 > 0$. For $\mu_4 = 0$, Eq.~(\ref{singlep4Mmu0}) becomes $M^* = k/\pi$, and using Eq.~(\ref{singlep4div}) the divergence point occurs at $\mu_4 = \frac{4}{k^2}\left(\frac{\pi}{3}\right)^3$, at which point $M^* = 3k/2\pi$.

Interestingly, the values of $M^*$ and $q$ all tend to the same values for $\mu_p = 0$ for each value of $p$. We have $M^* = k/\pi$ (see Eq.~(\ref{singlep2antiM}) for the same result for $\mu_2 = 0$) and $q = k^2\pi$ for all $p$.
These results are confirmed by the phase diagrams where there is no divergence point for $\gamma_p = -1/(p-1)$, for values of $\mu_p$ below the corresponding $\mu_d$, and the divergence boundary tends to the predicted values of $\mu_d$.

\subsection{Instability point}\label{singlepinstab}

To find the specific value of $\sigma_p$ where the system becomes linearly unstable ($\sigma_c$), we consider the condition in Eq.~(\ref{crit1}) which can be simplified using Eq.~(\ref{sigtot2}) to
\begin{align}\label{Hc}
    H^2 = \frac{I_0(p-1)\sigma_\Sigma^2}{q}
\end{align}
for a single value of $p$. Substituting the expression for $H^2$ from Eq.~(\ref{q2}) the condition becomes
\begin{align}\label{singlepcrit}
    I_2 = I_0(p - 1).
\end{align}
The value of $z_1$ for which this condition is satisfied can be found numerically.
By substituting Eq.~(\ref{singlepcrit}) into Eq.~(\ref{singlepH}) and simplifying, we obtain
\begin{align}\label{Hgam}
    H = \frac{1}{\gamma_p + 1}.
\end{align}
This can be substituted into expressions for $M^*$ and $\sigma_p$ for each value of $p$ to find these values at the instability point.

We find that for $p = 2$ the value of $z_1$ which satisfies this condition is $z_c = 0$. As this implies $I_0 = I_2 = 1/2$, the expression for $\sigma_c$ can be simplified to \cite{bunin2016,gallaepl2018},
\begin{align}
    \sigma_c^2 = \frac{2}{(\gamma + 1)^2}.
\end{align}
For antisymmetric interactions, $\gamma_2 = -1$, $z_c = 0$ is the same condition as for $\sigma_2 \to \infty$, so in this case there is no instability point and we find a unique stable fixed point for all values of $\sigma_2$, for values of $\mu_2$ below the divergence boundary.

For $p = 3$ we find $z_c \approx -0.84$. As $z_c<0$ and recalling that $\sigma_3\to\infty$ for $z_1\to 0$ for fully antisymmetric interactions, the linear instability is always observed in the model with $p=3$, even when couplings are fully antisymmetric. 

However, for $\gamma_3>-1/2$ and some values $\mu_3$, we observe that abundances diverge in the system before this this point is reached (this happens when $z_m<z_c$).

For $p = 4$ we find $z_c \approx -1.3259$, which again means that even a system with fully antisymmetric interactions would become unstable before $\sigma_4 \to \infty$. For values of $\gamma_4>-1/3$ the system may not reach this point as $z_m$ might again be lower than $z_c$.

\section{Further analyis of the model with a combination of second and third order interactions}\label{combinationpmethod}
We now specify parameters $\mu_2$, $\mu_3$, $\gamma_2$, $\gamma_3$ and $z_1$, and attempt to find the macroscopic quantities $M^*$, $q$, $\chi$, and $\sigma_2$ and $\sigma_3$. The macroscopic quantities can then be described as functions of $\sigma_2$ and $\sigma_3$ in parametric form as in the previous section. We note that $z_1 \in \mathbb{R}$ cannot map onto $\{\sigma_2, \sigma_3\} \in \mathbb{R}^2$; in fact we find each value of $z_1$ to correspond to a specific relationship between $\sigma_2^2$ and $\sigma_3^2$, therefore a further constraint on $\sigma_2$ or $\sigma_3$ is required to find their numerical values. For Figs.~\ref{3dplots}, \ref{3dplots2}, \ref{phasesim23}, \ref{plotz23} and \ref{phase23} in the main paper, the constraint is one of the following conditions, either $\sigma_2 = 0$, $\sigma_3 = 0$, or $\sigma_2$ is set at a specific value. Eqs.~(\ref{M2}, \ref{q2}, \ref{X2}, \ref{H2}, \ref{phi2}, \ref{F2}, \ref{S2}, and \ref{singlepz1}) hold from the previous section, but the summed quantities in Eqs.~(\ref{sigtot2}, \ref{mutot2}, \ref{gamtot2}) become
\begin{align}\label{sigsum}
    \sigma_\Sigma^2 &= \sigma_2^2q + \frac{3}{2}\sigma_3^2q^2,\\
    \label{musum}
    \mu_\Sigma &= \mu_2 M + \mu_3 M^2,
\end{align}
and
\begin{align}\label{gamsum}
    \gamma_\Sigma = \gamma_2\sigma_2^2 + 3\gamma_3\sigma_3^2q
\end{align}
for a combination of second and third order interactions.

\subsection{Method for solving the equations}\label{p23method}

In the previous section we were able to find a simplification in Eq.~(\ref{simp}) leading to an expression for $\chi$ in Eq.~(\ref{X(z1)}). However we are unable to do this if the model contains interactions of two or more different orders. We instead find $\chi$ via Eq.~(\ref{X2}) where $H$ is to be found. We first substitute the expression for $\sigma_3$ from Eq.~(\ref{sigsum}) into Eq.~(\ref{gamsum}) to find
\begin{align}
    \gamma_\Sigma = \left(\gamma_2 - 2\gamma_3\right)\sigma_2^2 + \frac{2\gamma_3\sigma_\Sigma^2}{q}.
\end{align}
Using Eq.~(\ref{q2}) this is simplified to
\begin{align}
    \gamma_\Sigma = \left(\gamma_2 - 2\gamma_3\right)\sigma_2^2 + \frac{2\gamma_3H^2}{I_2}.
\end{align}
We substitute for $\gamma_\Sigma$ using Eq.~(\ref{H2}), and then for $\chi$ using Eq.~(\ref{X2}) to obtain the quadratic equation in $H$
\begin{align}\label{Hquad}
    \frac{H - H^2}{I_0} = \left(\gamma_2 - 2\gamma_3\right)\sigma_2^2 + \frac{2\gamma_3H^2}{I_2}.
\end{align}
We now consider the three types of constraints that are enforced on $\sigma_2$ or $\sigma_3$.
\begin{itemize}
    \item For $\sigma_2 = 0$ we find Eq.~(\ref{Hquad}) to have solution
    \begin{align}\label{p23H3}
       H = \frac{I_2}{I_2 + 2\gamma_3I_0} \left(\equiv H_3\right),
    \end{align}
    which can be found from $z_1$ and $\gamma_3$. We note that we cannot have the solution $H = 0$ as this would imply infinite $\chi$ from Eq.~(\ref{X2}), which contradicts Eq.~(\ref{Xinf}).
    \item For $\sigma_3 = 0$, we find
    \begin{align}
        \sigma_2^2 = \frac{H^2}{I_2}
    \end{align}
    by setting $\sigma_3 = 0$ in Eq.~(\ref{sigsum}) and simplifying using Eq.~(\ref{q2}). Inserting this into Eq.~(\ref{Hquad}) we find
    \begin{align}\label{p23H2}
        H = \frac{I_2}{I_2 + \gamma_2I_0} \left(\equiv H_2\right).
    \end{align}
    \item For a predetermined value of $\sigma_2$, we find $H$ via
    \begin{align}\label{p23Hnon0}
        H = \frac{I_2 \pm \sqrt{I_2^2 - 4(I_2 + 2\gamma_3I_0)(\gamma_2 - 2\gamma_3)\sigma_2^2I_0 I_2}}{2(I_2 + 2\gamma_3I_0)}.
    \end{align}
    A similar expression can be found if $\sigma_3$ we to be the given parameter instead of $\sigma_2$.
\end{itemize}
For some cases of $\gamma_2$ and $\gamma_3$, we are able to find $H$ from Eq.~(\ref{Hquad}) without enforcing a constraint on $\sigma_2$ or $\sigma_3$. For example, in the case of non-correlated interactions where $\gamma_2 = \gamma_3 = 0$ we find $H = 1$, for the case of $\gamma_2 = 2\gamma_3$ we find that all of the above solutions for $H$ reduce to the same expression,
\begin{align}
    H = \frac{I_2}{I_2 + 2\gamma_3I_0} = \frac{I_2}{I_2 + \gamma_2I_0}.
\end{align}

After we have found $H$ we follow a similar process to the previous section, by substituting the expression for $\sigma_\Sigma$ from Eq.~(\ref{M2}), and $\mu_\Sigma$ from Eq.~(\ref{musum}) into Eq.~(\ref{singlepz1}). We find the following polynomial in $M^*$
\begin{align}
    \mu_3M^{*2} + \left(\frac{z_1H}{I_1} + \mu_2\right)M^* + k = 0,
\end{align}
which has the solution
\begin{align}\label{p23M}
    M^* = \frac{-\left(\frac{z_1H}{I_1} + \mu_2\right) - \sqrt{\left(\frac{z_1H}{I_1} + \mu_2\right)^2 - 4\mu_3k}}{2\mu_3},
\end{align}
unless $\mu_3 = 0$. For $\mu_3=0$ instead, we have
\begin{align}\label{p23M0}
    M^* = \frac{k}{\frac{-z_1H}{I_1} - \mu_2}.
\end{align}
After $M^*$ is found we use Eqs.~(\ref{M2}) and (\ref{q2}) to find $\sigma_\Sigma$ and $q$ via
\begin{align}\label{sigtot3}
    \sigma_\Sigma = \frac{HM^*}{I_1},
\end{align}
and
\begin{align}\label{q3}
    q = I_2\left(\frac{M^*}{I_1}\right)^2,
\end{align}
which gives a relationship between $\sigma_2$ and $\sigma_3$ from Eq.~(\ref{sigsum}).
We now use our constraint of either $\sigma_2 = 0$, $\sigma_3 = 0$, or a chosen value for $\sigma_2$ to find the other parameter. If a value of $\sigma_2$ is chosen, it must be within the bounds
\begin{align}\label{bound}
    0 \leq \sigma_2 \leq \frac{\sigma_\Sigma}{\sqrt{q}},
\end{align}
and $\sigma_3$ is found using
\begin{align}\label{p23sig3}
    \sigma_3 = \frac{I_1}{I_2M^*}\sqrt{\frac{2}{3}\left(H^2 - I_2\sigma_2^2\right)},
\end{align}
which is from Eq.(\ref{sigsum}) simplified using Eq.(\ref{sigtot3}) and Eq.(\ref{q3}).

\subsection{Divergence point}\label{p23methodiv}

The point at which the solution for $M^*$ ceases to be both real and finite is determined by the condition
\begin{align}\label{p23divcond}
    (\mu_{2d} - \mu_2)^2 = 4\mu_{3d}k,
\end{align}
where
\begin{align}\label{p23mu2d}
    \mu_{2d} = \frac{-z_1H}{I_1},
\end{align}
from setting the discriminant to zero in Eq.~(\ref{p23M}).
Substituting Eq.~(\ref{p23mu2d}) into Eq.~(\ref{p23M}) leads to
\begin{align}\label{p23Mcase1}
    M^* = \frac{(\mu_{2d} - \mu_2) - \sqrt{(\mu_{2d} - \mu_2)^2 - 4\mu_3k}}{2\mu_3}.
\end{align}
For the case of $\mu_2 > \mu_{2d}$, the numerator of Eq.~(\ref{p23Mcase1}) is negative; as $M^*$ must be a positive value, this requires $\mu_3$ in the denominator to also be negative. It is therefore not possible for $\mu_3$ to attain the value of $\mu_{3d}$ in Eq.~(\ref{p23divcond}) and therefore this is not valid as the divergence condition. Therefore, for $\mu_2 > \mu_{2d}$, as we require $\mu_3 < 0$, in order for $M^*$ to be positive, the divergence condition becomes $\mu_{3d} = 0$.
For the case of $\mu_2 < \mu_{2d}$, Eq.~(\ref{p23divcond}) is valid for the divergence condition, substituting this into Eq.~(\ref{p23Mcase1}) leads to
\begin{align}\label{p23Mcase2}
    M^* = \frac{\sqrt{k}}{\mu_3}\left(\sqrt{\mu_{3d}} - \sqrt{\mu_{3d} - \mu_3}\right)
\end{align}
for $\mu_3 \neq 0$, and
\begin{align}
    M^* = \frac{1}{2}\sqrt{\frac{k}{\mu_{3d}}} \left(= \frac{k}{\mu_{2d} - \mu_2} \text{for $\mu_2 \leq \mu_{2d}$}\right)
\end{align}
for $\mu_3 = 0$. We note that these two cases of $\mu_2$ are continuous as for $\mu_2 = \mu_{2d}$, Eq.~(\ref{p23divcond}) becomes $\mu_{3d} = 0$, and both cases lead to
\begin{align}
    M^* = \sqrt{\frac{k}{-\mu_3}},
\end{align}
which is both real and positive for negative $\mu_3$, as $\mu_3$ is necessarily below its divergence value of $\mu_{3d} = 0$.
Combining these two cases, we find the divergence boundary is the half-parabola in Eq.~(\ref{p23divcond}) for $\mu_2 \leq \mu_{2d}$, and continues along the half-line $\mu_{3d} = 0$ for $\mu_2 \geq \mu_{2d}$. On the half-parabola, where $\mu_3 = \mu_{3d}$, $M^* = \sqrt{k/\mu_3}$ from Eq.~(\ref{p23Mcase2}), which becomes infinite where it meets the half-line at $\mu_3 = 0$.
For the case of both $\sigma_2 = \sigma_3 = 0$, and $z_1 \to -\infty$, $\mu_{2d} = 1$ using Eq.~(\ref{sig0lim}).
The divergence boundary from Eq.~(\ref{p23divcond}) becomes the half parabola
\begin{align}\label{p23div}
    \mu_2 = 1 - 2\sqrt{\mu_{3d}k},
\end{align}
valid for $\mu_2 \leq 1$, and continues along the half-line $\mu_{3d} = 0$ for $\mu_2 \geq 1$.
This can be seen at the base of the 3D phase diagrams in Figs.~\ref{3dplots} and \ref{3dplots2}. The expression for $M^*$ in Eq.~(\ref{p23Mcase1}) becomes
\begin{align}
    M^* = \frac{(1 - \mu_2) - \sqrt{(1 - \mu_2)^2 - 4\mu_3k}}{2\mu_3},
\end{align}
which holds for both cases of $\mu_2$. For $\mu_2 < 1$ where Eq.~(\ref{p23div}) is valid, this results in the same expression for $M^*$ as in Eq.~(\ref{p23Mcase2}) for $\mu_3 \neq 0$, and for $\mu_3 = 0$,
\begin{align}
    M^* = \frac{1}{2}\sqrt{\frac{k}{\mu_{3d}}} \left(= \frac{k}{1 - \mu_2} \text{for $\mu_2 \leq 1$}\right).
\end{align}
As before, Eq.~(\ref{q3}) leads to $q = M^{*2}$ for $\sigma_2 = \sigma_3 = 0$. On the divergence boundary, $M^* = \sqrt{k/\mu_3} = 2k/(1 - \mu_2)$, which remains finite on the half-parabola $\mu_2 \leq 1$, but becomes infinite where it meets the half-line at $\mu_2 = 1$, $\mu_3 = 0$.

\subsection{Anti-symmetric interactions}\label{p23methodanti}
Fully anti-symmetric interactions can be realised in different ways: when $\sigma_3=0$, we require  $\gamma_2 = -1$ and the value of $\gamma_3$ is irrelevant, for $\sigma_2 = 0$ we require $\gamma_3 = -1/2$ and the value of $\gamma_2$ is irrelevant, and when $\sigma_2$ and $\sigma_3$ are both non-zero, then we require both $\gamma_2 = -1$ and $\gamma_3 = -1/2$. In any of these cases we find using Eq.~(\ref{p23H3}) for $\sigma_2 = 0$, Eq.~(\ref{p23H2}) for $\sigma_3 = 0$, and Eq.~(\ref{p23Hnon0}) for both $\sigma_2$ and $\sigma_3$  non-zero,
\begin{align}
    H = \frac{I_2}{I_2 - I_0}.
\end{align}
As in model with a single order of fully antisymmetric interactions (Sec.~\ref{sec:singlepanti}), $H$ diverges for $z_1 = 0$, and then consequently Eq.~(\ref{singlepz1H/F}) holds. The solution for $M^*$ in Eq.~(\ref{p23M}) becomes
\begin{align}\label{p23Manti}
    M^* = \frac{(\pi - \mu_2) - \sqrt{\left(\pi - \mu_2\right)^2 - 4\mu_3k}}{2\mu_3}
\end{align}
unless $\mu_3 = 0$, in which case we have from Eq.~(\ref{p23M0})
\begin{align}
    M^* = \frac{k}{\pi - \mu_2}.
\end{align}
In either of the two cases, $M^*$ remains finite for any values of $\sigma_2$ and $\sigma_3$, as long as $\mu_2$ and $\mu_3$ are below the divergence boundary for $\sigma_2=\sigma_3=0$. At the limit $z_1 \to 0^-$ we have from Eqs.~(\ref{sigtot3}) and (\ref{sig0lim}) that
\begin{align}
    \sigma_\Sigma = \frac{-\pi M^*}{z_1},
\end{align}
which diverges as $z_1 \to 0^-$. Further, in the limit $z_1\to 0^-$ we also have [from Eq.~(\ref{q3})]
\begin{align}
    q = \pi M^{*2},
\end{align}
which remains finite below the divergence boundary.

In order for $\sigma_\Sigma$ to diverge while $q$ being finite, we require one of $\sigma_2$ or $\sigma_3$ or both to tend to infinity, see Eq.~(\ref{sigsum}). This means that there is no upper bound on $\sigma_p$ beyond which the system diverges. If one or both of $\sigma_2$ and $\sigma_3$ are infinite, i.e., when $\sigma_\Sigma$ becomes infinite, the divergence boundary depends on $\mu_2$ and $\mu_3$. By setting the discriminant to zero in Eq.(\ref{p23Manti}), we find for this limit the divergence boundary becomes
\begin{align}
    (\pi - \mu_2)^2 = 4\mu_{3d}k
\end{align}
for $\mu_2 \leq \pi$, and $\mu_3 = 0$ for $\mu_2 > \pi$. Comparing with Eq.~(\ref{p23div}) (combined with further numerical study) indicates that the divergence region becomes smaller with increasing $\sigma_\Sigma$. On the divergence boundary $M^* = \sqrt{k/\mu_3} = 2k/(\pi - \mu_2)$ remains finite on the half-parabola but becomes infinite when the boundary meets the line segment at $\mu_2 = \pi$ and $\mu_3 = 0$.
\subsection{Instability point}\label{p23methodcrit}
To find the critical values for $\sigma_2$ and $\sigma_3$ where the unique fixed point becomes unstable, we consider the instability point condition in Eq.~(\ref{crit1}) which becomes
\begin{align}\label{critt}
    \frac{H^2}{I_0} = \sigma_2^2 + 3\sigma_3^2q,
\end{align}
for the combination of second and third order interactions. We substitute the expression for $\sigma_\Sigma$ in Eq.~(\ref{sigtot3}) and the expression for $q$ in Eq.~(\ref{q3}) into Eq.~(\ref{sigsum}) to obtain
\begin{align}\label{sigtott}
    H^2 = I_2\sigma_2^2 + \frac{3}{2}\frac{I_2^2M^2}{I_1^2}\sigma_3^2,
\end{align}
and substitute for $H^2$ from Eq.~(\ref{critt}) to find
\begin{align}\label{crit23}
    3\left(I_0 - \frac{I_2}{2}\right)\frac{I_2M^2}{I_1^2}\sigma_3^2 = (I_2 - I_0)\sigma_2^2.
\end{align}
We now consider the three types of constraints that are enforced on $\sigma_2$ or $\sigma_3$.
\begin{itemize}
    \item For $\sigma_2 = 0$, Eq.~(\ref{crit23}) reduces to the condition $2I_0 = I_2$, which is the instability condition for third-order interactions only, from setting $p = 3$ in Eq.~(\ref{singlepcrit}). As before, this condition is satisfied at $z_c \approx -0.84$. In this case, we find the solution for $H$ to be the same as for the model with third-order interactions only as in Eq.~(\ref{Hgam}), and $\sigma_c$ to satisfy the same expression as in Eq.~(\ref{singlepsig}), but the value of $M^*$ also depends on $\mu_2$.
    \item For $\sigma_3 = 0$, Eq.~(\ref{crit23}) reduces to the condition $I_0 = I_2$, which is the instability condition for second-order interactions only, from setting $p = 2$ in Eq.~(\ref{singlepcrit}). As before, this condition is satisfied at $z_c = 0$. In this case, we find the solution for $H$ in Eq.~(\ref{Hgam}) and $\sigma_c$ in Eq.~(\ref{singlepsig}) to be the same as for the model with second-order interactions only.
    \item For a chosen non-zero value of $\sigma_2$, we substitute the expression for $\sigma_3$ from Eq.~(\ref{sigtott}) into Eq.~(\ref{crit23}) to obtain
    \begin{align}\label{p23instab}
        \sigma_2^2 = H^2\left(\frac{2}{I_2} - \frac{1}{I_0}\right),
    \end{align}
    substituting this expression for $\sigma_2^2$ back into Eq.~(\ref{Hquad}) and rearranging for $H$ leads to a formula for finding $H$ along the instability transition boundary,
    \begin{align}\label{p23Hc}
        H = \frac{I_2}{I_2 + \gamma_2I_0 + (I_0 - I_2)(\gamma_2 - 2\gamma_3)} \left(\equiv H_c\right).
    \end{align}
    This can be used to find the instability condition for any specified value of $\sigma_2$ within the bounds given in Eq.~(\ref{bound}). We find that the critical value of $z_1$ is within the range $-0.84 \lesssim z_c \leq 0$ depending on the value of $\sigma_2$.
\end{itemize}

\section{Simulations}\label{smsims}

\subsection{Finite-size considerations}

The coloured spots in the phase diagrams in Figs.~\ref{phasesim2}, \ref{phasesim3}, \ref{phasesim4}, and \ref{phasesim23} show the average behaviour of 20 simulations run for a maximum of 10000 units of time, being stopped before then if the system reached a fixed point. A simulation of the system of $N$ species with order-$p$ interactions required the memory to store $N{N-1 \choose p-1}$ non-zero interaction coefficients and the computational power to perform tensor multiplication between the interaction coefficients and a vector of the $N$ population sizes. Therefore increasing the order of interactions decreases the maximum number of species the computer is able to cope with. The simulations were run with 500 species for second order interactions, 200 species for third order interactions, and 50 species for fourth order interactions. A combination of second and third order interactions requires roughly the same amount of computational power as having third order interactions only, so 200 species were also used for the combination of second and third order.

The random interaction coefficients have to be scaled with the number of species, so that after interacting with all the other species present, the total effect of interactions [the last term in Eq.~(\ref{lv})] has the correct mean variance and covariance. In previous work on second-order interactions, the interaction coefficients are scaled with $N$, even though there are $N-1$ other species present. This difference of one has little effect when the number of species is large, which is reasonable in simulations for second-order interactions. However, as the order of interactions is increased the corresponding difference increases while the number of species able to be simulated decreases. For example, for fourth order interactions, the coefficients would scale with $N^3$ in the limit of $N \to \infty$ but each species interacts with $(N-1)(N-2)(N-3)$ sets of other species. For $50$ species this difference is much bigger (125,000 vs. 110,544) and has a noticeable effect on the simulation results. During an integration step for fourth-order interactions, each species would receive the same payoff 6 times over, from interacting with the same set of 3 other species 6 times, i.e. $\alpha_{ijkl} = \alpha_{ijlk} = \alpha_{ikjl} = \cdots$. To simplify the simulation, 5 of these coefficients were set to zero, and the one non-zero coefficient was increased by a factor of 6 [see the factor $p!$ in the first relation in Eq.~(\ref{eq:alpha})]. These two contributions leads of the overall scaling with $N-1 \choose p-1$, the number of distinct sets of other species it can interact with, and the number of different payoffs it receives.

\subsection{Determination of behaviour}

\begin{figure}[b]
    \centering
    \includegraphics[width = 0.75\textwidth]{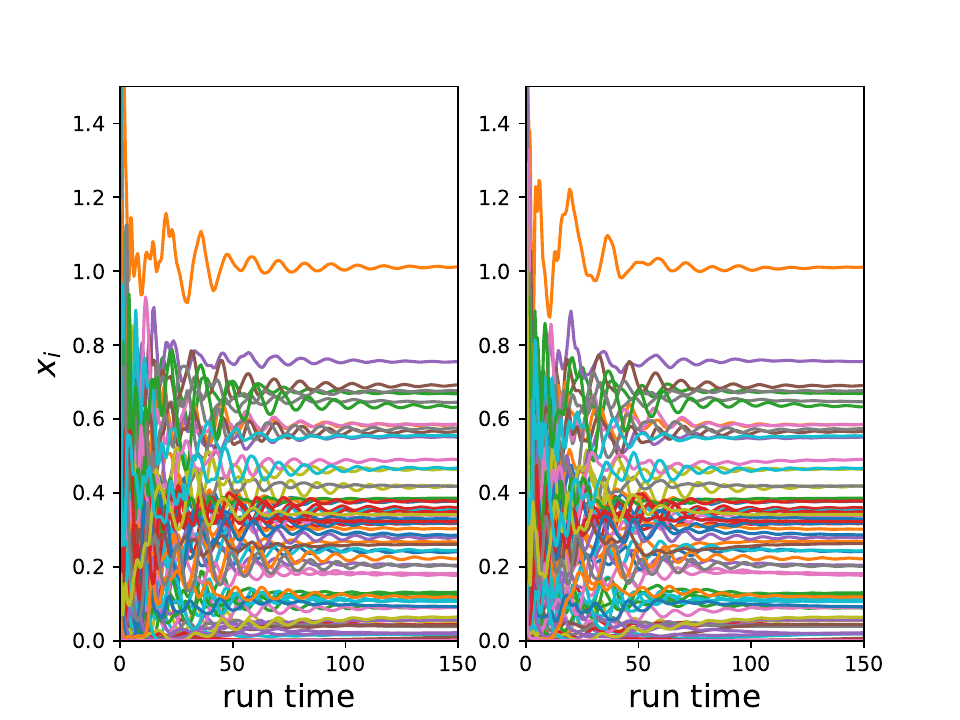}
    \caption{Trajectories of two realisations from different initial states of the same set of second and third order interaction coefficients (matrix elements) with $\gamma_2 = -1$, $\gamma_3 = -0.5$, $\mu_2 = -2$, $\mu_3 = -4$, $\sigma_2 = 5$, $\sigma_3 = 7$. The system displays decaying oscillations towards the unique fixed point.}
    \label{traj23_red}
\end{figure}

\begin{figure}
    \centering
    \includegraphics[width = 0.75\textwidth]{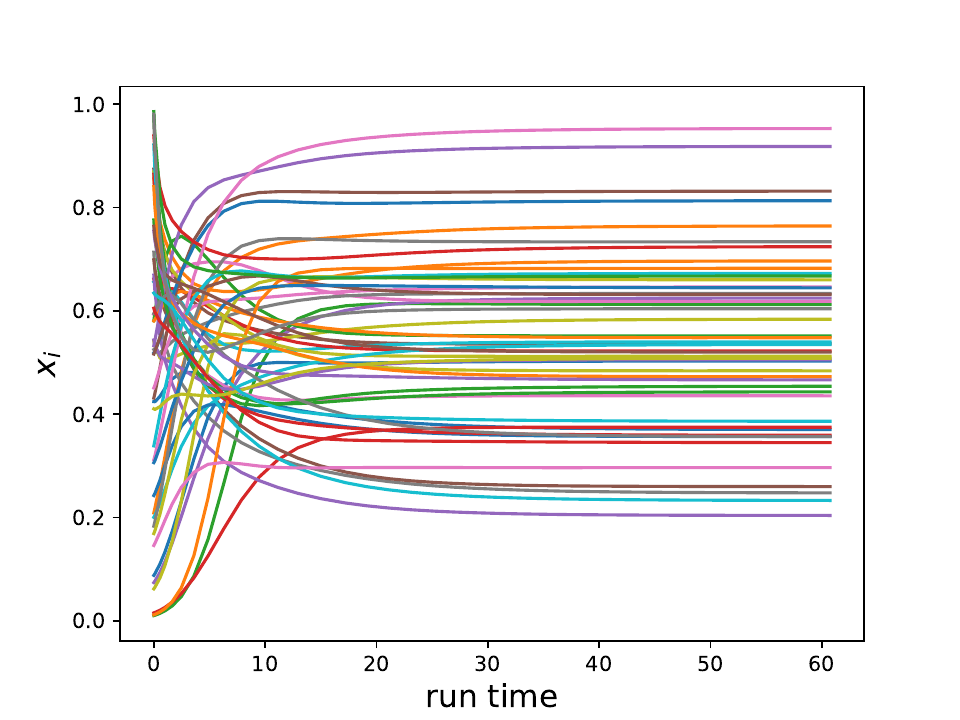}
    \caption{Trajectories for the system with fourth order interactions with parameters $\gamma_4 = 1$, $\mu_4 = -4$, $\sigma_4 = 10^{-0.3}$. The system attained the fixed point almost immediately.}
    \label{traj4_red}
\end{figure}

\begin{figure}
    \centering
    \includegraphics[width = 0.75\textwidth]{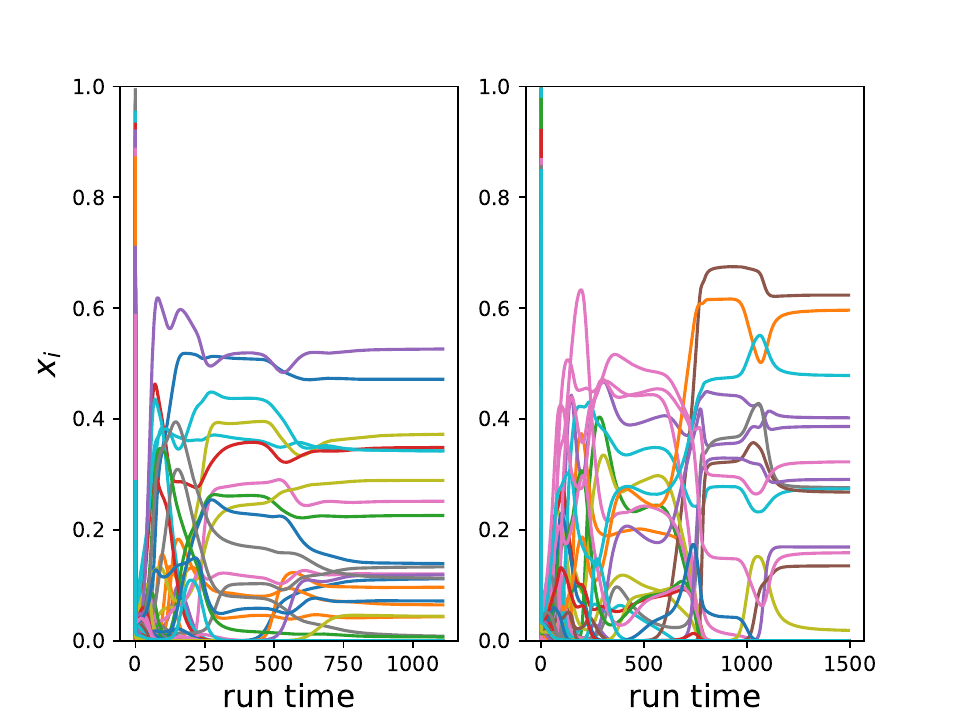}
    \caption{Trajectories of two realisations from different initial states of the same set of second and third order interaction coefficients with $\gamma_2 = 1$, $\gamma_3 = 0$, $\mu_2 = -55$, $\mu_3 = -15$, $\sigma_2 = 2$, $\sigma_3 = 10$. The system displays periods with many species remaining at constant abundances, where it is close to one of many fixed points. It moves gradually between the vicinities of fixed points until it remains at one of them for enough time to be classified as fixed. The simulation was ended after this but the system could eventually leave this point. The two realisations settled at different fixed points.}
    \label{traj23_green}
\end{figure}

\begin{figure}
    \centering
    \includegraphics[width = 0.75\textwidth]{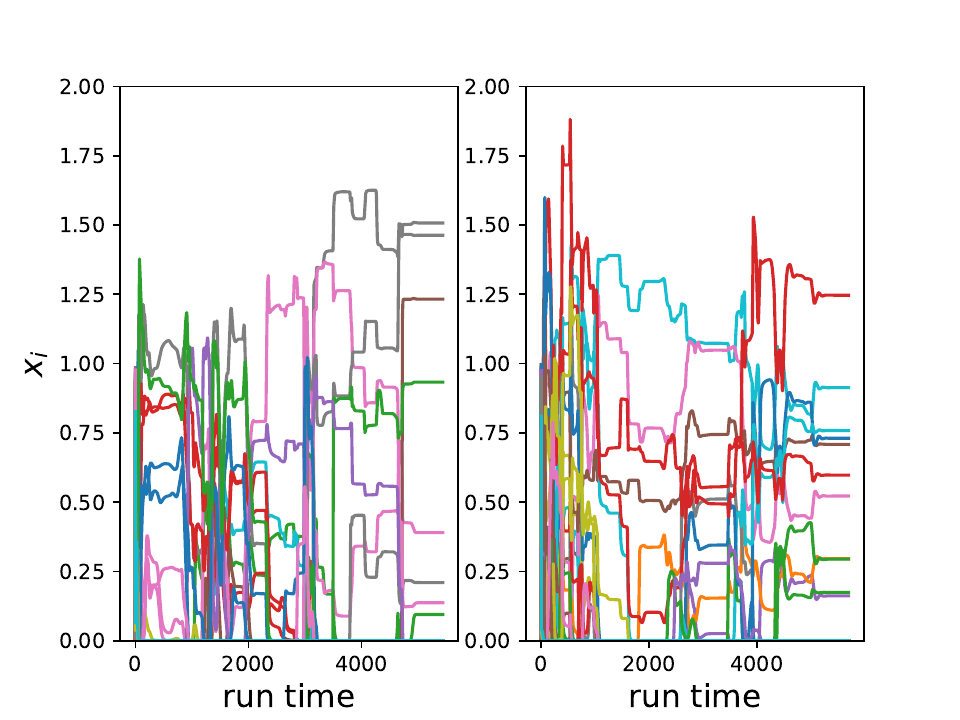}
    \caption{Trajectories for the system with second order interactions with parameters $\gamma_2 = 1$, $\mu_2 = -2.5$, $\sigma_2 = 10^{0.5}$. The systems move suddenly between fixed points, where many species remain at constant abundance for long periods of time, the two realisations eventually settled at different fixed points.}
    \label{traj2_green}
\end{figure}

During the simulation, if the abundance of each species had varied by less than 0.01 over at least 500 units of time, then the simulation was ended and classified as having reached a fixed point. In this case, another simulation was run with the same random interaction matrix, but a different randomly drawn initial set of species abundances, and if this also reached a fixed point, and the abundance of each species was within 0.01 of the same species from the previous run, the fixed point was classified as being unique. If there are multiple possible fixed points, it is assumed to be unlikely that two realisations that began in different independent places would reach the same fixed point. The difference in the behaviour of systems with a unique fixed point and multiple fixed points can be seen clearly in their trajectories. Examples are shown in Figs.~\ref{traj23_red} and \ref{traj4_red} (unique fixed point), \ref{traj23_green} and \ref{traj2_green} (multiple fixed points).

If the system has a globally stable unique fixed point, the abundances vary continuously until they come close to the fixed point values, after which they may converge gradually or display decaying oscillations towards them. However, the trajectories of a system with multiple fixed points will visit the vicinity of many points until it finally reaches one of them and is able to remain constant for at least 500 units of time. The parameters used as thresholds to classify behaviour, like the requirement of remaining fixed for at least 500 units of time, and staying within 0.01 of the same value, were chosen by plotting multiple trajectories ensuring the classification matched the displayed behaviour.

\begin{figure}
    \centering
    \includegraphics[width = 0.75\textwidth]{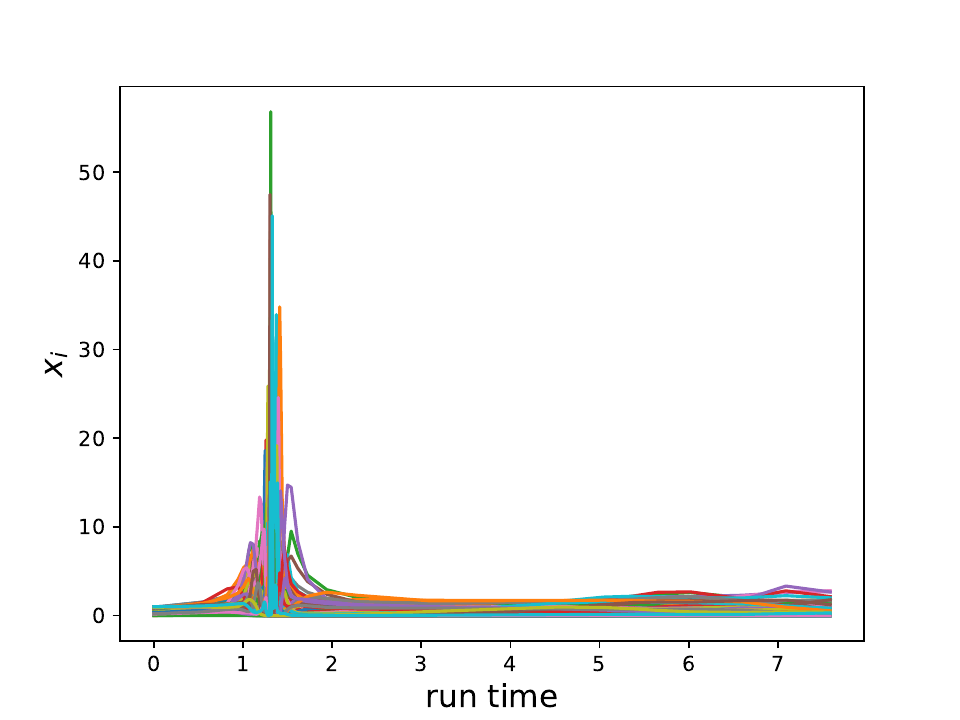}
    \caption{Trajectories for the system with fourth order interactions with parameters $\gamma_4 = -1$, $\mu_4 = 2$, $\sigma_4 = 10^{0.2}$. The system displays the beginnings of unbounded growth, but did not continue to grow.}
    \label{traj4_spike}
\end{figure}

\begin{figure}
    \centering
    \includegraphics[width = 0.75\textwidth]{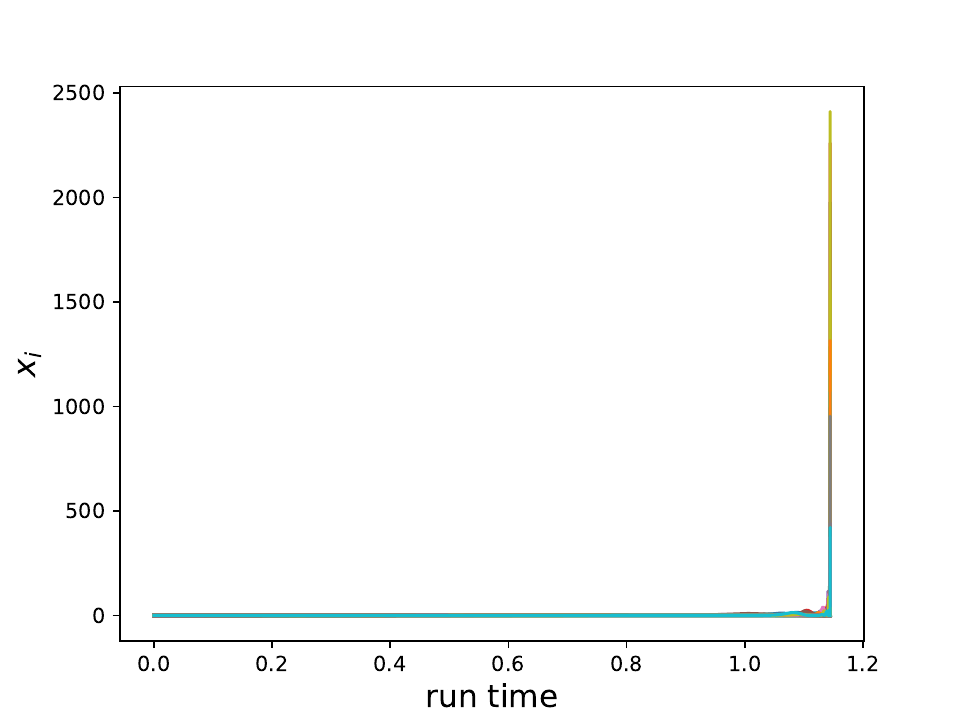}
    \caption{Trajectories for the system with fourth order interactions with parameters $\gamma_4 = -1$, $\mu_4 = 2$, $\sigma_4 = 10^{0.2}$. This time the abundances continued to grow without stopping.}
    \label{traj4_white}
\end{figure}

Another behaviour that had to be identified was unbounded growth of the species abundances. A simulation run was classified as divergent if the abundance of any of the species grew to above $10^6$. This behaviour was observed for simulations integrated with a finite time step, however it was efficient to carry out the computationally demanding integration of the equations with an adaptive time step that takes large time steps during periods of little movement, and much smaller time steps when the abundances are changing quickly. For systems that displayed unbounded growth, the asymptotic nature of the growth meant that the time steps became increasingly small, so small that it could sometimes result in the dynamics of the system being captured incorrectly (as non divergent). This effect was more pronounced when simulating interaction with higher orders, and therefore a lower number of species. 
This behaviour can again been seen clearly from plotting the trajectories, an example is shown in Fig.~\ref{traj4_spike} where the dynamics initially grow increasingly quickly, but then suddenly stop and  return to much smaller values, whereas Fig.~\ref{traj4_white} shows another realisation of the same parameters where the growth continues indefinitely. As the number of species was increased, this unbounded growth became less likely to stop suddenly and instead grew to $10^6$. For second and third order interactions, simulations with a high enough number of species were able to be run to reduce this effect but fourth-order interactions were limited to 50 species. Therefore to account for this effect an additional condition was used to identify divergent behaviour: if the value of $\dot{x}_i$ was found to be above 100 for any species within the first 2 units of time then this was classified as divergence.

\begin{figure}
    \centering
    \includegraphics[width = 0.75\textwidth]{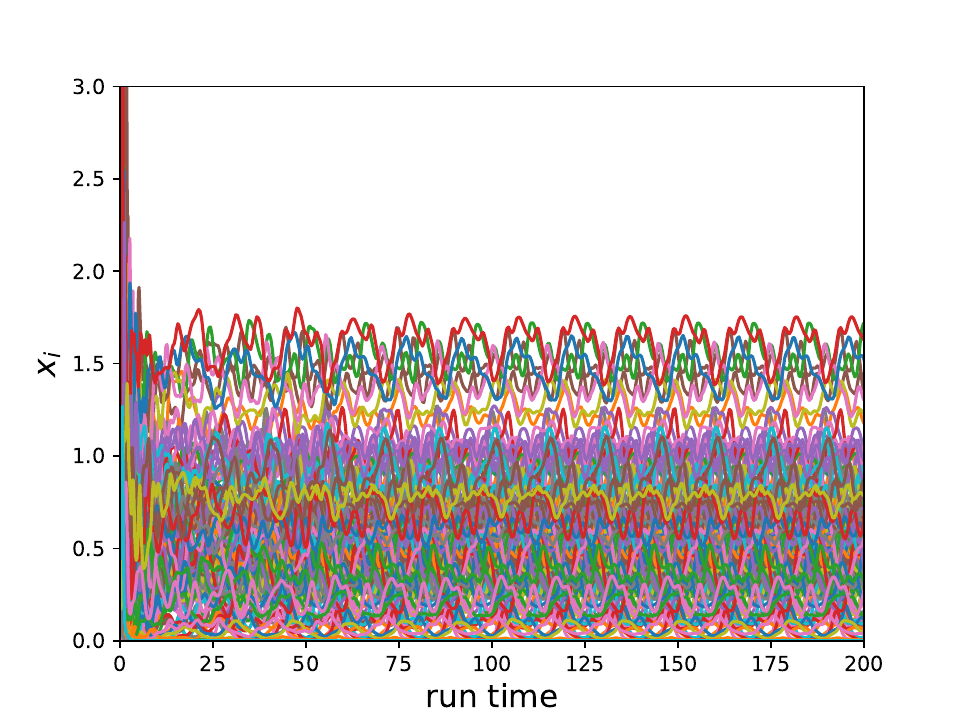}
    \caption{Trajectories for the systems with second and third order interactions with $\gamma_2 = -1$, $\gamma_3 = -0.5$, $\mu_2 = -25$, $\mu_3 = 4$, $\sigma_2 = 4$, $\sigma_3 = 3.6$. The system displays a periodic cycle.}
    \label{traj23_cycle}
\end{figure}

\begin{figure}
    \centering
    \includegraphics[width = 0.75\textwidth]{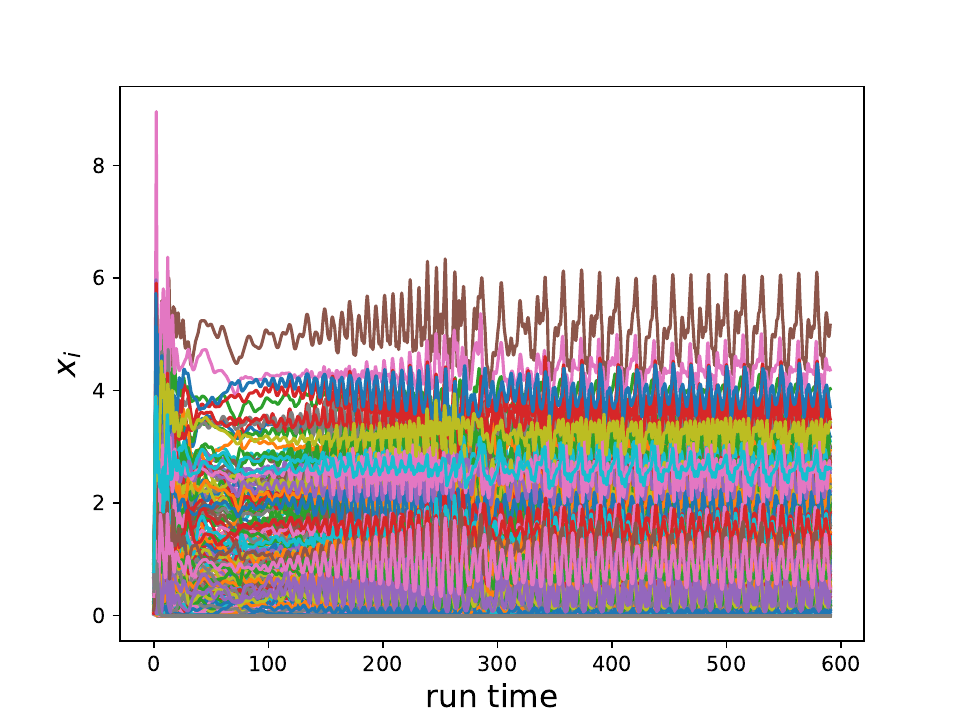}
    \caption{Trajectories for the system with third order interactions with $\gamma_3 = -0.5$, $\mu_3 = 1$, $\sigma_3 = 10$. After a transient the system settles to a quasi-periodic cycle or potentially chaotic attractor.}
    \label{traj3_blue}
\end{figure}

\begin{figure}
    \centering
    \includegraphics[width = 0.75\textwidth]{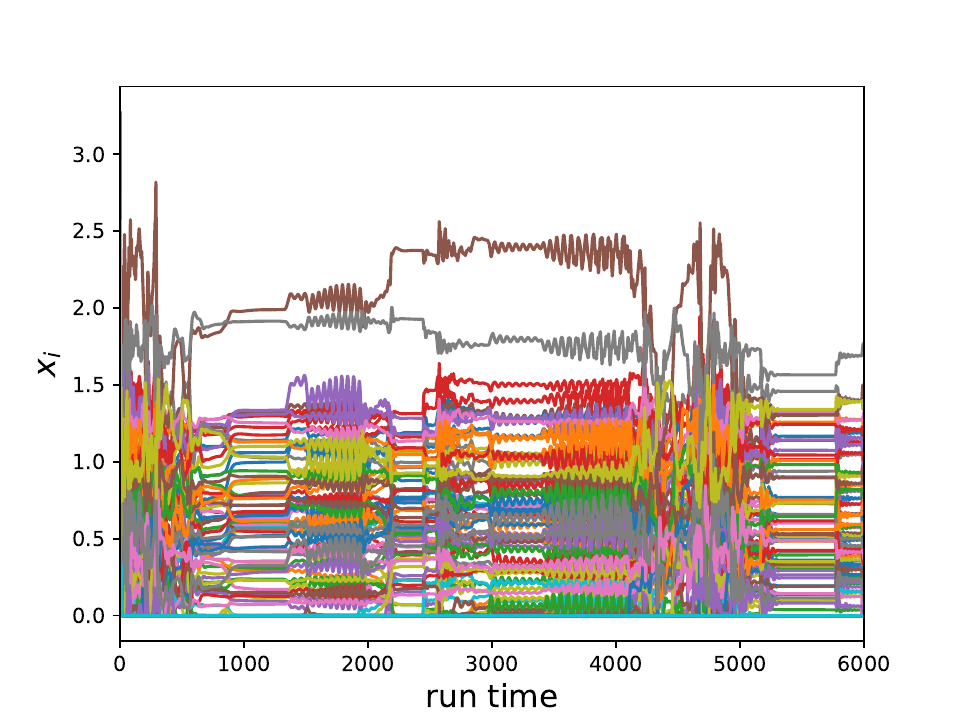}
    \caption{Trajectories of the systems with second and third order interactions with $\gamma_2 = -1$, $\gamma_3 = -0.5$, $\mu_2 = -25$, $\mu_3 = 4$, $\sigma_2 = 15$, $\sigma_3 = 2$. The system displays changing intervals of seemingly periodic, constant, and chaotic behaviour.}
    \label{traj23_blue}
\end{figure}

\begin{figure}
    \centering
    \includegraphics[width = 0.75\textwidth]{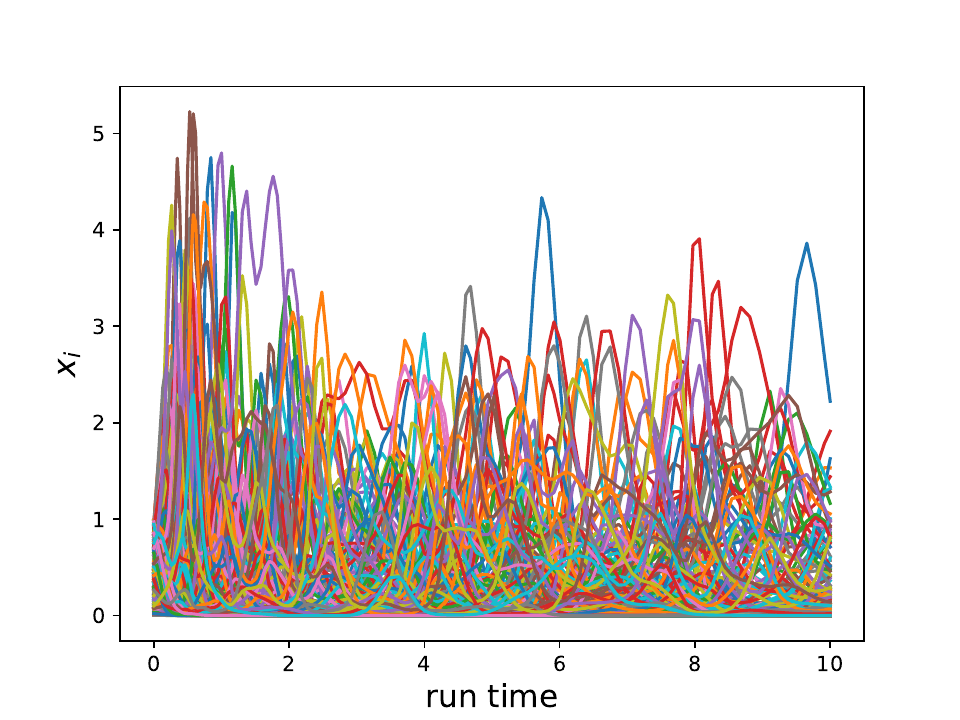}
    \caption{Trajectories for the system with third order interactions with $\gamma_3 = -0.5$, $\mu_3 = 1$, $\sigma_3 = 100$. With a large variance of interaction coefficients, the system displays highly irregular behaviour.}
    \label{traj3_chaos}
\end{figure}

If neither of the above conditions were met, then the system was classified as displaying persistent dynamics, which was found to sometimes consists of periodic cycles (Fig.~\ref{traj23_cycle}), quasi-periodic or possibly chaotic attractors (Fig.~\ref{traj3_blue}), or some combination of multiple behaviours (Fig.~\ref{traj23_blue}), or other irregular behaviour (Fig.~\ref{traj3_chaos}).

\begin{figure}[H]
    \centering
    \includegraphics[width = 0.75\textwidth]{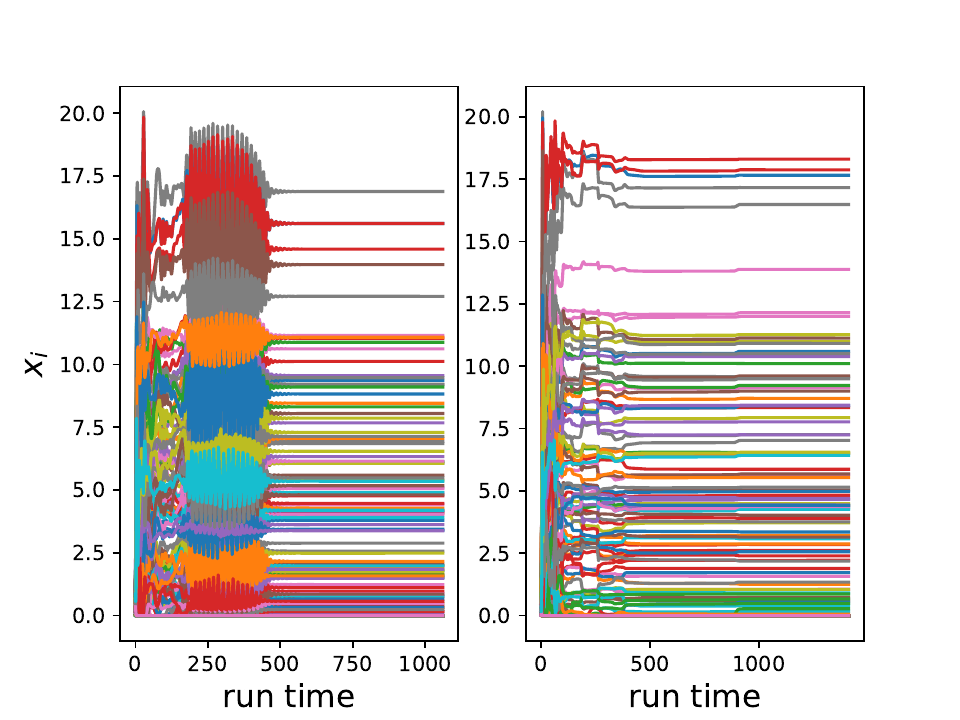}
    \caption{Trajectories of two realisations of the system with second order interactions with with the same interaction coefficients, but started from different initial states. Parameters are $\gamma_2 = 0$, $\mu_2 = -1$, $\sigma_2 = 10^{0.2}$. The system displays a long period of transient behaviour until it eventually finds a fixed point, the second run found a different fixed point in a shorter time.}
    \label{traj2_blue}
\end{figure}

\begin{figure}
    \centering
    \includegraphics[width = 0.75\textwidth]{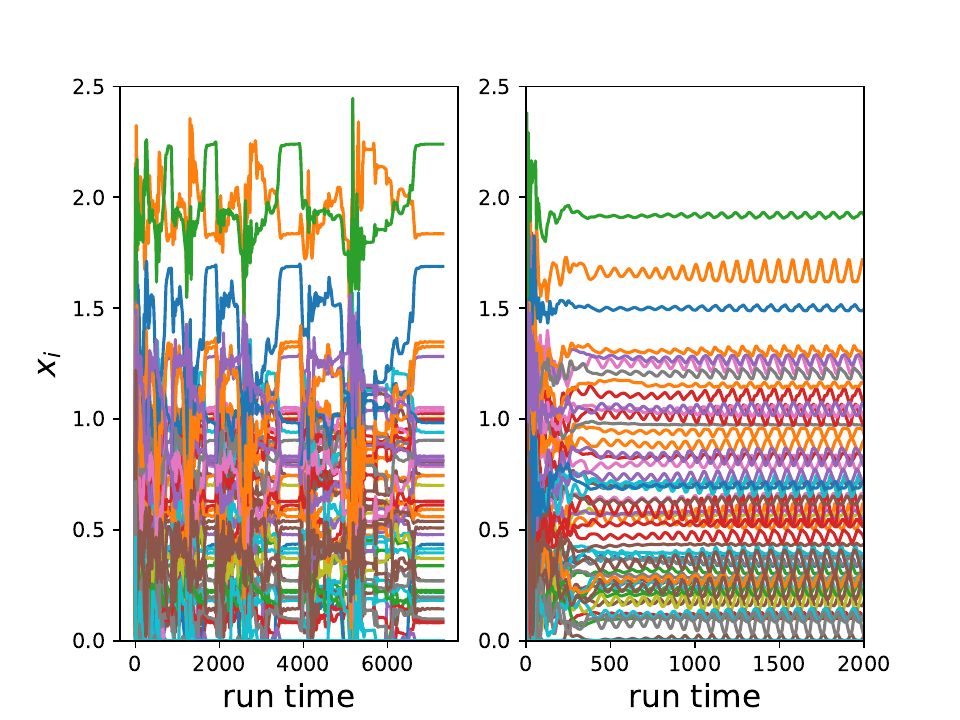}
    \caption{Trajectories of two realisations from different initial states of the system with second order interactions with the same interaction coefficients. Parameters are $\gamma_2 = -0.5$, $\mu_2 = -4$, $\sigma_2 = 10^{0.5}$. The first run alternates between transient behaviour and spending short periods close to a unique fixed point, each time remaining there for longer, unit it eventually stays there long enough to be classed as fixed. It could potentially move away from this point if the simulation had been continued. The second run has a much shorter period of transience before it comes very close to the same fixed point, it then oscillates about the point with initially very small oscillations, but these oscillations grow and eventually reach a limit cycle. The fixed points reached by the two realisations are close to each other, as these parameters are close to the unique fixed point phase. The same colour corresponds to the same species, and the sequence of colours is similar in both panels.}
    \label{traj2_cycle}
\end{figure}

For parameters close to the transition boundary, the transient phase can last for a long period of time until it starts to approach any fixed point(s), an example is shown in Fig.~\ref{traj2_blue}. In some cases the transient phase could last longer than the length of the simulation, so no fixed point is observed and the behaviour would have been classified differently if the simulation were to be run for longer. In some cases, the system can display  small oscillations about a unique fixed  (Fig.~\ref{traj2_cycle}), and if the oscillations stay within a range 0.01, this could be classified as reaching a fixed point even though the oscillations continue indefinitely. This can explain why the colours are mixed close to transition boundaries in the phase diagrams.

\subsection{Colour in phase diagrams}\label{sec:colour}
After the 20 simulations were run for each position in the phase diagram, the behaviour was determined and classified, and the number of each type was counted. They were classified as either having a unique fixed point ($N_U$), multiple fixed points ($N_M$), persistent dynamics ($N_P$), or divergence of abundance ($N_D$). These numbers were then converted to a RGB colour code with
\begin{align}
    R &= \frac{N_U + N_D}{20} \nonumber \\
    G &= \frac{N_M + N_D}{20} \nonumber \\
    B &= \frac{N_P + N_D}{20}.
\end{align}
By doing this, the colour in the phase diagram represents the mixture and proportions of behaviour observed for the given parameters.

\end{document}